\documentclass[11pt,a4paper]{article}

\usepackage{microtype}
\usepackage{amsmath, amsfonts,amssymb,mathtools,nicefrac,nccmath,cases,physics}
\usepackage{bm}
\usepackage{soul}
\usepackage{dsfont}
\usepackage{booktabs,multirow}
\usepackage{tabularx}
\usepackage[usenames,dvipsnames,table]{xcolor}
\usepackage{colortbl}
\usepackage{caption}
\usepackage{subcaption}
\usepackage{float}
\usepackage{hhline}
\usepackage{enumitem}
\usepackage{cite}
\numberwithin{equation}{section}
\numberwithin{table}{section}
\usepackage[linktocpage=true]{hyperref}
\hypersetup{colorlinks=true,linkcolor=colorloc3,citecolor=colorloc3,urlcolor=colorloc3}
\usepackage{afterpage}

\def\hybrid{\topmargin -20pt    \oddsidemargin 0pt
	\headheight 0pt \headsep 0pt
	\textwidth 6.5in        
	\textheight 9in         
	\textwidth 6.25in       
	\textheight 9.1 in       
	\marginparwidth .875in
	\parskip 5pt plus 1pt 
	\jot = 1.5ex
}
\hybrid

\definecolor{colorloc1}{RGB}{164,42,46} 
\definecolor{colorloc2}{RGB}{100,100,100} 
\definecolor{colorloc3}{RGB}{204,119,34}  
\definecolor{colorloc4}{RGB}{25,25,112}  
\definecolor{colorloc5}{RGB}{0,128,128}  

\usepackage{sectsty}

\sectionfont{\color{colorloc4}\Large} 

\subsectionfont{\color{colorloc2}\large} 

\subsubsectionfont{\color{colorloc5}} 

\usepackage[linktocpage=true]{hyperref}
\hypersetup{colorlinks=true,linkcolor=colorloc5,citecolor=colorloc4,urlcolor=colorloc4}

\usepackage[framemethod=default]{mdframed}

\newmdenv[skipabove=10pt,
skipbelow=7pt,
rightline=false,
leftline=true,
topline=false,
bottomline=false,
linecolor=colorloc4,
backgroundcolor=colorloc2!5,
innerleftmargin=4pt,
innerrightmargin=0pt,
innertopmargin=0pt,
leftmargin=2pt,
rightmargin=0pt,
linewidth=2pt,
innerbottommargin=0pt,
frametitlebackgroundcolor=colorloc2]{lbBox}
\newenvironment{importantbox}{\begin{lbBox}\vspace{3 mm}
	} {\vspace{1.5 mm}\end{lbBox}}

\newmdenv[skipabove=10pt,
skipbelow=7pt,
rightline=false,
leftline=true,
topline=false,
bottomline=false,
linecolor=colorloc2,
backgroundcolor=colorloc2!5,
innerleftmargin=4pt,
innerrightmargin=4pt,
innertopmargin=0pt,
leftmargin=0pt,
rightmargin=0pt,
linewidth=2pt,
innerbottommargin=0pt,
frametitlebackgroundcolor=colorloc2]{lcBox}

\usepackage{fancyhdr}
\def\Re           {{\rm Re\hskip0.1em}}
\def\Im           {{\rm Im\hskip0.1em}}

\newcommand{\im}{\mathbf{i}}
\newcommand{\mpl}{M_{\rm P}}

\begin{document}


	\baselineskip=14pt
	\parskip 5pt plus 1pt

	\vspace*{-1.5cm}
	\begin{flushright}    
		{\small 
				DESY-24-199
			}
	\end{flushright}
	
	\vspace{1.5cm}
	\begin{center}        

		{\color{colorloc4} \bf \huge Uplifts in the Penumbra:\\[0.35cm]
        Features of the Moduli Potential away from\\[0.5cm]
        Infinite-Distance Boundaries}   
	\end{center}
	
	\vspace{0.5cm}
	\begin{center}        
		{\large \bf Stefano Lanza}${}^a$\footnote{E-mail:
    stefano.lanza@desy.de} and {\large \bf Alexander Westphal}${}^b$\footnote{E-mail:
    alexander.westphal@desy.de}
        
	\vspace{0.2cm}

	\emph{${}^a$II. Institut f\"ur Theoretische Physik, Universit\"at Hamburg,\\
	Luruper Chaussee 149, 22607 Hamburg, Germany}
	\\
    \vspace{0.2cm}
    \emph{${}^b$Deutsches Elektronen-Synchrotron DESY, Notkestr. 85, 22607 Hamburg, Germany}
	\end{center}
	
	\vspace{0.5cm}
	
	
	\begin{abstract}
		\noindent The construction of meta-stable four-dimensional de Sitter vacua in type IIB string compactifications represents an important question and an ongoing area of work. There is considerable support both for stringy de Sitter vacua in the interior of moduli space and for their scarceness in the strict asymptotic regime towards infinite-distance boundaries of the compactification moduli space. Here, we present evidence for the existence of uplifting vacua in the three-form flux-induced scalar potential of the complex structure moduli of type IIB string theory on Calabi-Yau orientifolds in the cross-over region between the interior of the moduli space and its strictly asymptotic infinite-distance regions. Moreover, we also exhibit the existence of long-range axion valleys which, while not yet supporting slow-roll inflation, do show a flattened scalar potential from complex structure moduli backreaction and axion monodromy. We further illustrate how such regions hosting axion valleys may be obtained by using machine learning techniques. 
	\end{abstract}

	\thispagestyle{empty}
	\clearpage
	
	\setcounter{page}{1}
	
	
\newpage

\tableofcontents

\newpage

\section{Introduction}
\label{sec:Introduction}

Cosmological inflation remains so far the most compelling mechanism accessible in effective quantum field theory to describe an epoch of extremely fast accelerated expansion of the very early universe. If present, inflation solves various initial condition problems of the hot big bang and provides a mechanism for generating the primordial nearly scale-invariant spectrum of curvature perturbations needed for structure formation.

Similarly, the observed late-time accelerated expansion of the universe is most simply described in terms of a positive yet tiny cosmological constant producing an asymptotic de Sitter (dS) state of the universe to the future.

The simplest mechanism to realize inflation consists of a scalar field with a positive scalar potential satisfying the slow-roll conditions ensuring a regime of slow-roll: The two slow-roll conditions ensure sustained dominance of the scalar potential over the kinetic energy of the slowly-rolling inflaton scalar field. The near-dS quantum vacuum of slow-roll inflation generates a nearly scale-invariant spectrum of primordial gravitational waves, the tensor modes, besides the scalar curvature perturbation. It turns out, that the relative strength of these tensor modes (the tensor-to-scalar ratio $r$) is tied to the field range $\Delta\phi$ which the inflaton scalar traverses during the observable 50...60 e-folds of inflationary expansion. Our current level of technological ability to detect these tensor modes as B-mode polarization patterns in the cosmic microwave background (CMB) radiation implies a reach of $r\gtrsim 10^{-3}$. This corresponds to an inflationary field range $\Delta\phi\gtrsim \mpl$. 

It is a curious coincidence that such a  field range of $\Delta\phi\sim \mpl$ correlates with the change from the need for suppression of dimension-six operators to suppression of an infinite series of higher-dimension operators as a condition to maintain smallness of the slow-roll parameters against quantum corrections induced by heavy states. Hence, $\Delta\phi\sim \mpl$ separates large-field models of inflation ($\Delta\phi\gtrsim \mpl$) producing detectable tensor modes and needing suppression of an infinite series of corrections from small-field models ($\Delta\phi < \mpl$) which do and need neither of those.

Clearly, realizing large-field inflation models is of special interest as the detectable tensor-mode signal of such models offers a strong test of the slow-roll inflation paradigm. The required suppression of an infinite series of corrections has been argued to arise from the scalar field being axion-like, possessing a shift symmetry which is softly broken predominantly by the inflationary scalar potential itself. 

However, the technical naturalness of such an assumption rests on the quality of the underlying source of the shift symmetry as well as the suppression of the axion-inflaton couplings to heavy states in the theory. Realizing such an axion with an approximate shift symmetry and controlled sources of soft breaking generating the large-field scalar potential is possible in string theory compactifications using the mechanism of axion monodromy~\cite{Silverstein:2008sg,McAllister:2008hb}. The protection of the shift symmetry there arises from underlying gauge symmetries, which were discovered in~\cite{Kaloper:2008fb,Kaloper:2011jz} to allow for a fully four-dimensional effective description axion monodromy inflation in terms of the spontaneously broken gauge symmetries of an axion coupled to an effective four-form field strength. For a review of this idea and much subsequent related work see, for instance,~\cite{Cicoli:2023opf}.

Already the initial string theory models of axion monodromy revealed that energetic backreaction~\cite{Flauger:2009ab,Dong:2010in,Conlon:2011qp} from the finitely massive moduli of the stabilized extra dimensions as well as from energy connected to the monodromy charge accrued by an axion displaced over multiple axion periods generically lead to a flattening of the large-field axion scalar potential. This effect is present in both ten-dimensional string models~\cite{Dong:2010in,Blumenhagen:2014nba,McAllister:2014mpa,Hebecker:2014kva,Baume:2016psm,Landete:2016cix,Valenzuela:2016yny,Blumenhagen:2017cxt,Landete:2017amp,Kim:2018vgz,Buratti:2018xjt,Grimm:2020ouv} and in the effective four-dimensional descriptions~\cite{Kaloper:2011jz,Kaloper:2015jcz,Kaloper:2016fbr,DAmico:2017cda}.

The underlying starting point for models of inflation in critical string theory is the realization of string vacua describing meta-stable dS space with small positive cosmological constant and stabilized moduli. Following the path-breaking initial mechanisms in~\cite{Kachru:2003aw,Saltman:2004jh,Balasubramanian:2005zx}, considerable evidence was assembled over many works, as recently reviewed in~\cite{Cicoli:2023opf}, that moduli stabilization produces AdS vacua from a two-term structure with a third `slower' term providing an uplift to positive vacuum energy.

Again, similar to the situation for large-field axion monodromy inflation, the uplift part of string constructions of meta-stable dS vacua shows a certain susceptibility to backreaction effects from moduli stabilization and the field energy connected to charges tied to localized uplift sources such as anti-D3-branes. In this context, it is relevant to note the increasing evidence for uplift sources `distributed in the extra dimensions' from F-term SUSY breaking minima of the flux-induced complex structure moduli scalar potential of type IIB Calabi-Yau orientifold compactifications~\cite{Gallego:2017dvd,Krippendorf:2023idy} based on the initial idea in~\cite{Saltman:2004sn}. Similar F-term breaking uplifts arise from the quasi-axion partners of the complex structure moduli under the combined effects of three-form fluxes and instanton corrections~\cite{Hebecker:2020ejb,Carta:2021sms}.

The susceptibility of both string inflation mechanisms and the uplift part of stringy dS construction to backreaction effects has led to the formulation of two conjectures, that are now part of the so called \emph{Swampland program}. 
Such a program ambitiously tries to infer which lower-dimensional effective field theories (EFTs) admit an ultraviolet completion within string theory, by formulating a set of criteria, the \emph{Swampland conjectures}, that the effective theories need to obey.

Among the proposed conjectures within the program, are the \emph{Distance conjecture}~\cite{Ooguri:2006in}, and the \emph{de Sitter conjecture}~\cite{Garg:2018reu,Ooguri:2018wrx,Hebecker:2018vxz}.
The Distance conjecture states that, upon approaching a field space boundary located at infinite geodesic distance in the moduli space of string compactifications, an infinite tower of states become light, with the mass scale decreasing exponentially with the geodesic distance towards the boundary.
Instead, the de Sitter conjecture asserts that it should be impossible to find meta-stable four-dimensional dS solutions in the asymptotic regions of moduli space (the original and stronger conjecture~\cite{Obied:2018sgi}, which applied to the interior of moduli space as well, was contradicted by nature itself~\cite{Denef:2018etk}).

At the level of string compactifications with four-dimensional ${\cal N}=2$ effective theories, and increasingly also for four-dimensional ${\cal N}=1$ effective theories, there is considerable evidence~\cite{Grimm:2018ohb,Grimm:2018cpv,Corvilain:2018lgw,Marchesano:2019ifh,Joshi:2019nzi,Font:2019cxq,Grimm:2019wtx,Gendler:2020dfp} for the general validity of the distance conjecture, and the closely related \emph{Emergent string conjecture}~\cite{Lee:2019xtm,Lee:2019wij} in the parametric limit, that is, at distances parametrically large compared to a displacement of ${\cal O}(\mpl)$. Similarly, evidence suggests that the de Sitter conjecture is realized towards strict asymptotic regimes \cite{Grimm:2019ixq,Calderon-Infante:2022nxb}.

However, things are less clear when crossing from the interior region of the moduli space, to which we refer as `\emph{umbra}', into the `\emph{penumbra}'. 
The latter is the cross-over regime at finitely large field displacements between the umbra and the strict asymptotic regime, where distances are parametrically large and all terms in the flux-induced complex structure moduli scalar potential have been dropped, except for the strictly leading ones. In the penumbra the appearance of the first tower state of the Distance conjecture depends both on the exponential coefficient determining the fall-off of the infinite tower of states, and on the initial gap between the characteristic scale of the effective theory and the tower mass scale~\cite{Bielleman:2016olv,Valenzuela:2016yny,Blumenhagen:2017cxt,Dias:2018ngv,Etheredge:2022opl}.

This paper delves into the features that the flux-induced scalar potential displays in such a penumbra region.
Indeed, we will provide evidence for the following: 
\begin{itemize}
\item we move inwards from the strict asymptotic regions of the complex structure moduli space into the penumbra, where the moduli vevs are only moderately large;

\item within the penumbra, in the expansion of the periods, which govern the effective field theory interactions, we also retain the subleading contributions, those that would not survive in the strictly asymptotic limit;

\item then, we find meta-stable uplifting complex structure vacua realizing the proposal of~\cite{Saltman:2004sn} at moderately large moduli vevs; moreover, we also find long-range axion valleys showing axion monodromy and flattening from complex structure moduli backreaction.

\end{itemize}

The plan of the paper goes as follows. 
In Section~\ref{sec:Swamp_Inflation} we review some basic features of multi-field slow-roll inflation, highlighting how some of the recently formulated Swampland conjectures may hinder the realization of inflation in consistent effective theories.

Section~\ref{sec:IIB} presents the four-dimensional type IIB effective field theories, defined towards either a large complex structure or a Tyurin boundary, that are the main objects of the analysis that follows.

In Section~\ref{sec:dS_uplift} we study concrete examples of four-dimensional type IIB effective field theories where the complex structure modulus delivers a de Sitter uplift.
The analysis is expanded in Section~\ref{sec:Infl_Examples}, where we compute the scalar potential valleys for some selected examples, and study the field backreaction; therein, we further present a family of toy models that could host viable inflationary valleys.

In Section~\ref{sec:ML} we illustrate how four-dimensional type IIB effective field theories exhibiting axion monodromy valleys can be searched via machine learning algorithms.

Finally, Appendix~\ref{sec:IIB_summ} presents a brief overview of the moduli space of four-dimensional effective field theories, obtained compactifying type IIB string theory over an orientifolded Calabi-Yau three-fold, and Appendix~\ref{sec:Hodge} contains an overview of the main concepts of asymptotic Hodge theory that are employed in the main text.

\section{Swampland constraints and Inflation}
\label{sec:Swamp_Inflation}

In order to lay the foundations for the upcoming sections, here we briefly review the basic features that an effective theory should be endowed with in order to accommodate for multi-field inflation. 
We then overview to what extent some of the proposed Swampland conjectures may obstruct inflation in effective field theories that admit an ultraviolet completion within string theory.

\subsection{A brief review of multi-field inflation}

Preliminarily, let us specify the class of effective field theories which we will be focused on.
In order to keep the discussion general, we will consider effective theories formulated in $d$ spacetime dimensions, and coupled to gravity.
The effective theory is additionally endowed with $N$ scalar fields $\varphi^A$, with $A = 1, \ldots, N$.
Since the effective theories that we will be working with stem from the compactification of higher, $D$-dimensional string, or M-theory compactified over some $(D-d)$-dimensional internal manifold, the scalar fields $\varphi^A$ serve as \emph{moduli} of the internal manifold.
Specifically, the internal manifold is assumed to be endowed with a moduli space $\mathcal{M}_{\rm mod}$, and a local patch within $\mathcal{M}_{\rm mod}$ can be parametrized with the coordinates $\varphi^A$ and is equipped with the positive-definite metric $G_{AB}(\varphi)$.
However, we shall assume that the scalar fields $\varphi^A$ are not free within the moduli space $\mathcal{M}_{\rm mod}$; rather, they are constrained to vary only within some submanifolds thereof due to the presence of a scalar potential $V(\varphi)$.
With these ingredients, the two-derivative effective field theories locally encoding the dynamics of gravity and the scalar fields $\varphi^A$ which we will consider are described by the following action:
\begin{equation}
	\label{eq:Sw_Infl_Sgen}
	S_{\text{\tiny{EFT}}}^d = \int {\rm d}^dx\, \sqrt{-g} \left( \frac12 M_{\rm P}^{d-2} R - \frac{1}2 M_{\rm P}^{d-2}\, G_{AB}(\varphi) \partial \varphi^A \cdot  \partial \varphi^B - V(\varphi)   \right)\,,
\end{equation}
where $M_{\rm P}$ denotes the $d$-dimensional Planck mass, and $R$ is the Ricci scalar.

The explicit functional forms of both the field metric $G_{AB}(\varphi)$ and the scalar potential $V(\varphi)$ critically depend on the choice of the geometric data of the internal manifold.
Indeed, we additionally assume that the scalar potential displays a critical locus and, along such a locus, the value of the scalar potential is very close to zero.
Using the field space metric and the scalar potential, it is convenient to introduce
\begin{equation}
	\label{eq:Sw_Infl_gamma}
	\gamma = \frac{\| \nabla V \|}{V}\,,  \qquad  \text{with} \qquad \| \nabla V \|^2 := G^{AB} \partial_A V \partial_B V\,,
\end{equation}
where we have employed the shorthand notation $\partial_A := \frac{\partial}{\partial \varphi^A}$.
As in \cite{Calderon-Infante:2022nxb}, we will refer to $\gamma$ as `\emph{de Sitter coefficient}', and it plays a pivotal role in determining whether inflation can be realized within string theory-originated effective theories.

Let us now overview how inflation takes place within effective theories of the form \eqref{eq:Sw_Infl_Sgen}. 
Assuming that the universe is isotropic and homogeneous, its evolution can be described with the FLRW metric
\begin{equation}
	\label{eq:Sw_Infl_FLRW}
	{\rm d}s^2 = - {\rm d}t^2 + a(t)^2 {\rm d} \ell_{d-1}^2
\end{equation}
with ${\rm d} \ell_{d-1}$ being the (time-independent) line element of a $(d-1)$-dimensional space with Euclidean signature and constant curvature.
In turn, the scale factor $a(t)$ defines the Hubble parameter $H = \frac{\dot{a}}{a}$, with the dot denoting the time derivative. 
Additionally, we shall assume that, over the FLRW background, the scalar fields $\varphi^A$ evolve only in time, namely $\varphi^A := \varphi^A(t)$.

Given these hypotheses, the equations of motion determining the dynamics of the scalar fields $\varphi^A$ reduce to
\begin{equation}
	\label{eq:Sw_Infl_eom_phi}
	\ddot{\varphi}^A + \Gamma^A_{BC} \dot{\varphi}^B \dot{\varphi}^C + (d-1) H \dot{\varphi}^A + \frac{1}{M_{\rm P}^{d-2}} G^{AB} \partial_B V = 0
\end{equation}
with $\Gamma^A_{BC}$ being the Christoffel symbols computed out of the scalar metric $G_{AB}$.
On the other hand, the Einstein equations lead to the following equation that governs the evolution of the scale factor $a(t)$:
\begin{equation}
	\label{eq:Sw_Infl_eom_grav_a}
	(d-1)(d-2) \dot{H} + (d-1) G_{AB} \dot{\varphi}^A \dot{\varphi}^B - \frac{\kappa}{a^2} = 0\,,
\end{equation}
where $\kappa$ is the constant curvature of the Euclidean $(d-1)$-dimensional space. 
Alternatively, the scale factor $a(t)$ can be computed via the equation
\begin{equation}
	\label{eq:Sw_Infl_eom_grav_b}
	\frac{(d-1)(d-2)}{2} H^2 - \frac12 G_{AB} \dot{\varphi}^A \dot{\varphi}^B + \frac{\kappa}{ 2 a^2} - \frac{1}{M_{\rm P}^{d-2}} V = 0\,.
\end{equation}
In the following, we shall assume the Euclidean $(d-1)$-dimensional space to be flat, thus implying $\kappa = 0$.

The inflationary equations \eqref{eq:Sw_Infl_eom_phi} can be recast in a more useful form by introducing the number of e-folds $N := \log a(t)$; indeed, by expressing the time derivatives appearing in \eqref{eq:Sw_Infl_eom_phi} in terms of the number of e-folds and employing \eqref{eq:Sw_Infl_eom_grav_a} and \eqref{eq:Sw_Infl_eom_grav_b}, the equations \eqref{eq:Sw_Infl_eom_phi} become 
\begin{equation}
	\label{eq:Sw_Infl_eom_phi_N}
	{\varphi^{A\prime\prime}} + \frac12 \left[ (d-1)(d-2) - G_{BC} {\varphi^{B\prime}} {\varphi^{C\prime}} \right] \left( \frac{2}{d-2} {\varphi^{A\prime}} + G^{AB} \frac{\partial_B V}{V} \right) + \Gamma^{A}_{BC} {\varphi^{B\prime}} {\varphi^{C\prime}}= 0\,,
\end{equation}
where we have introduced the shorthand notation $(\quad)':= \frac{{\rm d}}{{\rm d} N}$.

It is worth remarking that, although the evolution of the scalar fields happens in the physical spacetime, we can think of this path as being threaded in (a local patch of) the moduli space.
Concretely, we can define the inflationary path $\mathcal{P}_{\text{\tiny{infl}}}$ within the moduli space $\mathcal{M}_{\rm mod}$ as
\begin{equation}
	\label{eq:Sw_Infl_Pinfl}
	\mathcal{P}_{\text{infl}} = \{ \varphi^A(t) \in \mathcal{M}_{\rm mod} \,|\, t \in \Delta T_{\text{infl}} \}
\end{equation}
where $\Delta T_{\text{\tiny{infl}}}$ is the interval of time during which inflation take place.

Inflation is defined as a period of accelerated expansion, characterized by $\ddot{a} > 0$. 
By employing the Einstein equations \eqref{eq:Sw_Infl_eom_grav_a} and the equations of motion for the scalar fields $\varphi^A$ \eqref{eq:Sw_Infl_eom_phi}, it can be shown that this condition can be recast as
 \begin{equation}
 	\label{eq:Sw_Infl_eps}
 	\varepsilon := - \frac{\dot{H}}{H^2}  = \frac{1}{(d-2) H^2} G_{AB} \dot{\varphi}^A \dot{\varphi}^B  < 1 \,,
 \end{equation}
where we have introduced the \emph{first slow-roll parameter} $\varepsilon$.
However, most of the proposed inflationary models are of the \emph{slow-roll} kind, obeying more stringent conditions.
Firstly, slow-roll inflation requires the Hubble parameter to be constant, so as to deliver an exponential growth of the scale factor $a(t)$; this can be achieved by requiring
\begin{equation}
	\label{eq:Sw_Infl_slowroll_1}
	\varepsilon \ll 1 \,,
\end{equation}
which, by \eqref{eq:Sw_Infl_eom_grav_b}, implies that
\begin{equation}
	\label{eq:Sw_Infl_eom_slowroll_1b}
	V \simeq M_{\rm P}^{d-2} \frac{(d-1)(d-2)}{2} H^2 \,.
\end{equation}
Secondly, slow-roll inflation requires that inflation lasts for enough time to be consistent with the cosmological observations. 
This can be achieved by imposing the following constraint on the \emph{second slow-roll parameter} $\eta$:
\begin{equation}
	\label{eq:Sw_Infl_slowroll_2}
	\eta := \frac{\dot{\varepsilon}}{H \varepsilon} = 2 \varepsilon + \frac{2}{(d-2) H^2} G_{AB} \dot{\varphi}^A \ddot{\varphi}^B  \ll 1 \,.
\end{equation}

Now, although the de Sitter coefficient \eqref{eq:Sw_Infl_gamma} does not deliver direct information about whether slow-roll inflation can be realized in a direct way, it can be related to the slow-roll parameter $\varepsilon$ as follows.
Preliminarily, we define the unit-vector $T^A$, tangent at every point to the path $\mathcal{P}_{\text{infl}}$ as
\begin{equation}
	\label{eq:Sw_Infl_Ta}
	T^A := \frac{\dot{\varphi}^A}{\|\dot{\varphi}\|}\,, \qquad \text{with} \qquad \| T \|^2 := G_{AB} T^A T^B = 1\,.
\end{equation}
Then, we can introduce a basis of vectors $\{N^A_{(a)}\}$, $a = 1, \ldots, n-1$,  orthogonal to $T^A$ along every point of the inflationary path and obeying $G_{AB} N^A_{(a)} T^B = 0$.  
Thus, the gradient of the scalar potential can be split along the directions of $T^A$ and $N^A_{(a)}$ as 
\begin{equation}
	\label{eq:Sw_Infl_dV_dec}
	G^{AB} \partial_B V =  V_T  T^A + \sum\limits_{a = 1}^{n-1} V_{N_{(a)}}  N^A_{(a)} \,, \qquad \text{with} \qquad V_T := T^A \partial_A V \,, \quad  V_{N_{(a)}} := N^A_{(a)} \partial_A V\,.
\end{equation}
Employing the newly introduced local basis of vectors, we can also project the scalar field equations \eqref{eq:Sw_Infl_eom_phi} along their directions, yielding
\begin{subequations}
	\label{eq:Sw_Infl_eom_phi_b}
	\begin{align}
		\label{eq:Sw_Infl_eom_phi_b1}
		& \frac{\partial}{\partial t} \| \dot{\varphi} \| + (d-1) H \| \dot{\varphi} \| + \frac{1}{M_{\rm P}^{d-2}} V_T = 0 \,,
		\\
		\label{eq:Sw_Infl_eom_phi_b2}
		& N^A_{(a)} G_{AB} D_t T^B + \frac{1}{M_{\rm P}^{d-2} \| \dot{\varphi} \|} V_{N_{(a)}} = 0\,,
	\end{align}
\end{subequations}
where we have introduced $D_t T^A := \dot{T}^A + \Gamma^A_{BC} T^B \dot{\varphi}^C$. Finally, by introducing the \emph{turning rate} $\Omega$, defined via
\begin{equation}
	\label{eq:Sw_Infl_Omega}
	\Omega^2 :=  \frac{\sum\limits_{a = 1}^{n-1} V_{N_{(a)}}^2}{M_{\rm P}^{2d-4} \| \dot{\varphi} \|^2} =  \| D_t T \|^2 \,,
\end{equation}
one can show that the de Sitter coefficient \eqref{eq:Sw_Infl_gamma} can be recast as
\begin{equation}
	\label{eq:Sw_Infl_gamma_eps}
	\gamma^2 \simeq \frac{4 \varepsilon}{d-2} \left( 1 + \frac{\Omega^2}{(d-1)^2 H^2}\right)\,,
\end{equation}
where we have exploited the decomposition \eqref{eq:Sw_Infl_dV_dec}, alongside \eqref{eq:Sw_Infl_eom_phi_b1}, \eqref{eq:Sw_Infl_eps}, \eqref{eq:Sw_Infl_eom_grav_b} and the definition \eqref{eq:Sw_Infl_Omega}.

\subsection{Swampland obstructions to inflation}

With the advent of the Swampland program, whose start can be traced back to \cite{Ooguri:2006in,Arkani-Hamed:2006emk}, a novel viewpoint for exploring consistent effective field theories has been pursued. 
Namely, rather than obtaining effective field theories methodically, by compactification of the higher dimensional string, or M-theory, one can \emph{start} with the lower dimensional effective field theories, and investigate which conditions these ought to obey in order to be obtainable from string, or M-theory.
These conditions, named \emph{Swampland conjectures}, are guided by string theory predictions in their formulation, and are intended to apply to any consistent effective field theory.

Crucially, some of the conjectures postulate that the scalar potential of any effective field theory that has to be consistent with quantum gravity needs to obey certain conditions, which we will review below.
\begin{description}
	\item[de Sitter conjecture] Formulated in \cite{Obied:2018sgi,Ooguri:2018wrx}, the \emph{de Sitter conjecture} asserts that the de Sitter coefficient \eqref{eq:Sw_Infl_gamma} is lower-bounded as
	\begin{equation}
		\label{eq:Sw_Infl_dS}
		\gamma \geq c_d
	\end{equation}
	with $c_d$ a positive constant, allegedly of order one. 
	It is worth stressing that, since the de Sitter coefficient \eqref{eq:Sw_Infl_gamma} is moduli-dependent, the statement \eqref{eq:Sw_Infl_dS} offers a lower bound over $\gamma$ that supposedly holds throughout $\mathcal{M}_{\text{mod}}$.
	
	The de Sitter conjecture seems a common feature that scalar potentials obtained from string theory obey \emph{asymptotically} in the moduli space, namely sufficiently close to boundaries of the moduli space \cite{Grimm:2019ixq}.
	However, the realization of the de Sitter conjecture towards the bulk of the moduli space, where the conjecture is harder to check, is an open question.
	
	\item[Strong de Sitter conjecture] Based on the consistency of the de Sitter conjecture under dimensional reduction, the \emph{strong de Sitter conjecture} \cite{Rudelius:2021oaz,Rudelius:2021azq} proposes the following value for the constant $c_d$ unspecified in \eqref{eq:Sw_Infl_dS}:
	\begin{equation}
		\label{eq:Sw_Infl_c_strong}
		c_d^{\text{strong}} = \frac{2}{\sqrt{d-2}}\,.
	\end{equation}

    \item[Distance conjecture] The Distance conjecture, originally proposed in \cite{Ooguri:2006in}, states that any effective field theory that is consistent with string theory breaks down when the moduli fields approach a field space boundary located at infinite field distance.
    The reason of such a break down is rooted in the appearance of an infinite tower of states, whose masses become exponentially light in the geodesic distance.
    Namely, consider the masses $M_n(\varphi_0)$ of the states composing the tower at a given point $\varphi^A_0$, close enough to an infinite-distance boundary.
    Then, the masses $M_n(\varphi_1)$ of the states within such a tower, evaluated at a different point $\varphi^A_1$, behave, with respect to the former, as
	\begin{equation}
		\label{eq:Sw_Infl_Dc}
		M_n(\varphi_1) \sim e^{- \alpha d(\varphi_0, \varphi_1) }M_n(\varphi_0)\,.
	\end{equation}
    Here, $d(\varphi_0, \varphi_1)$ is the geodesic distance, measured in Planck units, between the points $\varphi_0^A$, $\varphi_1^A$, and $\alpha$ is a positive constant, supposedly of order one. 
    Thus, as $d(\varphi_0, \varphi_1) \to \infty$, the masses of the states within the tower become massless, and ought to be included within the effective field theory, rendering the effective description invalid.
\end{description}

The conjectures inevitably influence the possibility of realizing slow-roll inflation in consistent effective field theories. 
Among the most pressing possible obstructions to inflation that the above Swampland conjectures imply are the following:

\noindent{\textbf{Geodesic inflationary paths.}} The equation \eqref{eq:Sw_Infl_gamma_eps}, relating the de Sitter coefficient with inflationary quantities, is the crucial one for examining the interplay between the de Sitter conjecture and inflation. Indeed, if the inflationary path $\mathcal{P}_{\text{infl}}$, obtained after solving \eqref{eq:Sw_Infl_eom_phi} and \eqref{eq:Sw_Infl_eom_grav_a}, is geodesic within $\mathcal{M}_{\rm mod}$, then $\Omega = 0$ by the very definition of geodesicity, and the slow-roll parameter $\varepsilon$ is related to $\gamma$ via the simpler relation $\varepsilon = \frac{\gamma^2}{2}$.
In turn, if the Swampland conjectures stated above are true, this implies that the slow-roll parameter $\varepsilon$ is lower-bounded by an $\mathcal{O}(1)$-constant, in stark contrast with the first slow-roll condition \eqref{eq:Sw_Infl_slowroll_1}.

\noindent{\textbf{Non-geodesic inflationary paths.}} Inflation may still be allowed, given the Swampland conjectures above, if the inflationary path is non-geodesic. 
Indeed, the generic constraint \eqref{eq:Sw_Infl_dS} may still be consistent with the first slow-roll condition \eqref{eq:Sw_Infl_slowroll_1} provided that the turning rate $\Omega$ is sufficiently large.
However, such `\emph{rapid-turn inflation}' seems to be disfavored in supergravity \cite{Aragam:2021scu} for light fields; otherwise, one could achieve rapid-turn inflation by including heavy fields, with masses above the Hubble scale, as in the `\emph{fat inflation}' scenarios \cite{Chakraborty:2019dfh}.

\noindent{\textbf{Effective theory breakdown along the inflationary path.}} In order to guarantee the realization of inflation within a given effective theory, it is crucial that, along the inflationary path $\mathcal{P}_{\text{infl}}$, the effective field theory hosting inflation remains well-defined.
However, owing to the Distance conjecture \eqref{eq:Sw_Infl_Dc}, moving in the moduli space may lead to a breakdown of the effective theory: if the inflationary path $\mathcal{P}_{\text{infl}}$, stretching from the field configuration $\varphi_0^A$ to $\varphi_1^A$, has a geodesic length $d(\varphi_0, \varphi_1)$ of order one, several states composing the tower \eqref{eq:Sw_Infl_Dc} may become lighter than the cutoff.
As such, these need to be included within the effective theory, potentially causing either controlled backreaction leading to flattening of the effective scalar potential or even a breakdown of slow-roll.
In this sense, the Distance Conjecture \eqref{eq:Sw_Infl_Dc} constrains the path that fields can draw before a breakdown occurs; however, the length, in Planck units, of the path that scalar fields can consistently traverse before an effective theory breakdown critically depends on the constant $\alpha$ that appears in \eqref{eq:Sw_Infl_Dc} \cite{Bielleman:2016olv,Valenzuela:2016yny,Blumenhagen:2017cxt,Dias:2018ngv,Etheredge:2022opl} as well as the gap between the EFT cutoff and the mass scale of the tower at the origin of scalar field space.

\subsection{Difficulties and possibilities of inflation in strictly and not-so-strictly asymptotic regions of moduli space}
\label{sec:Swamp_Inflation_Tame}

The realization of the de Sitter conjecture may appear discomforting for the realization of inflation in string theory scenarios. 
However, as we have mentioned earlier, in string theory-originated effective theories, concrete checks of the de Sitter conjecture \eqref{eq:Sw_Infl_dS} have been performed in \emph{asymptotic regions} of the field space, namely when scalar fields are so close to the field space boundary, that effective theories are under arbitrarily good parametric control, for they are arbitrarily weakly coupled.
Towards the `\emph{bulk}' of the moduli space, where effective theories can 
often enough be under sufficient (but not full) parametric control the validity of the de Sitter conjecture is questionable.
In the following, we quickly overview the arguments against the realization of inflation in the asymptotic regions of the moduli space, indeed suggesting that achieving inflation in string theory effective field theories may require a thorough analysis of the theories towards the bulk of the moduli space.

\noindent\textbf{Why inflation cannot happen in strict asymptotic regimes}. Firstly, we provide some very general arguments about why inflation is not expected to be realized in strict asymptotic regimes, where some scalar fields acquire \emph{too} large values.
Recently, it has been postulated that \emph{all} the couplings that appear in a consistent effective theory ought to be \emph{tame} functions of the fields and the parameters \cite{Grimm:2021vpn} -- see \cite{Grimm:2021vpn,Grimm:2022sbl} for a mathematical definition of \emph{tameness}, and for concrete applications to stringy effective field theory.
In particular, the scalar potential $V(\varphi)$ that appears in a string theory-originated effective theory has to be a tame function of the scalar fields $\varphi^A$, regarded as geometric \emph{moduli} of the compactification manifold leading to the given effective theory.
Indeed, the scalar fields $\varphi^A$ may be understood as local coordinates of
a patch $\mathcal{U}$ of the moduli space $\mathcal{M}_{\text{mod}}$.
Whenever $\mathcal{U}$ describes an asymptotic, or near-boundary regime, it is convenient to split the $N$ moduli $\varphi^A$ that enter the effective field theory into the following two sets:
\begin{description}
	\item[Saxions $s^i$] ($i = 1, \ldots, n$) which describe non-compact domains, which we assume to be of the form $s^i > 1$;
	\item[Axions $a^\alpha$] ($\alpha = 1, \ldots, N - n$), which span compact domains that we conventionally assume to be $[0,1[$.
\end{description}
The near-boundary local patch $\mathcal{U}$ can then be described as
\begin{equation} 
	\label{eq:Sw_Infl_cU}
	\mathcal{U} = \{ |a^\alpha| < 1,\ \  s^1 ,  s^2 , \ldots , s^n > 1 \}\,.
\end{equation}
Roughly speaking, the saxions $s^i$ tell `\emph{how close}' to the field space boundaries the associated field space point is, with a boundary reached as any of the saxions $s^i \to \infty$, whereas the axions $a^\alpha$ parametrize the orthogonal field directions.

The tameness of $V(a,s)$ implies that its tails, in the saxion directions, have to be solely of three kinds: $V(a,s)$ is either constant, or monotonically decreasing, or  monotonically increasing for large enough values of the saxions.
Such a property excludes, for instance, that the scalar potential $V(a,s)$ can be a periodic function of the saxions.
However, these asymptotic behaviors cannot accommodate for an inflationary scenario.
Indeed, assume that the inflationary path \eqref{eq:Sw_Infl_Pinfl} drives the saxions within any of these asymptotic regions towards value of larger saxions.
In the case in which the scalar potential $V(a,s)$ is constant asymptotically, it means that some saxions are strictly flat directions for the scalar potential; thus, in this region, there is no vacuum towards which the saxions can roll, ending the inflationary phase.
In the case where the scalar potential is monotonically decreasing, a vacuum could at most be an `asymptotic' vacuum, formally located at the boundary of the field space, where the effective theory is expected to break down.
The last case, where the scalar potential is monotonically increasing is the least interesting one for the inflation towards larger values of the saxions: indeed, the scalar potential should push the scalar fields towards smaller values of the saxions, not larger ones.

In sum, given an enough large positive constant $C$, whenever $s^i > C$ for some $i$, then there is no nearby vacuum to fall into, and the associated field space region is disfavored for accommodating inflation.

\noindent\textbf{String theory lamppost evidence of no-strict asymptotic inflation}. The above general argument is consistent with what we expect from string theory--originated scalar potentials.
Indeed, let us consider the four-dimensional effective field theories that originate from the compactification of F-theory over a Calabi-Yau four-fold $Y$.
Such effective theories are endowed with a closed string-sector complex muduli space, spanned by the complex fields $z^i$, which we split in axion and saxion components as  $z^i = a^i + \im s^i$, with $i = 1, \ldots, h^{3,1}(Y)$.
The presence of internal $G_4$--fluxes induces a scalar potential for the fields $z^i$ that, in the near-boundary regime, for large $s^i$, acquires the following form \cite{Grimm:2019ixq}\footnote{Note, that flux-induced power-law scalar potentials for several moduli of this type were also obtained in Riemann surface compactifications of type IIB string theory in~\cite{Saltman:2004jh}.}
\begin{equation}
	\label{eq:Sw_Infl_VF}
	V_{\text{F}} (a,s) = \sum\limits_{{\bm \ell} \in \mathcal{E}} \rho_{{\bm \ell}}(a,{\bf f})^2 \prod\limits_{j = 1}^n (s^j)^{\Delta \ell_j}\,.
\end{equation}
Here we have denoted ${\bm \ell} = (\ell_1, \ldots, \ell_n)$, and $\mathcal{E}$ is the set of allowed vectors ${\bm \ell}$; moreover,  $\rho_{{\bm \ell}}(a,f)$ are analytic, bounded functions of the axions $a^i$, and of the internal fluxes, generically collected in the vector ${\bf f}$.
Both the explicit form of the functions $\rho_{{\bm \ell}}(a,{\bf f})$ and the set $\mathcal{E}$ depend on the specific boundary around which the effective field theory is defined.

In the very strict asymptotic regime, for $s^i \gg 1$, the scalar field metric over the field space typically acquires the diagonal form
\begin{equation}
	\label{eq:Sw_Infl_metr_lead}
	G_{AB}^{\text{strict}} = \begin{pmatrix}
		G^{\text{strict}}_{ij} & 0 \\ 0 & G^{\text{strict}}_{ij} 
	\end{pmatrix}\,, \qquad \text{with} \quad G^{\text{strict}}_{ij} = \frac{d_i}{(s^i)^2} \delta_{ij}\,,
\end{equation}
with the fields ordered as $\varphi^A = (a^i, s^i)$, and for some positive constants $d_i$.
On the other hand, after choosing a \emph{growth hierarchy} among the saxion fields $s^i$, such as $s^1 > s^2 > \ldots > s^n > 1$, or permutations thereof, and using the boundedness of the functions $\rho_{{\bm \ell}}(a,{\bf f})$, one can show that, within the scalar potential \eqref{eq:Sw_Infl_VF}, one can single out a leading term as 
\begin{equation}
	\label{eq:Sw_Infl_VF_lead}
	V_{\text{F}}^{\text{strict}}(s) = C \prod\limits_{j = 1}^n (s^j)^{\Delta \ell_j^{\text{lead}}}\,.
\end{equation}
for some leading $\Delta \ell_j^{\text{lead}}$ and a positive constant $C$.

In other words, for large values of the saxions, the scalar potential is well-approximated by a monomial that has definite growth, and exhibits one of the three behaviors that tame functions are expected to be endowed with in the strict asymptotic regime \cite{Grimm:2022sbl}.
Additionally, it is then clear that, exploiting the strict asymptotic relations \eqref{eq:Sw_Infl_metr_lead} and \eqref{eq:Sw_Infl_VF_lead}, the de Sitter coefficient \eqref{eq:Sw_Infl_gamma} trivially obeys the de Sitter conjecture \eqref{eq:Sw_Infl_dS}, hindering the possibility to realize inflation asymptotically, for large values of the saxions $s^i$.

Similar arguments can be reproduced for the complex structure sector scalar potential of four-dimensional theories stemming from type IIB string theory, or for the K\"ahler sector scalar potential of four-dimensional theories originating from Type IIA string theory compactified over Calabi-Yau three-folds \cite{Grimm:2019ixq}.
These observations lead to the suggestion that, if string theory effective field theories ought to support inflationary paths, then these need to lie towards the bulk of the moduli space, away from strict asymptotic regimes.

\section{The scalar potential of type IIB effective theories}
\label{sec:IIB}

In this section we introduce the main objects of our investigations, namely the effective field theories obtained from the compactification of type IIB string theory over an orientifolded Calabi-Yau three-fold.
Our focus will be on effective theories defined towards regions of the complex structure moduli space close enough to a boundary located at infinite-field distance. 
While we defer the technical details to the Appendices~\ref{sec:IIB_summ} and~\ref{sec:Hodge}, here we outline the general features that such near-boundaries effective field theory share, highlighting, in particular, the structure of the scalar potential towards such infinite-field distance boundaries.
For simplicity, we restrict our attention to effective field theories characterized by a single complex structure modulus, for which the general structure of the scalar potentials can be generically computed.

\subsection{The asymptotic structure of the scalar potential}
\label{sec:IIB_Vgen}

The compactification of type IIB string theory over a Calabi-Yau three-fold $Y$ equipped with ${\rm O}3$, ${\rm O}7$-orientifolds yields four-dimensional effective theories endowed with $\mathcal{N} = 1$ supersymmetry.
As reviewed in Appendix~\ref{sec:IIB_summ}, restricting to the closed string sector, the four-dimensional effective theories so obtained are characterized by three sectors of moduli, residing within $\mathcal{N} = 1$ chiral superfields: the axio-dilaton sector, the K\"ahler moduli sector, and the complex structure sector.
In this section, we disregard the K\"ahler moduli sector and the axio-dilaton, and focus solely on the complex structure sector.

The complex structure moduli sector is described by a special K\"ahler geometry.
As such, parametrizing the complex structure moduli with the complex coordinates $z^i$, with $i = 1, \ldots, h^{2,1}_-(Y)$, all the interactions involving the complex structure moduli can be fully expressed in terms of holomorphic \emph{periods} which, in a symplectic basis, can be recast as
\begin{equation}
	\label{eq:HT_Pi_sym}
	{\bm \Pi}(z) = \begin{pmatrix}
		X^I(z) \\ -\mathcal{F}_I(z)
	\end{pmatrix}\,,
\end{equation}
where $X^I(z)$, $\mathcal{F}_I(z)$, with $I = 1, \ldots, h^{2,1}_-(Y)+1$, are holomorphic functions of the complex structure moduli $z^i$.

The K\"ahler potential determining the kinetic terms for the complex structure moduli is
\begin{equation}
	\label{eq:HT_Kcsb}
	K^{\rm cs} = - \log \im(\bar{X}^I \mathcal{F}_I -  X^I \bar{\mathcal{F}}_I)\, ,
\end{equation}
with the K\"ahler metric defined as $K^{\rm cs}_{i\bar\jmath} := \frac{\partial^2 K^{\rm cs}}{\partial z^i \partial \bar{z}^{\bar\jmath}}$.
Further interactions among the complex structure moduli stem from their scalar potential.
Here we assume that the scalar potential is generated by internal Ramond-Ramond fluxes, and it can be written in the following form: 
\begin{equation}
	\label{eq:HT_Vfluxquad}
	V_{\text{flux}} =: M^4_{\rm P} e^{\hat{K}}\, V^{\rm cs} = \frac12 M^4_{\rm P} e^{\hat{K}}\, {\bf f}^T\, \mathcal{T}(z,{\bar z}) \, {\bf f}\,, \qquad \text{with}\quad {\bf f} = \begin{pmatrix}
		m^I \\ - e_I
	\end{pmatrix}\,,
\end{equation}
where $e_I, m^I \in \mathbb{Z}$ denote the Ramond-Ramond flux quanta, and $\mathcal{T}(z,{\bar z})$ is a real, positive semi-definite matrix that can be expressed in terms of the periods \eqref{eq:HT_Pi_sym} as in \eqref{eq:IIB_Tmat}.
Furthermore, in \eqref{eq:HT_Vfluxquad}, we have generically denoted with $\hat{K}$ the K\"ahler potential involving the K\"ahler moduli and the axio-dilaton.

In a complete candidate four-dimensional vacuum of type IIB string theory compactified on a CY orientifold, the full scalar potential including a mechanism of K\"ahler moduli stabilization like KKLT or LVS would read
\begin{equation}
V=V_{\hat K}+V_{\rm flux}\quad.
\end{equation}
Valid setups of K\"ahler moduli stabilization would generically stabilize the K\"ahler moduli in and AdS minimum such that we get $\langle V_{\hat K}\rangle < 0$ and $|\langle V_{\hat K}\rangle | \ll 1$. In addition, K\"ahler moduli stabilization mechanisms like KKLT or LVS involve non-perturbative effects which suppresses the K\"ahler moduli masses compared the the flux-induced masses of the complex structure moduli and the axo-dilaton. Hence, for  KKLT and LVS K\"ahler moduli stabilization approximately decouples from complex structure stabilization by flux, and will only produced suppressed shifts of the complex structure moduli vevs compared to the vevs the acquire from $V_{\rm flux}$ alone.

As we shall see in the upcoming sections, the scalar potential \eqref{eq:HT_Vfluxquad} admits non-trivial minima in the complex structure moduli space, along which the scalar potential  acquires a positive value $V^{\text{cs}}_{\text{min}}$.

Hence, since $\exp(\hat K)\sim 1/{\cal V}^2$ (where $\cal V$ is the CY volume) for such complex structure minima with $V^{\text{cs}}_{\text{min}}>0$ the complex structure moduli potential
\begin{equation}
    \langle V_{\rm flux}\rangle= M^4_{\rm P} \langle e^{\hat{K}}\rangle\, \langle V^{\rm cs}\rangle \sim \frac{1}{{\cal V}^2}\underbrace{V^{\text{cs}}_{\text{min}}}_{> 0}
\end{equation}
acts as an anti-D3-brane like uplift to potentially four-dimensional de Sitter. It is for this reason, that we call the pure complex structure contribution $V^{\rm cs}$ a `de Sitter uplift' and label quantities depending solely on $V^{\rm cs}$ with the subscript `de Sitter uplift'.

For the following discussion, it is convenient to define as one such quantity the `de Sitter uplift coefficient' \eqref{eq:Sw_Infl_gamma} restricted to the complex structure sector only as
\begin{equation}
    \label{eq:HT_dScoeff_cs}
    \gamma^{\text{cs}}_{\text{uplift}} = \frac{\sqrt{2 K^{z\bar z}_{\rm cs} \partial_z V^{\rm flux} \partial_{\bar z} V^{\rm flux}}}{V^{\rm flux}} =\frac{\sqrt{2 K^{z\bar z}_{\rm cs} \partial_z V^{\text{cs}} \partial_{\bar z} V^{\text{cs}}}}{V^{\text{cs}}}\,,
\end{equation}
to which we will refer as \emph{uplift de Sitter coefficient}, due to the positive semi-definiteness of the scalar potential.

Furthermore, in the following, we would like to study the viability of scalar potentials of the form \eqref{eq:HT_Vfluxquad} for inflation. For this purpose, we need to impose the combined vacuum of K\"ahler and complex structure moduli after inflation to have nearly-zero vacuum energy.

To mimic this without inserting manifest K\"ahler moduli stabilization, we subtract the minimum value $V^{\text{cs}}_{\text{min}}$ that the scalar potential displays from \eqref{eq:HT_Vfluxquad} from $V_{\rm cs}$, effectively assuming that K\"ahler moduli stabilization produces $\langle V_{\hat K}\rangle\simeq -V^{\text{cs}}_{\text{min}}$.

Consequently, for such a normalized scalar potential, we introduce the quantity
\begin{equation}
    \label{eq:HT_dScoeff_cs_corr}
    \tilde{\gamma}^{\text{cs}}_{\text{late-dS}} =\frac{\sqrt{2 K^{z\bar z}_{\rm cs} \partial_z V^{\text{cs}} \partial_{\bar z} V^{\text{cs}} }}{V^{\text{cs}}- V^{\text{cs}}_{\text{min}}} \,,
\end{equation}
to which we will refer as \emph{late de Sitter coefficient}, where `late' refers to the late-time near-zero vacuum energy.

The periods of a Calabi-Yau three-fold are notoriously hard to compute in general, rendering difficult to grasp the generic structure of both the complex structure K\"ahler potential \eqref{eq:HT_Kcsb} and the scalar potential \eqref{eq:HT_Vfluxquad} across the entirety of the complex structure moduli space.
Therefore, the investigation carried out in this work will focus on regions of the complex structure moduli space moderately close to a given complex structure boundary.
In such regions, it is possible to infer the general structures that the periods \eqref{eq:HT_Pi_sym} may acquire by employing the powerful techniques developed in Hodge theory \cite{MR0382272,MR840721}.
While we refer to Appendix~\ref{sec:Hodge} for a brief overview of Hodge theory, or \cite{Grimm:2021ckh} for a more detailed review thereof, here we limit ourselves to highlighting the salient implications of Hodge theory for the asymptotic structure of the scalar potential \eqref{eq:HT_Vfluxquad}.

For simplicity, we will focus on the case where the four-dimensional effective field theory contains only a single complex structure modulus $z$, which we decompose as $z = a + \im s$, with $a$ denoting the \emph{axion} and $s$ the \emph{saxion} field.
Indeed, as in Section~\ref{sec:Swamp_Inflation_Tame}, we assume that the axion $a$ spans the bounded interval
\begin{equation}
	\label{eq:HT_a}
	0 \leq a < 1\,,
\end{equation}
playing the role of a fundamental domain for the axion $a$, whereas the saxion $s$ spans the non-compact domain $s > 1$.
In such conventions, we further assume that the boundary of the complex structure moduli space is reached as $s \to \infty$.

A key feature that characterizes a boundary is the associated \emph{monodromy}.
The \emph{monodromy matrix} $T$ around a given boundary is defined by how the periods \eqref{eq:HT_Pi_sym} transform under a unit shift of the axion $a$ as follows\footnote{In this work, we shall focus on the \emph{unipotent} part of monodromy matrices. Namely, the matrix $T$ is such that $(T-\mathds{1})^n \neq 0$, $(T-\mathds{1})^{n+1} = 0$ for some $n$. Instead, we disregard the semi-simple part \cite{Grimm:2018ohb}.}
\begin{equation}
	\label{eq:HT_TPi}
	{\bm \Pi}(a,s) \xrightarrow{a \to a + 1}  {\bm\Pi}'(a,s) =: T {\bm \Pi} (a,s)\,.
\end{equation}
It is worth stressing that the monodromy matrix $T$ is symplectic, and belongs to the larger duality group $\text{Sp}(4,\mathbb{Z})$ of the four-dimensional effective action \cite{Andrianopoli:1996cm,Craps:1997gp,Lanza:2022zyg}. 
In turn, the monodromy matrix define the \emph{log-monodromy matrix} $N$ as follows
\begin{equation}
    \label{eq:HT_logmon}
    N = \log T \,.
\end{equation}
Furthermore, the matrix $N$ obeys $N^T \eta = - \eta N$, as a consequence of the fact that $T$ is symplectic, and nilpotent, namely $N^n \neq 0$, $N^{n+1} = 0$ for some $n \in \mathbb{N}$.

The log-monodromy matrix play a pivotal role in finding an asymptotic expression for the periods \eqref{eq:HT_Pi_sym}.
In fact, towards a moduli space boundary, for $s > 1$, the periods can \eqref{eq:HT_Pi_sym} enjoy the following expansions \cite{MR0382272,MR840721,Grimm:2018ohb}
\begin{equation}
	\label{eq:HT_Per_exp}
	{\bm \Pi}  = e^{z N} \left({\bf a}_0 + e^{2\pi \im z} {\bf a}_{1}  + e^{4\pi \im z} {\bf a}_{2} + \ldots \right).
\end{equation}
where ${\bf a}_0, {\bf a}_1, {\bf a}_2, \ldots$ are four-dimensional vectors.

For large enough values of the saxion $s$, the leading contribution in the expansion \eqref{eq:HT_Per_exp} constitutes the \emph{nilpotent orbit approximation} of the periods \eqref{eq:HT_Pi_sym}:
\begin{equation}
	\label{eq:HT_Per_nil}
	{\bm \Pi}_{\rm nil} := e^{z N} {\bf a}_0 \, .
\end{equation}
In particular, it is worth remarking that the nilpotency of the matrix $N$ implies that, at leading order, the components of the periods are polynomials in the complex structure $z$, of at most of degree $n$.
Furthermore, the sole ingredients entering the leading behavior \eqref{eq:HT_Per_nil} are the nilpotent matrix $N$ and the vector ${\bf a}_0$, and their specific forms are tied to the type of boundaries towards which the expansion \eqref{eq:HT_Per_exp} is performed. 

Clearly, in order for the nilpotent orbit approximation \eqref{eq:HT_Per_nil} to serve as a good enough approximation for the general expansion \eqref{eq:HT_Per_exp}, we need that the exponential corrections in the are negligible with respect to \eqref{eq:HT_Per_nil}, namely $|e^{2\pi \im z}| = e^{-2\pi s} \ll 1$. 
Such a condition can be achieved for moderately large values of the saxion $s$.

Equipped with the nilpotent orbit approximation \eqref{eq:HT_Per_nil}, it is possible to extract the form that the scalar potential \eqref{eq:HT_Vfluxquad} towards a moduli space boundary by simply employing the leading contribution \eqref{eq:HT_Per_nil} for computing the matrix $\mathcal{T}(z, \bar{z})$ entering \eqref{eq:HT_Vfluxquad}.
In particular, due to the structure of \eqref{eq:HT_Per_nil}, the general form of a nilpotent orbit-approximated scalar potential is
\begin{equation}
	\label{eq:HT_Vfluxquad_nil}
	V^{\rm nil} =  \frac12 M^4_{\rm P} e^{\hat{K}} \, {\bm \rho}^T(a) \, \mathcal{Z}(s) \, {\bm \rho}(a)\,, \qquad \text{with}\quad {\bm \rho}(a) = e^{- a N} {\bf f}\,,
\end{equation}
where $\mathcal{Z}(s)$ is a saxion-only dependent matrix.

While the scalar potential \eqref{eq:HT_Vfluxquad_nil} could be most readily obtained out of the nilpotent-orbit-approximated periods \eqref{eq:HT_Per_exp}, in this work we will follow another route to determine the scalar potential.
Indeed, we will employ the ${\rm sl}(2)$-approximation \cite{MR0382272,MR840721,Grimm:2019ixq,Grimm:2021ckh}.
This approximation scheme generically differs from the nilpotent orbit approximation \eqref{eq:HT_Per_nil}; however, for values of the saxion $s$ large enough, it offers a good approximation of the nilpotent-orbit estimation \eqref{eq:HT_Per_nil}. 
However, working with the ${\rm sl}(2)$-approximation has several advantages: firstly, the procedure to infer the ${\rm sl}(2)$-approximation for the periods \eqref{eq:HT_Pi_sym} is algorithmic \cite{Grimm:2021ckh}; secondly, within such an approximation, we may ensure that the fluxes are integer, namely ${\bf f} \in \mathbb{Z}^{2h^{2,1}+2}$; moreover, it allows for immediate estimations of the growths, or fall-offs of key physical quantities \cite{Grimm:2019ixq}.

Prominently, the ${\rm sl}(2)$-approximation offers a clear approximation for the matrix $\mathcal{Z}(s)$ appearing in \eqref{eq:HT_Vfluxquad_nil}, for integral choices of fluxes entering the vector ${\bf f}$.
In fact, as illustrated in \cite{Grimm:2019ixq,Grimm:2020ouv,Grimm:2022sbl}, the scalar potential \eqref{eq:HT_Vfluxquad_nil} may be interpreted as an \emph{Hodge norm} for the flux vector ${\bf f}$.
As such, assuming that axion $a$ resides in the bounded domain \eqref{eq:HT_a}, the scalar potential \eqref{eq:HT_Vfluxquad_nil} can be estimated as
\begin{equation}
	\label{eq:HT_Vfluxquad_sl2}
	V^{\rm nil} = \frac12 M^4_{\rm P} e^{\hat{K}} \| {\bf f} \|^2 \sim s^\ell\,, \qquad \text{for some $\ell \in \mathbb{Z}$\,,}
\end{equation}
where we have introduced the Hodge norm $\| \cdot \|$.
Thus, in order to estimate the saxion-dependent behavior of the scalar potential \eqref{eq:HT_Vfluxquad} it is just enough to know the geometric data of the boundary near which the effective theory is defined, necessary to determine the explicit expression of the norm \eqref{eq:HT_Vfluxquad_sl2}, and the flux vector ${\bf f}$ that determines the scalar potential.

\subsection{The scalar potential towards infinite-distance boundaries}
\label{sec:IIB_Vbound}

In this work, we will be interested in effective field theories defined close to field space boundaries located at infinite-field distance. 
In the case of a single, dynamical complex structure modulus $z$, only two kinds of such boundaries can be constructed \cite{Grimm:2018ohb, Grimm:2018cpv,Grimm:2020ouv,Grimm:2022xmj}: the \emph{Type IV} boundaries, or large complex structure (LCS) boundaries, and the \emph{Type II} boundaries, or Tyurin degenerations. 
These can be most readily computed out of the boundary data listed in \cite{Green:2008}, by using the techniques outlined in Appendix~\ref{sec:Hodge}.
Below, we summarize the relevant quantities that are necessary to estimate the scalar potential \eqref{eq:HT_Vfluxquad} in the ${\rm sl}(2)$-approximation.
We will follow closely the notation of \cite{Green:2008,Grimm:2022xmj}.

\subsubsection{The scalar potential towards a large complex structure boundary}
\label{sec:IIB_Vbound_LCS}

Firstly, let us consider the Type IV boundaries, or large complex structure boundaries.
The nilpotent matrix $N$ that enters the determination of the nilpotent orbit approximation of the periods \eqref{eq:HT_Per_nil} is \cite{Green:2008}:
\begin{equation}
	\label{eq:IIB_1mod_IV_N}
	N=\begin{pmatrix}
		0 & 0 & 0 & 0\\
		m & 0 & 0 & 0\\
		c & b & 0 & -m\\
		b & n & 0 & 0
	\end{pmatrix},\qquad \text{with} \quad \begin{cases}
		m,n\in\mathbb{Z}\,,\\
		b+\frac{mn}2\in\mathbb{Z}\,,\\
		c-\frac{m^2n}6\in\mathbb{Z}\,,\\
		m\neq 0, \,n>0\,,
	\end{cases}
\end{equation}
with the vector ${\bf a}_0$ being
\begin{equation}
	\label{eq:IIB_1mod_IV_a0}
	{\bf a}_0 = \left(1,0,\xi, \frac{c}{2m}\right)^T\,,
\end{equation}
At the nilpotent order approximation, the periods \eqref{eq:HT_Pi_sym} can be computed out of the holomorphic prepotential
\begin{equation}
	\label{eq:IIB_1mod_IV_prepot}
	\mathcal{F}_{\text{\tiny{LCS}}}^{\rm nil} = -\frac{n}{6m} \frac{(X^1)^3}{X^0}-\frac{b}{2m} (X^1)^2-\frac{c}{2m} X^1X^0-\frac{1}{2}\xi (X^0)^2
\end{equation}
whence we obtain the periods
\begin{equation}
	\label{eq:IIB_1mod_IV_period}
	{\bm \Pi}_{\text{\tiny{LCS}}}^{\rm nil} =e^{zN}\bf{a}_0=\begin{pmatrix}
		1\\
		m z\\
		-\frac16 m^2nz^3+ \frac12 cz +\xi\\
		\frac12 mnz^2 +bz+\frac{c}{2m}
	\end{pmatrix}\,,
\end{equation}
once we gauge-fix $X^0 = 1$, $X^1 = m z$.

The K\"ahler potential \eqref{eq:HT_Kcsb} is then given by
\begin{equation}
    \label{eq:IIB_1mod_IV_Kcs}
    K^{\rm cs}_{\text{\tiny{LCS}}} \simeq - \log \left( \frac{4}{3} m^2 n s^3 + 2\, \Im \xi \right) \,,
\end{equation}
and the scalar potential, in the ${\rm sl}(2)$-approximation, is as in \eqref{eq:HT_Vfluxquad_nil}, with
\begingroup
\renewcommand*{\arraystretch}{1.5}
\begin{equation}
    \label{eq:IIB_1mod_IV_Z}
    \mathcal{Z}_{\text{\tiny{LCS}}}(s) \simeq 
    \begin{pmatrix}
        \frac{3 c^2 s^2 + m^4 n^2 s^6 + 36 (\Re \xi)^2}{6 m^2 n s^3} & \frac{c}{m^3 n s^3} (b m s^2 + \Re \xi) & -\frac{6 \Re \xi}{m^2 n s^3} & -\frac{c}{m n s}
        \\
        \frac{c}{m^3 n s^3} (b m s^2 + \Re \xi) & \frac{3 c^2 + 4 b^2 m^2 s^2 + m^4 n^2 s^4}{2m^4 n s^3} & -\frac{3c}{m^3 n s^3} & -\frac{2b}{mns}
        \\
        -\frac{6 \Re \xi}{m^2 n s^3} & -\frac{3c}{m^3 n s^3} & \frac{6}{m^2 n s^3} & 0
        \\
        -\frac{c}{m n s} & -\frac{2b}{mns} & 0 & \frac{2}{ns}
    \end{pmatrix} \,.
\end{equation}
\endgroup
For the following sections, it will be convenient to further introduce the integral parameters
\begin{equation}
    \label{eq:IIB_1mod_IV_par}
    \beta = b + \frac{mn}{2} \in \mathbb{Z} \,, \qquad \chi = c - \frac{m^2n}{6} \in \mathbb{Z} \,.
\end{equation}

\subsubsection{The scalar potential towards a Tyurin boundary}
\label{sec:IIB_Vbound_Tyurin}

The second kind of complex structure boundary located at infinite field distance is the \emph{Tyurin boundary}\cite{arxiv.math/0302101,Bastian:2021eom}.
Such a boundary is specified by the log-monodromy matrix \cite{Green:2008,Grimm:2022xmj}
\begin{equation}
\label{eq:IIB_1mod_II_N}
    N=\begin{pmatrix}
    0 & 0 & 0 & 0\\
    0 & 0 & 0 & 0\\
    m & 0 & 0 & 0\\
    0 & n & 0 & 0\\
    \end{pmatrix},\qquad \text{with}\quad m,n\in \mathbb{N}\,,
\end{equation}
and the vector
\begin{equation}
\label{eq:IIB_1mod_II_a0}
    {\bf a}_0 = \left(1,0,\im\, \alpha, \im\,\alpha c + d\right)^T\,,\qquad \text{with } \alpha = \sqrt{\frac{m}{n}}\quad \text{and}\quad c,d \in \mathbb{R}.
\end{equation}
Thus, the nilpotent-orbit approximated period vector \eqref{eq:HT_Per_nil} is
\begin{equation}
\label{eq:IIB_1mod_II_period}
    {\bm \Pi}_{\text{\tiny{Tyurin}}}^{\rm nil}=e^{zN}\bf{a}_0=\begin{pmatrix}
    1\\
    \im\, \alpha \\
    m z\\
    d + \im\, c \alpha + \im\, n \alpha z
    \end{pmatrix}.
\end{equation}
In \cite{Grimm:2022xmj} it has been shown that the periods \eqref{eq:IIB_1mod_II_period} can be equivalently obtained from the prepotential
\begin{equation}
\label{eq:IIB_1mod_II_prep}
    \mathcal{F}_{\text{\tiny{Tyurin}}} = \im X^0 X^1 + \im \frac{\alpha d}{2} (X^0)^2\,,
\end{equation}
upon performing a symplectic transformation of the periods \eqref{eq:IIB_1mod_II_period} and for a suitable gauge-fixing.

Inserting the nilpotent orbit-approximated periods in the general expression of the complex structure K\"ahler potential \eqref{eq:HT_Kcsb}, we obtain
\begin{equation}
    \label{eq:IIB_1mod_II_Kcs}
    K^{\rm cs}_{\text{\tiny{Tyurin}}} \simeq - \log \left( 4 ms - 2 \sqrt{\frac{m}{n}} d \right) \,.
\end{equation}
In particular, we notice that, with respect to the large complex structure K\"ahler potential \eqref{eq:IIB_1mod_IV_Kcs}, it grows slower, as $K^{\rm cs}_{\text{\tiny{Tyurin}}} \sim - \log s$ for large saxion $s$.
Additionally, in order to ensure that \eqref{eq:IIB_1mod_IV_Kcs} is well-defined within our approximation, we need that $s > \frac{d}{2 \sqrt{mn}}$.

In the ${\rm sl}(2)$-approximation, the scalar potential acquires the form \eqref{eq:HT_Vfluxquad_nil}, with the saxion-dependent matrix
\begingroup
\renewcommand*{\arraystretch}{1.5}
\begin{equation}
    \label{eq:IIB_1mod_II_Z}
    \mathcal{Z}_{\text{\tiny{Tyurin}}}(s) \simeq 
    \begin{pmatrix}
        ms + \frac{d^2}{4ns} & \frac{cd}{2ns} & 0 & - \frac{d}{2ns} 
        \\
        \frac{cd}{2ns} & ns + \frac{c^2}{ns} + \frac{d^2}{4ms} & - \frac{d}{2 ms} & -\frac{c}{ns}
        \\
        0 & - \frac{d^2}{2ms} & \frac{1}{ms} & 0 
        \\
         - \frac{d}{2ns} & -\frac{c}{ns} & 0 & \frac{1}{ns}
    \end{pmatrix} \,.
\end{equation}
\endgroup

\section{de Sitter uplifts from the penumbra}
\label{sec:dS_uplift}

A crucial ingredient for the realization of inflation is the presence of a meta-stable de Sitter vacuum, towards which the inflationary paths \eqref{eq:Sw_Infl_Pinfl} ends, thus leading to the reheating phase of the universe. 
But where, in the field space, can such a de Sitter vacuum be located, if it exists?

Momentarily, let us focus only on the complex structure sector.
The scalar potential \eqref{eq:HT_Vfluxquad} of the type IIB effective field theories that we are considering is positive semi-definite.
Hence, if it admits a vacuum, then it must be a Minkowski, or a de Sitter vacuum.
However, for a given choice of RR fluxes ${\bf f}$ and within the axion fundamental domain \eqref{eq:HT_a}, such a vacuum cannot be located too close to a field space boundary, where the saxion $s$ acquires very large vevs, namely $s \gg 1$.
In fact, in such a strict asymptotic approximation, the scalar potential \eqref{eq:HT_Vfluxquad} is well approximated by a monomial, as in \eqref{eq:HT_Vfluxquad_sl2}, and cannot display non-trivial critical points.
As observed in \cite{Grimm:2019ixq}, such a feature is shared by large classes of scalar potentials stemming from the compactification of string theory over a Calabi-Yau three-fold, and is a reflection of the famous Dine-Seiberg problem \cite{Dine:1985he}.

Thus, on the one hand, in the hope to find critical points for the scalar potential \eqref{eq:HT_Vfluxquad} we must move away from the strict asymptotic region, and consider regions of the field space where the saxion vevs are not necessarily very large.
On the other hand, small values of the saxion invalidate the approximation \eqref{eq:HT_Per_nil} for the periods, and consequently the approximation \eqref{eq:HT_Vfluxquad_nil} for the scalar potential, for the exponential corrections of order $\mathcal{O}(e^{-2n\pi s})$ appearing in \eqref{eq:HT_Per_exp} become important.
Hence, we ought to focus on regions of the moduli space where the saxion vev is just \emph{moderately large} in such a way that, in this regime, the scalar potential \eqref{eq:HT_Vfluxquad} is \emph{not} well-approximated by a monomial, and several competing terms may deliver critical point, while keeping the exponential corrections of \eqref{eq:HT_Per_exp} suppressed.
Thus, we define:

\begin{figure}[H]
    \centering
    \includegraphics[width=7cm]{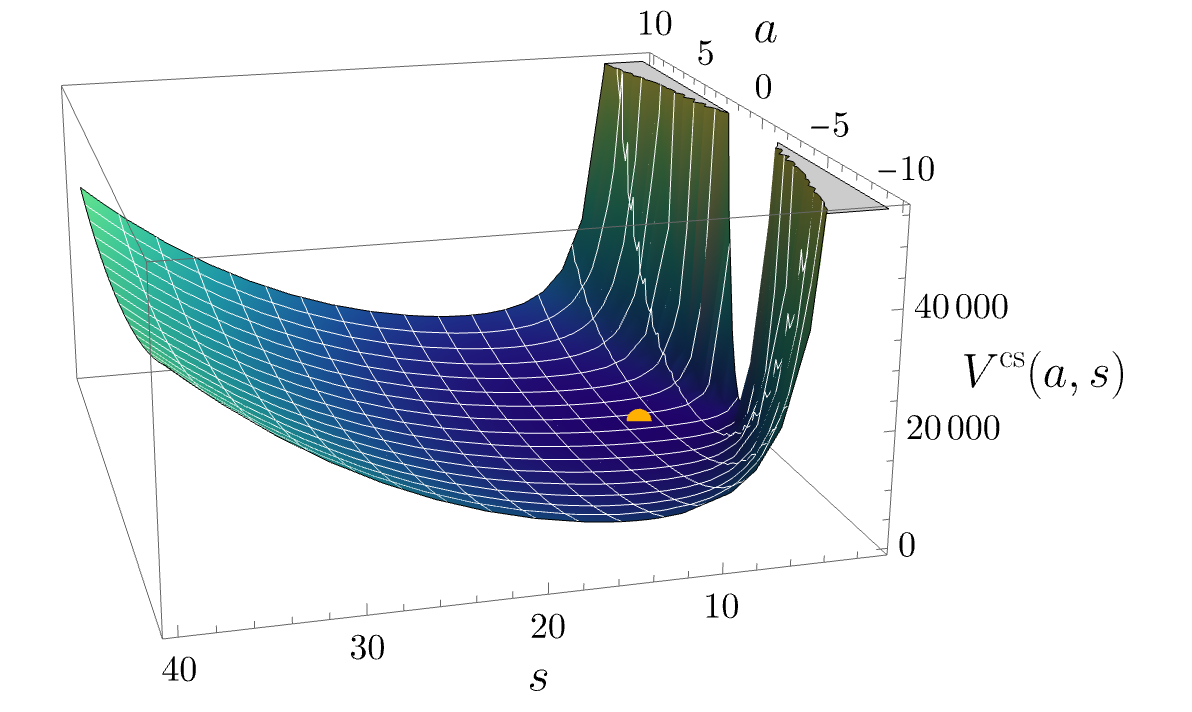} \includegraphics[width=7cm]{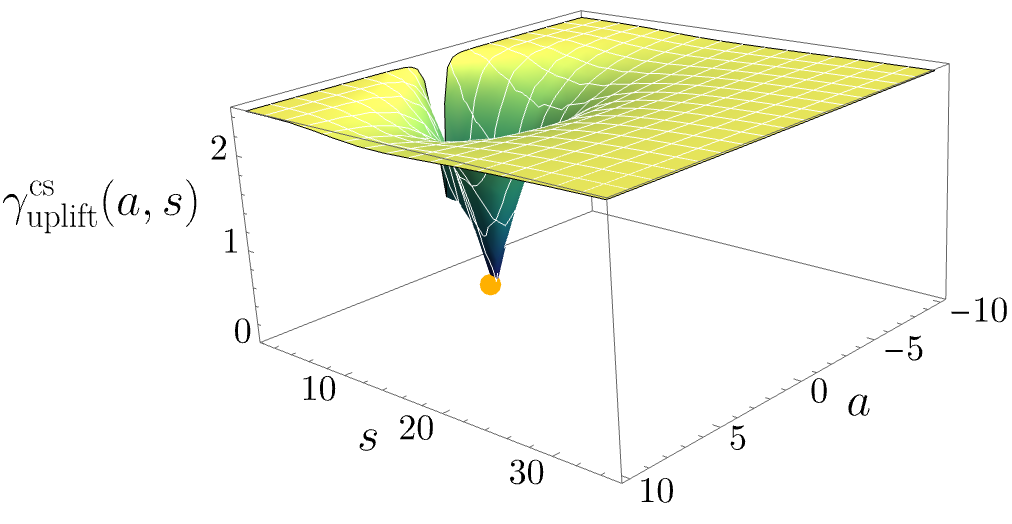} 
    \caption{\footnotesize The scalar potential (left) \eqref{eq:HT_Vfluxquad_nil} 
    and the associated uplift de Sitter coefficient (right) \eqref{eq:HT_dScoeff_cs} 
    towards the large complex structure boundary obtained employing the log-monodromy matrix \eqref{eq:IIB_1mod_IV_N} and the ${\rm sl}(2)$-approximated matrix \eqref{eq:IIB_1mod_IV_Z}, particularized to the choice of geometric parameters \eqref{eq:LCS_Ex_Ib_par}. 
    The orange dot denotes the location of the minimum of the scalar potential.
    \label{Fig:LCS_Ex_Ib_min}}
\end{figure}

\begin{importantbox}
    \noindent\emph{Strict asymptotic region:} A field space region where the fields acquire very large vevs, such that, within the scalar potential, it is always possible to single out a leading term, with definite asymptotic behavior,
\end{importantbox}
which is distinguished from 
\begin{importantbox}
    \noindent\emph{Penumbra (or penumbral region):} A field space region where the fields acquire moderately large vevs, so that the scalar potential is given by several, competing terms which now include the first subleading terms that have been dropped in the strict asymptotic region.
\end{importantbox}

In general, it may still be hard to obtain critical points for the scalar potential $V^{\rm cs}$ within the penumbra region, since it may be nontrivial to have several competing contributions to the scalar potential that could consistently deliver a critical point. 
However, the effective field theories defined towards infinite-distance boundaries are not unique, and they come in large families: indeed, both the effective theories defined towards the LCS and the Tyurin boundaries introduced in Section~\ref{sec:IIB_Vbound_LCS} and~\ref{sec:IIB_Vbound_Tyurin} depend on several geometric parameters, namely those entering the periods \eqref{eq:IIB_1mod_IV_period}, \eqref{eq:IIB_1mod_II_period}, or the scalar potential-defining matrices \eqref{eq:IIB_1mod_IV_Z}, \eqref{eq:IIB_1mod_II_Z}.
As will become clear in the explicit examples below, assuming that these geometric parameters can acquire large, or just moderately large values may offset the strength of the different contributions to the scalar potential \eqref{eq:HT_Vfluxquad}, consequently leading to nontrivial critical points.

Finally, it is worth remarking that the complex structure modulus is not the sole modulus entering the scalar potential \eqref{eq:HT_Vfluxquad}.
In fact, the scalar potential also depends on the K\"ahler moduli and the axio-dilaton, and, in the framework that we are considering, their dependence in only encoded in the prefactor $e^{\hat{K}}$.
As such, the K\"ahler moduli and the axio-dilaton \emph{cannot} be stabilized with the scalar potential \eqref{eq:HT_Vfluxquad}.
Consequently, even though the contribution $V_{\bf f}^{\rm cs}$ entering the scalar potential \eqref{eq:HT_Vfluxquad} has a critical point located at $(a_*, s_*)$ with respect to the complex structure, such that $V^{\rm cs}(a_*, s_*) > 0$, the field space point $(a_*, s_*)$ is not a fully-fledged vacuum, for the K\"ahler moduli and the axio-dilaton are runaway directions.
Still, this indicates that the complex structure modulus $z$ offers a \emph{de Sitter uplift} for the full scalar potential \eqref{eq:HT_Vfluxquad}.
Whether such an uplift is enough to guarantee that a fully-fledged vacuum is still characterized by a positive value of the scalar potential is an open question.

Hence, to summarize, the working hypotheses that we will employ in the forthcoming discussion can be summarized as follows:
\begin{description}
    \item[Only a single, dynamical complex structure modulus:] we restrict our attention solely to a single, dynamical complex structure modulus, and we will disregard the remaining axio-dilaton and K\"ahler moduli.
    Concretely, we will be treating the prefactor $e^{\hat{K}}$ that appears in \eqref{eq:HT_Vfluxquad_nil} as a constant;
    \item[Monodromy-parameters hierarchy:] we shall assume that the parameters that enter the log-monodromy matrices \eqref{eq:HT_Per_nil} -- such as those appearing in \eqref{eq:IIB_1mod_IV_N} or \eqref{eq:IIB_1mod_II_N}  -- can be \emph{hierachically separated}. Namely, one, or some parameters entering the log-monodromy matrices \eqref{eq:HT_Per_nil} can be larger than the remaining by some order of magnitudes;
    \item[Explorable `penumbra' region:] we depart from the strict asymptotic regime, by considering \emph{moderately large} values of the saxion $s$, for which the approximation \eqref{eq:HT_Per_nil} still holds. 
\end{description}
Below, we shall see that, under these assumptions, the four-dimensional type IIB effective field theories defined towards LCS and Tyurin boundaries introduced in Section~\ref{sec:IIB_Vbound_LCS} and~\ref{sec:IIB_Vbound_Tyurin} offer \emph{large} families of possible de Sitter uplifts.

\subsection{de Sitter uplifts towards LCS boundaries}
\label{sec:dS_uplift_LCS}

As a first, concrete example of a de Sitter uplift, let us consider an effective field theory defined towards a large complex structure boundary, within the family of theories introduced in Section~\ref{sec:IIB_Vbound_LCS}.
Such an effective theory is endowed with the scalar potential of the form \eqref{eq:HT_Vfluxquad_nil}, with the log-monodromy matrix \eqref{eq:IIB_1mod_IV_N} and saxion-dependent matrix \eqref{eq:IIB_1mod_IV_Z}.
To further specify the model, we choose the following fluxes and monodromy parameters:
\begin{equation}
    \label{eq:LCS_Ex_Ib_par}
    e_0 = e_1 = m^0 =m^1 = 1\,, \qquad m = 1\,, \qquad n=6\,, \qquad \beta = \chi = -100 \,, \qquad \xi = 0\,.
\end{equation}

The scalar potential obtained this way and the associated de Sitter coefficient \eqref{eq:HT_dScoeff_cs} are plotted in Figure~\ref{Fig:LCS_Ex_Ib_min}.

\begin{figure}[H]
    \centering
    \includegraphics[width=15cm]{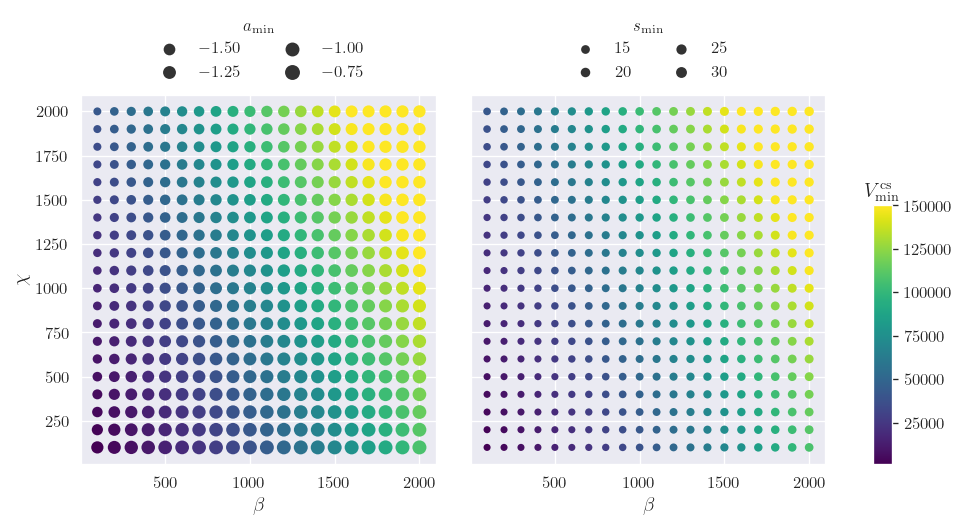} 
    \caption{\footnotesize 
    Values of the axion and saxion minima $a_{\rm min}$, $s_{\rm min}$ for different values of $\beta$, $\chi$.
    The fluxes are fixed as $e_0 = e_1 = m^0 =m^1 = 1$, and the additional geometric parameters are set to $m =1$, $n = 6$, $\xi = 0$.
    \label{Fig:LCS_min_scan}}
\end{figure}

Interestingly, at such a level of approximation, the scalar potential so constructed exhibits a de Sitter critical point located at 
$a_{\rm min} \simeq 0.3517$, $s_{\rm min} \simeq 7.183$, at which $V_{\text{flux}}(a_{\rm min},s_{\rm min}) \simeq 724.1 e^{\hat K} M_{\rm P}^4$. 
It is worth noticing that by appropriately stabilizing the K\"ahler moduli, the scalar potential can easily be small in Planck units. For instance, for K\"ahler moduli stabilization to be reliable typically requires the CY volume ${\cal V}\gtrsim 10^3$. This implies $e^{\hat K}\sim \frac{1}{{\cal V}^2}\lesssim 10^{-6}$ and thus we expect in a fully stabilized model $V_{\text{flux}}(a_{\rm min},s_{\rm min})\lesssim 10^{-2}$ for our example here.

Consequently, there are regions of the $(a,s)$-plane for which the uplift de Sitter coefficient \eqref{eq:HT_dScoeff_cs} is small: as we shall stress further in the following section, in particular, in the penumbra region, the de Sitter coefficient \eqref{eq:HT_dScoeff_cs} may avoid the strong de Sitter conjecture bound \eqref{eq:Sw_Infl_c_strong}.

The features just outlined for the specific LCS model with parameters chosen as \eqref{eq:LCS_Ex_Ib_par} are not unique.
In fact, de Sitter uplifts exist for a \emph{large} number of choices of parameters.
For instance, in Figure~\ref{Fig:LCS_min_scan} we have plotted a scan of de Sitter minima, performed with \textsf{python}, for $20$ different, positive values of the parameters $\beta$ and $\chi$ defined in \eqref{eq:IIB_1mod_IV_par}, with the fluxes set at $e_0 = e_1 = m^0 =m^1 = 1$, and choosing the remaining geometric parameters as $m =1$, $n = 6$, $\xi = 0$.
In particular, in Figure~\ref{Fig:LCS_min_scan}, each dot denotes the point in the parameter space $(\beta,\chi)$ analyzed, and the size of the dot indicates the vev that the axion and the saxion take at the de Sitter minimum on the left and the right plot, respectively.
Furthermore, the color of each point indicates the value of the scalar potential $V^{\text{cs}}$ at the given de Sitter minimum.

\begin{figure}[H]
    \centering
    \includegraphics[width=7cm]{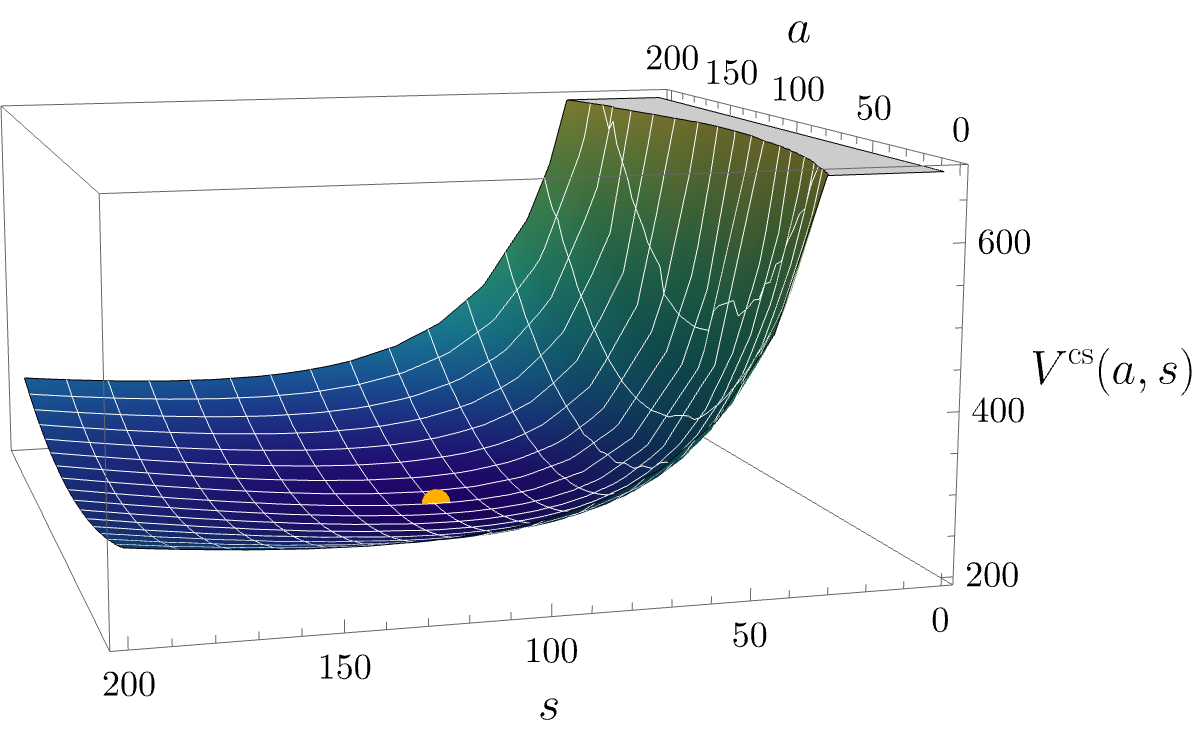} \includegraphics[width=7cm]{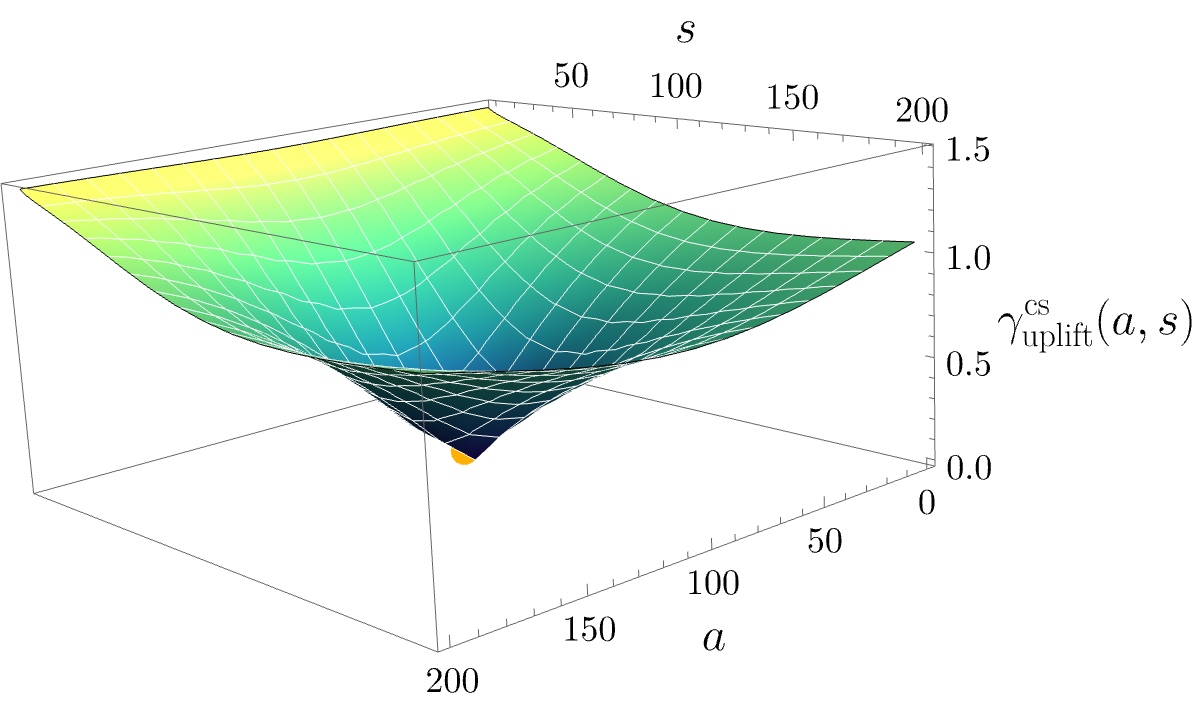} 
    \caption{\footnotesize The scalar potential (left) \eqref{eq:HT_Vfluxquad_nil} 
    and the uplift de Sitter coefficient (right) \eqref{eq:HT_dScoeff_cs}
    towards the Tyurin boundary obtained employing the log-monodromy matrix \eqref{eq:IIB_1mod_II_N} and the ${\rm sl}(2)$-approximated matrix \eqref{eq:IIB_1mod_II_Z}, particularized to the choice of geometric parameters \eqref{eq:Tyurin_Ex_I_par}. 
    The orange dot denotes the location of the minimum of the scalar potential.
    \label{Fig:Tyurin_Ex_I_min}}
\end{figure}

\subsection{de Sitter uplifts towards Tyurin boundaries}
\label{sec:dS_uplift_Tyurin}

The effective field theories defined towards the Tyurin boundaries introduced in Section~\ref{sec:IIB_Vbound_Tyurin} may also deliver de Sitter uplifts.
For example, let us choose the following fluxes and geometric parameters: 
\begin{equation}
    \label{eq:Tyurin_Ex_I_par}
    e_0 = e_1 = m^0 =m^1 = 1\,, \qquad m = n = 1\,, \qquad d = -1 \,, \qquad c = -200\,.
\end{equation}
The scalar potential, obtained using the log-monodromy matrix \eqref{eq:IIB_1mod_II_N} and the saxion-dependent matrix \eqref{eq:IIB_1mod_II_Z}, and the corresponding de Sitter coefficient \eqref{eq:HT_dScoeff_cs} are plotted in Figure~\ref{Fig:Tyurin_Ex_I_min}.

Indeed, within the approximation that we are considering, with respect to the complex structure moduli only, the scalar potential \eqref{eq:HT_Vfluxquad_nil} displays a de Sitter critical point located at $a_{\rm min} \simeq s_{\rm min} \simeq 100$, at which $V^{\text{cs}}(a_{\rm min},s_{\rm min}) \simeq 200\, e^{\hat K} M_{\rm P}^4$.
Hence, such an effective theory offers a potential de Sitter uplift for the full scalar potential \eqref{eq:HT_Vfluxquad_nil}. Again, from reliable regimes of K\"ahler moduli stabilization we expect  $e^{\hat K}\sim \frac{1}{{\cal V}^2}\lesssim 10^{-6}$ and thus $V_{\text{flux}}(a_{\rm min},s_{\rm min})\lesssim 10^{-2}$ for our example here.

\begin{figure}[H]
    \centering
    \includegraphics[width=15cm]{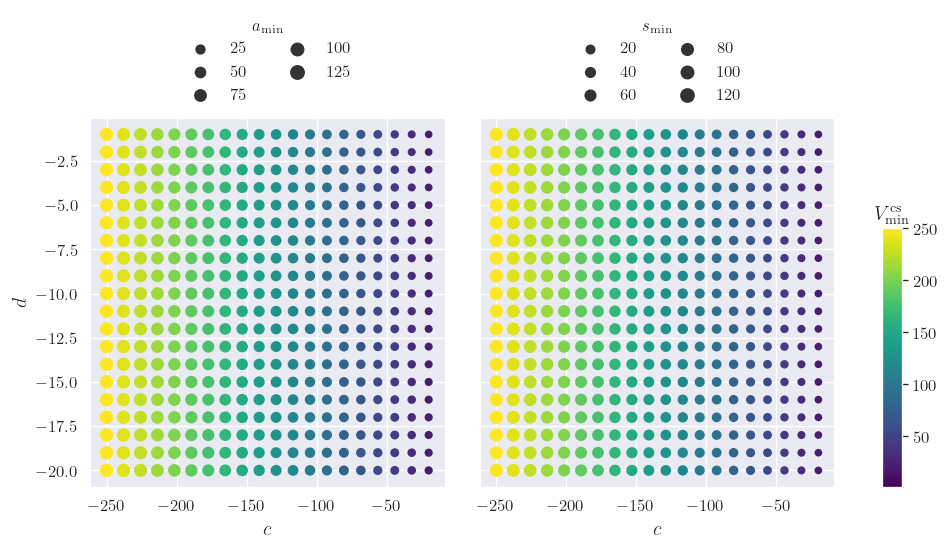} 
    \caption{\footnotesize 
    Values of the axion and saxion minima $a_{\rm min}$, $s_{\rm min}$ for different values of $c$, $d$.
    The fluxes are fixed as $e_0 = e_1 = m^0 =m^1 = 1$, and the additional geometric parameters are set to $m = n =1$.
    \label{Fig:Tyurin_min_scan}}
\end{figure}

As noticed for the families of effective theories defined towards an LCS boundary, the existence of such a de Sitter uplift is not uncommon.
Rather, they come in large families, since a large number of possible geometric parameter choices may lead to a de Sitter uplift.
Indeed, in Figure~\ref{Fig:Tyurin_min_scan} is a scan of complex structure minima, performed with \textsf{python}.
The scan is performed over the $(c,d)$ parameter space, and we have fixed the fluxes as $e_0 = e_1 = m^0 =m^1 = 1$, and the additional geometric parameters are set to $m = n =1$.
Each dot corresponds to a point in such a parameter space that is tested.
The size of each dot on the left and on the right plot denotes the vev that the axion and the saxion acquire at the minimum, respectively; furthermore, the color of each dot indicates the value that the complex structure scalar potential $V^{\text{cs}}$ acquires at the given minimum.

\section{Candidate axion monodromy valleys}
\label{sec:Infl_Examples}

If a given stringy effective field theory ought to accommodate axion monodromy inflation as it was originally conceived in \cite{Silverstein:2008sg,McAllister:2008hb,Kaloper:2011jz}, then the effective theory has to host appropriate, well-controlled paths in the axion space, while leaving the saxion fields fixed at their own vevs.
In the models that we are considering in this work, which involve solely a single complex structure modulus, this translates into the requirement of having scalar potential \emph{valleys} along which only the axion varies significantly.
Indeed, such a candidate valley can be obtained by solving the equation
\begin{equation}
    \label{eq:Ex_sv_def}
    \frac{\partial V}{\partial s} \Big|_{s = s_v(a)} = 0 \,.
\end{equation}
This is generically solved by a function $s_v(a)$, which determines the shape of said valley, and quantifies the \emph{backreaction} of the axion vev onto the saxion minimum.
If the axion $a$ ought to be the sole, dynamical field at low energies, playing the role of the inflaton, we need that $s_v(a)$ is approximately constant, at least along the inflationary path.

\begin{figure}[H]
    \centering
    \includegraphics[width=7cm]{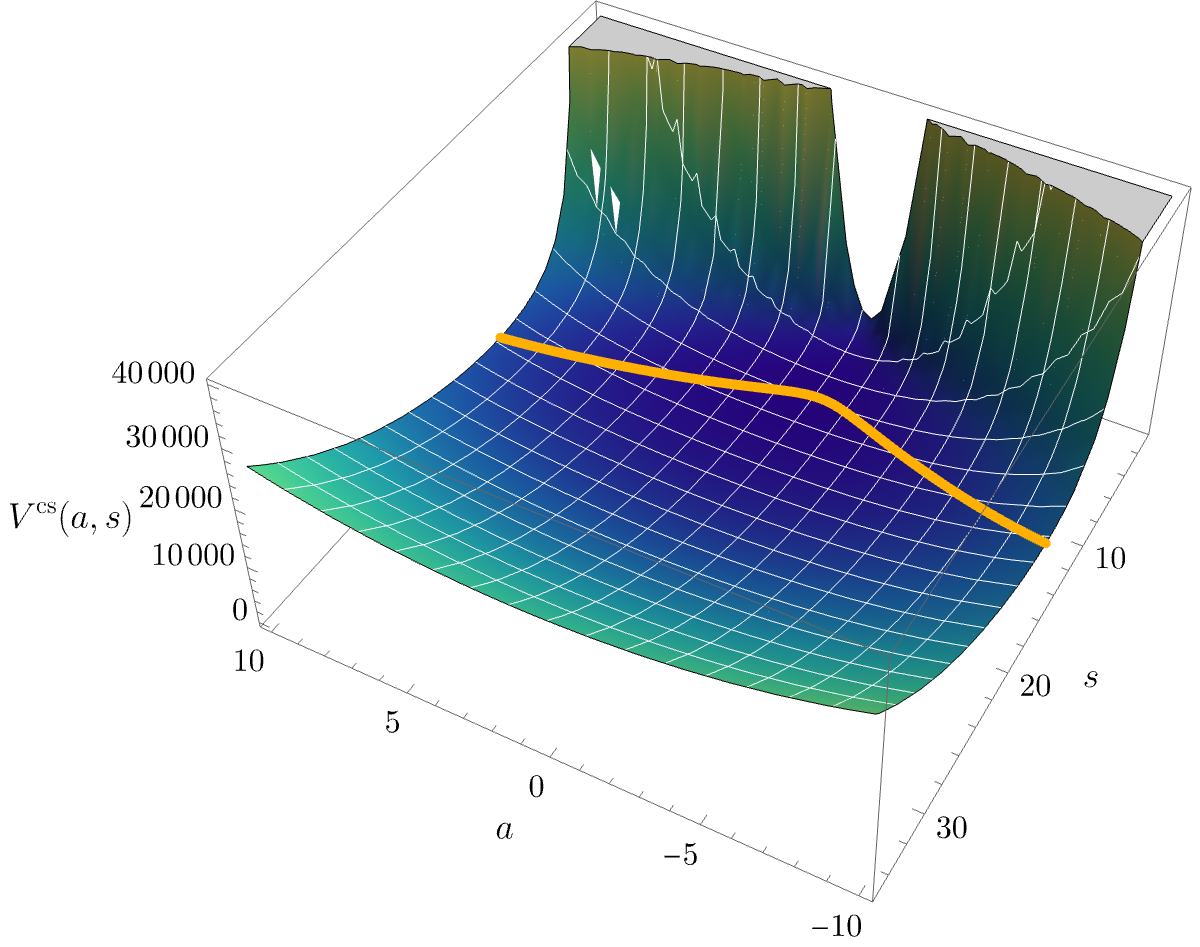} 
    \includegraphics[width=7cm]{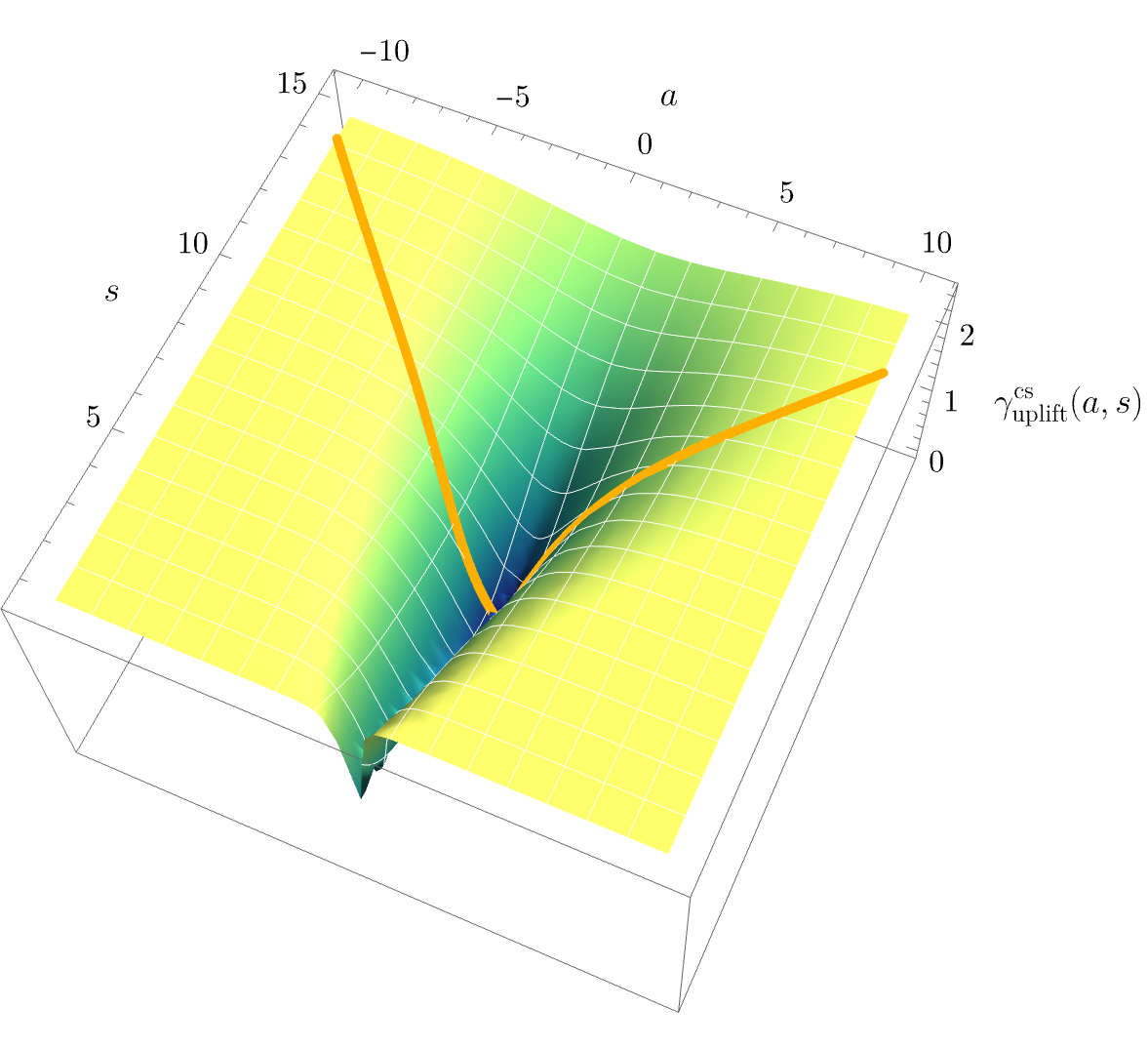} 
    \caption{\footnotesize The complex structure scalar potential (left) defined \eqref{eq:HT_Vfluxquad_nil} 
    and the associated uplift de Sitter coefficient (right) \eqref{eq:HT_dScoeff_cs}
    towards a large complex structure boundary, obtained employing the log-monodromy matrix \eqref{eq:IIB_1mod_IV_N} and the ${\rm sl}(2)$-approximated matrix \eqref{eq:IIB_1mod_IV_Z}, particularized to the choice of geometric parameters \eqref{eq:LCS_Ex_Ib_par}. 
    The orange, thick line denotes the backreaction of the axion vev onto the saxion minimum.
    \label{Fig:LCS_Ex_Ib_back}}
\end{figure}

However, four-dimensional EFTs arising from string theory are are subject to backreaction effects onto the saxion minimum caused when displacing the axion from its own minimum: namely, as has been observed in large classes of string theoretic EFTs
\cite{Blumenhagen:2014nba,Hebecker:2014kva,Baume:2016psm,Valenzuela:2016yny,Grimm:2020ouv}, whenever we displace an axion from its minimum,  the associated build-up of axion potential energy pushes the saxion vevs to inevitably change; in other words, string theory seems to consistently deliver scalar potentials whose valleys, if any, cannot lie in the axion directions only, for they have to involve saxion directions as well.
Concretely, \eqref{eq:Ex_sv_def} cannot be generically solved for a constant $s_v(a) = s_{\rm min}$.
In particular, asymptotically in the field space, certain classes of stringy examples exhibit \emph{linear} backreactions, with the saxion minima scaling linearly with some linear combination of the axions, for large enough axion vevs.

We note that in four-dimensional EFTs from string theory which can accommodate inflationary paths, these may not need to reside asymptotically in field space, and may well extend towards the bulk of the moduli space, entering the penumbral region. 
Indeed, the penumbral region of the moduli space seems to be rather attractive for searching possible candidate valleys that may realize axion monodromy inflation.
On the one hand, as we already observed in the previous section, it is within the penumbral regions of the type IIB effective theories that de Sitter uplifts can be found; furthermore, the penumbra hosts subregions of the moduli space where the uplift de Sitter coefficient \eqref{eq:Sw_Infl_gamma} may be arbitrarily small, possibly leading to a small enough first slow-roll parameter \eqref{eq:Sw_Infl_eps} via \eqref{eq:Sw_Infl_gamma_eps}.
On the other hand, as we shall see in this section, while backreaction effects of the axion vev onto the saxion minimum are still present in the penumbral region, these are \emph{milder} than those that characterize the asymptotic regions of the moduli space.
Indeed, by examining three examples of type IIB effective field theories -- two defined towards an LCS boundary, and one defined towards a Tyurin boundary -- we shall explicitly see how the backreaction effects are ameliorated when moving from the asymptotic region towards the penumbra region of the moduli space.

\begin{figure}[H]
    \centering
    \includegraphics[width=7cm]{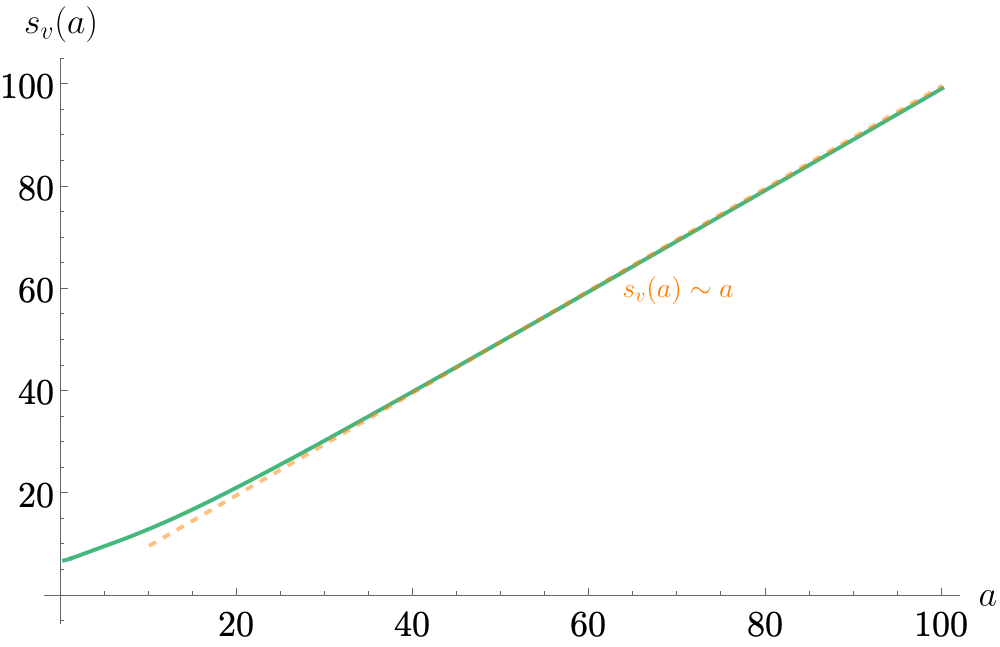} 
    \includegraphics[width=7cm]{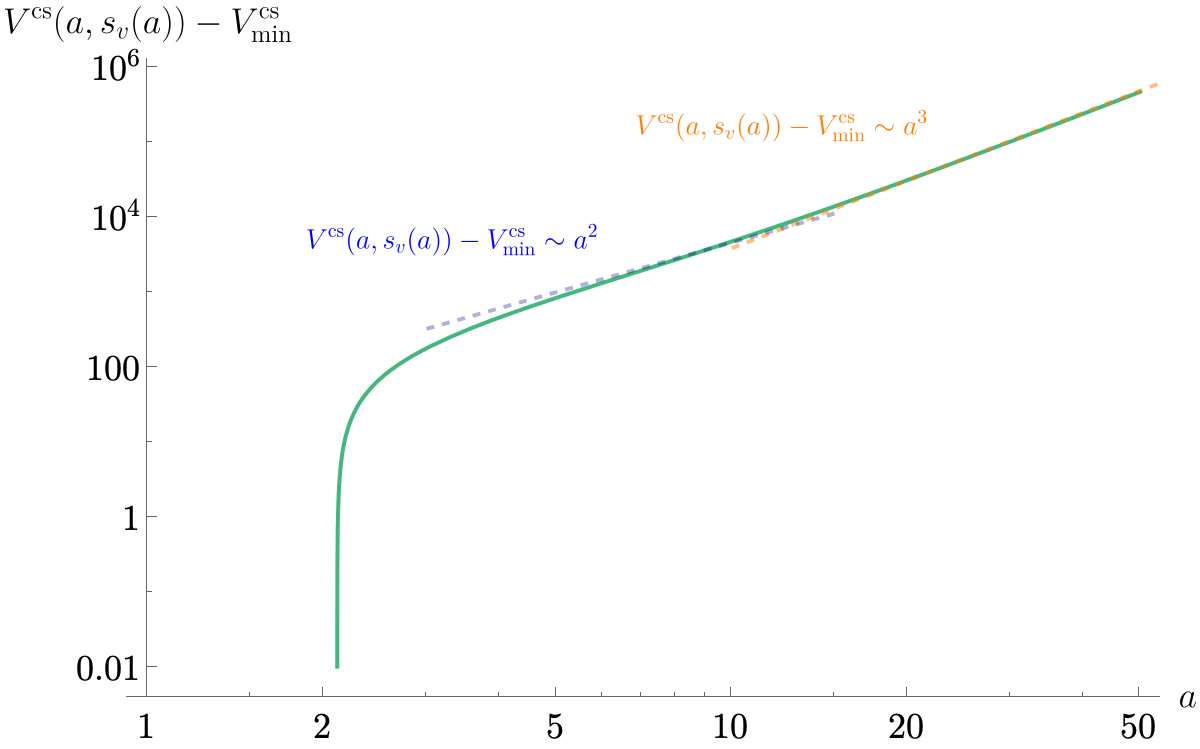} 
    \caption{\footnotesize On the left, the plot of the backreaction of the axion vev onto the saxion minimum dictated by \eqref{eq:Ex_sv_def}, for the effective theory defined towards a large complex structure boundary, specified by the parameters \eqref{eq:LCS_Ex_Ib_par}; on the right, the log-log plot of the associated complex structure scalar potential, to which we have subtracted its value at the minimum $V^{\rm cs}_{\text{min}}$.   
    For very large values of the axion $a$, $V^{\rm cs}(a,s_v(a)) \sim a^{3}$, while for smaller values of $a$ $V^{\rm cs}(a,s_v(a)) \sim a^{2}$.
    \label{Fig:LCS_Ex_Ib_valley_V}}
\end{figure}

\subsection{Examples of axion valleys towards LCS boundaries}
\label{sec:Infl_Examples_LCS}

As a first example of candidate axion valley lying in the penumbra region, let us consider the same effective field theory, defined towards an LCS boundary, as the one that we introduced in Section~\ref{sec:dS_uplift_LCS}. 
In Figure~\ref{Fig:LCS_Ex_Ib_back} we have plotted the backreaction of the axion vev onto the saxion minimum, as governed by \eqref{eq:Ex_sv_def}, along the surfaces of the scalar potential and the uplift de Sitter coefficient.

For the sake of clarity, in Figure~\ref{Fig:LCS_Ex_Ib_valley_V} are the plots of the backreacted saxion minimum $s_v(a)$ along the valley, defined via \eqref{eq:Ex_sv_def}, and the scalar potential along the valley as a function of the axion. The plot of the saxion backreaction on the left of Figure~\ref{Fig:LCS_Ex_Ib_valley_V} shows that, for large axion vevs, the backreaction behaves linearly with the axion, as expected.
However, for smaller values of the axion, within the penumbra region, the backreaction is milder, with smaller local slopes.
This behavior is indeed reflected on the plot of the scalar potential defined onto the valley on the right of Figure~\ref{Fig:LCS_Ex_Ib_valley_V}: while the scalar potential behaves as $V^{\rm cs}(a,s_v(a)) \sim a^{3}$ for large axion vevs, it displays the milder scaling $V^{\rm cs}(a,s_v(a)) \sim a^{2}$ in the penumbra region.

\begin{figure}[H]
    \centering
    \includegraphics[width=7cm]{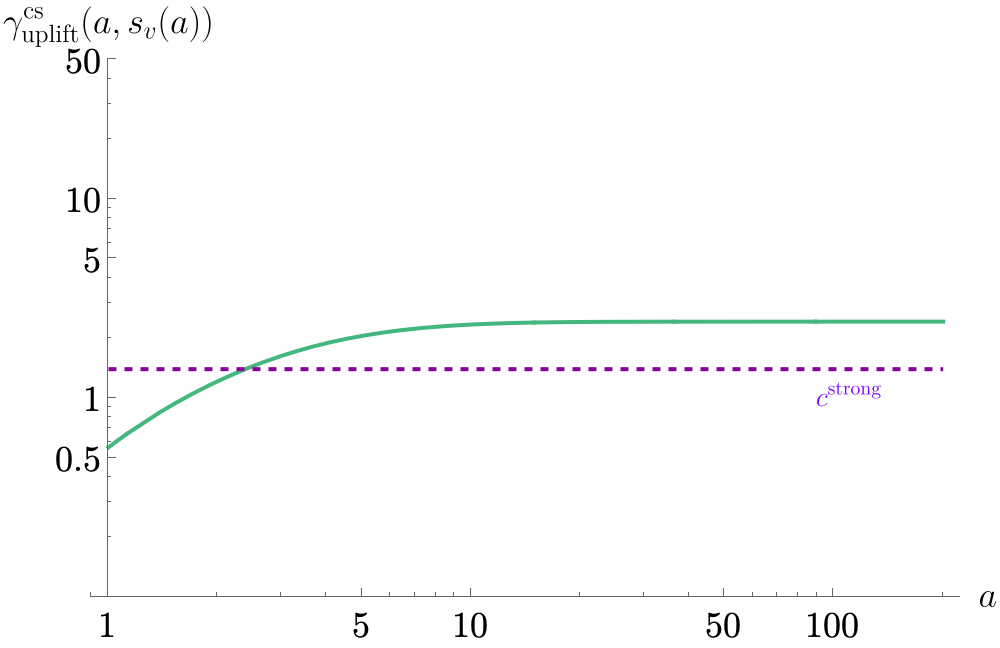} \qquad
    \includegraphics[width=7cm]{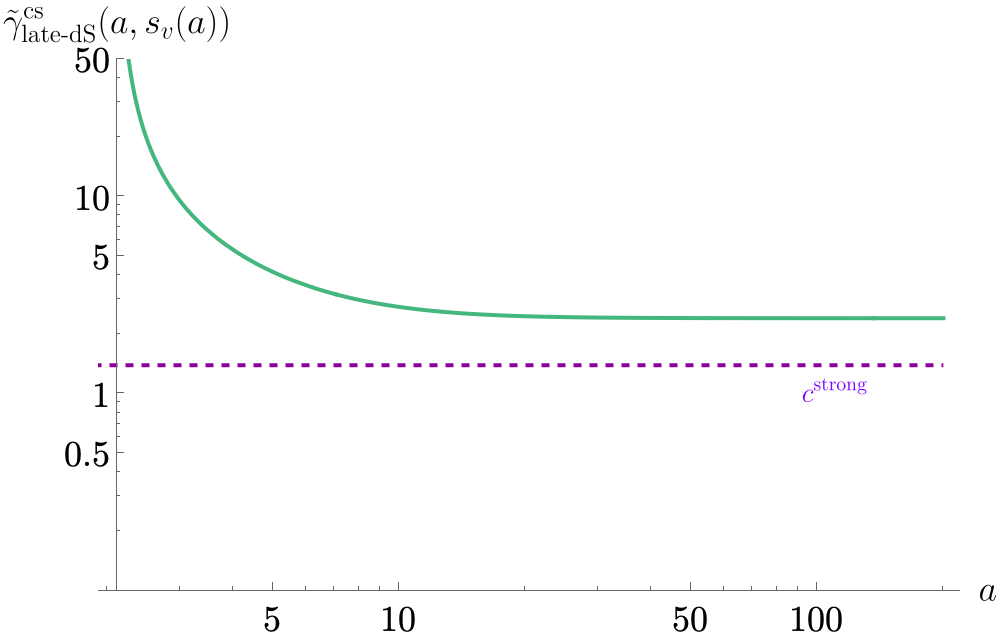} 
    \caption{\footnotesize On the left, the log-log plot of the uplift de Sitter coefficient \eqref{eq:HT_dScoeff_cs} and, on the right, the log-log plot of the late de Sitter coefficient \eqref{eq:HT_dScoeff_cs_corr} along the scalar potential valley plotted in Figure~\ref{Fig:LCS_Ex_Ib_back}.
    \label{Fig:LCS_Ex_Ib_valley_gamma}}
\end{figure}

Consequently, both the uplift de Sitter coefficient and late de Sitter coefficients \eqref{eq:HT_dScoeff_cs}, \eqref{eq:HT_dScoeff_cs_corr} vary when moving from the asymptotic region towards the penumbra region.
Indeed, on the left of Figure~\ref{Fig:LCS_Ex_Ib_valley_gamma} is plotted the uplift de Sitter coefficient \eqref{eq:HT_dScoeff_cs}: asymptotically, for large axion vevs, $\gamma^{\rm cs}_{\text{uplift}} \to \sqrt{6}$, as expected, whereas it becomes smaller towards the penumbra region.
In particular, the uplift de Sitter coefficient becomes arbitrarily small as the minimum is approached, escaping the strong de Sitter bound \eqref{eq:Sw_Infl_c_strong}.

However, the late de Sitter coefficient \eqref{eq:HT_dScoeff_cs_corr}, depicted on the right of Figure~\ref{Fig:LCS_Ex_Ib_valley_gamma}, unlike its uplift de Sitter coefficient counterpart, does not become arbitrarily small along the axion valley, and stays well above the strong de Sitter bound \eqref{eq:Sw_Infl_c_strong}.
As such, such valleys cannot be regarded as fully-fledged candidates of inflationary paths.

It is worth remarking that, while the scalar potential and saxion backreaction share universal features in the asymptotic region of the moduli space \cite{Grimm:2020ouv}, these quantities do not seem to exhibit general behaviors in the penumbra region.
Indeed, let us consider a second example of a type IIB effective field theory defined towards an LCS boundary, specified by the following geometric data:
\begin{equation}
    \label{eq:LCS_Ex_I_par}
    e_0 = e_1 = m^0 =m^1 = 1\,, \qquad m = 1\,, \qquad n=6\,, \qquad \beta = \chi = 10^3 \,, \qquad \xi = 0\,.
\end{equation}
As for the model introduced in Section~\ref{sec:dS_uplift_LCS}, the complex structure sector of this effective theory also leads to a de Sitter uplift. 
Indeed, the complex structure scalar potential, defined in \eqref{eq:HT_Vfluxquad_nil}, and computed with the log-monodromy matrix \eqref{eq:IIB_1mod_IV_N} and saxion-dependent matrix \eqref{eq:IIB_1mod_IV_Z}, with the parameters as in \eqref{eq:LCS_Ex_I_par}, is equipped with a minimum located at
$a_{\rm min} \simeq -1.000$, $s_{\rm min} \simeq 22.28$, where $V_{\text{flux}}(a_{\rm min},s_{\rm min}) \simeq 2.230 \cdot 10^4 e^{\hat K} M_{\rm P}^4$. Here as well, as reliable regimes of K\"ahler moduli stabilization typically imply  $e^{\hat K}\sim \frac{1}{{\cal V}^2}\lesssim 10^{-6}$ we expect $V_{\text{flux}}(a_{\rm min},s_{\rm min})\lesssim 10^{-2}$ for our example here.

\begin{figure}[H]
    \centering
    \includegraphics[width=7cm]{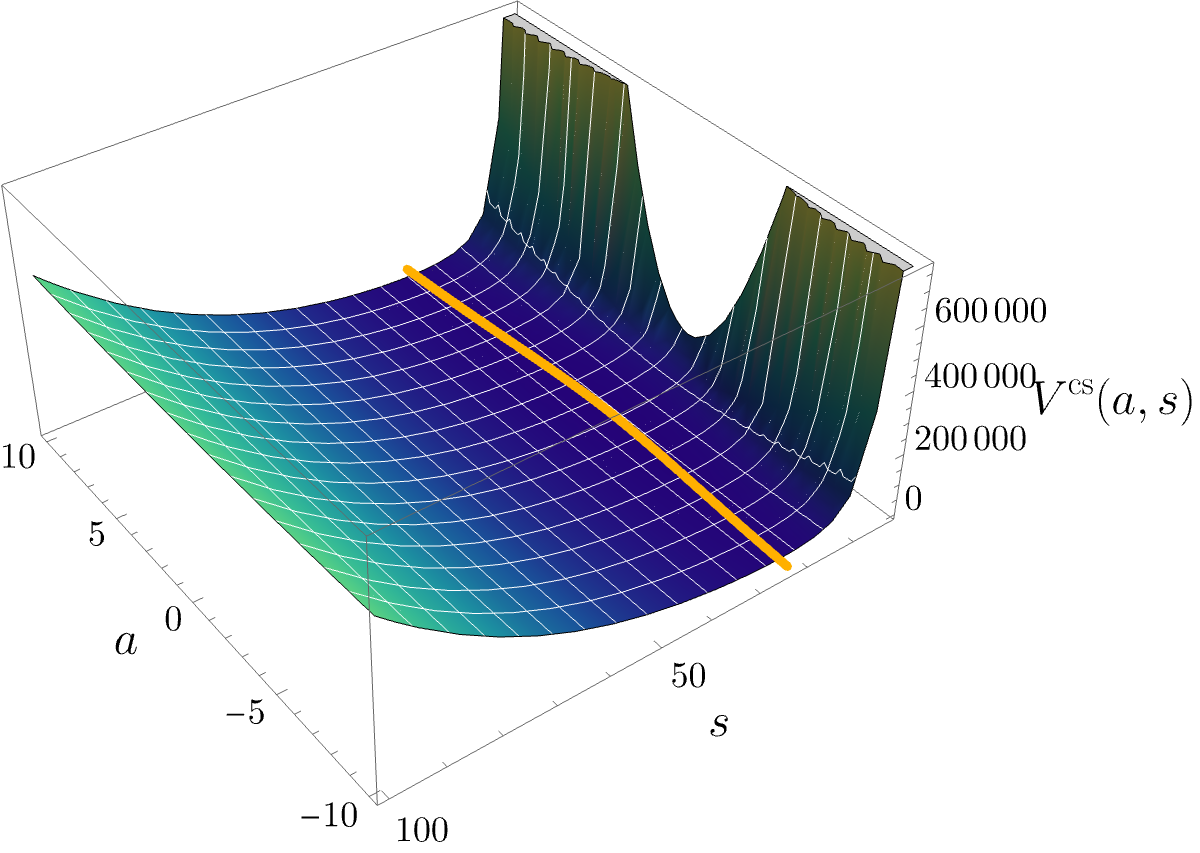} 
    \includegraphics[width=7cm]{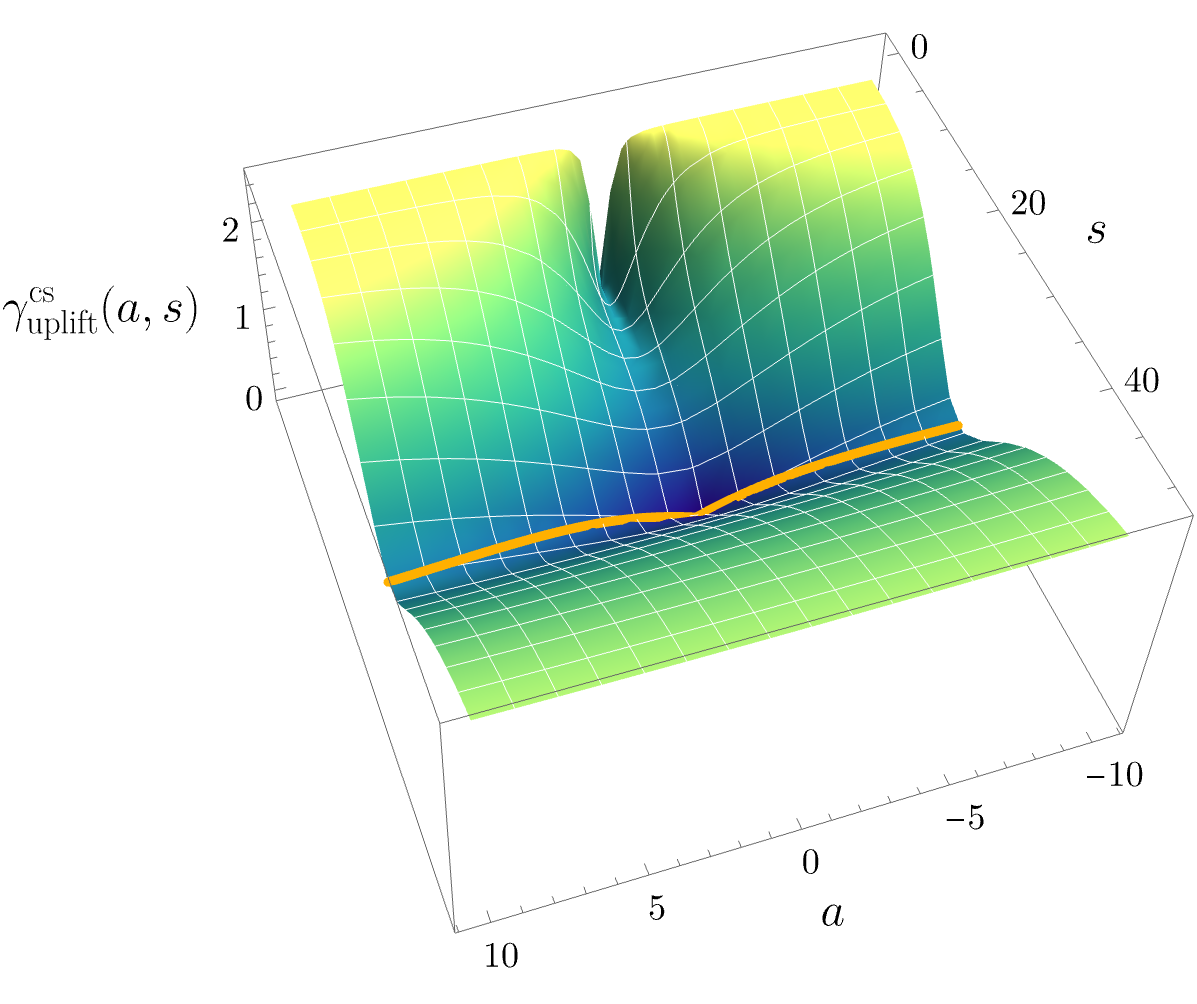} 
    \caption{\footnotesize The complex structure scalar potential (left) defined \eqref{eq:HT_Vfluxquad_nil} 
    and the associated uplift de Sitter coefficient (right) \eqref{eq:HT_dScoeff_cs}
    towards a large complex structure boundary, obtained employing the log-monodromy matrix \eqref{eq:IIB_1mod_IV_N} and the ${\rm sl}(2)$-approximated matrix \eqref{eq:IIB_1mod_IV_Z}, particularized to the choice of geometric parameters \eqref{eq:LCS_Ex_I_par}. 
    The orange, thick line denotes the backreaction of the axion vev onto the saxion minimum.
    \label{Fig:LCS_Ex_I_back}}
\end{figure}

\begin{figure}[H]
    \centering
    \includegraphics[width=7cm]{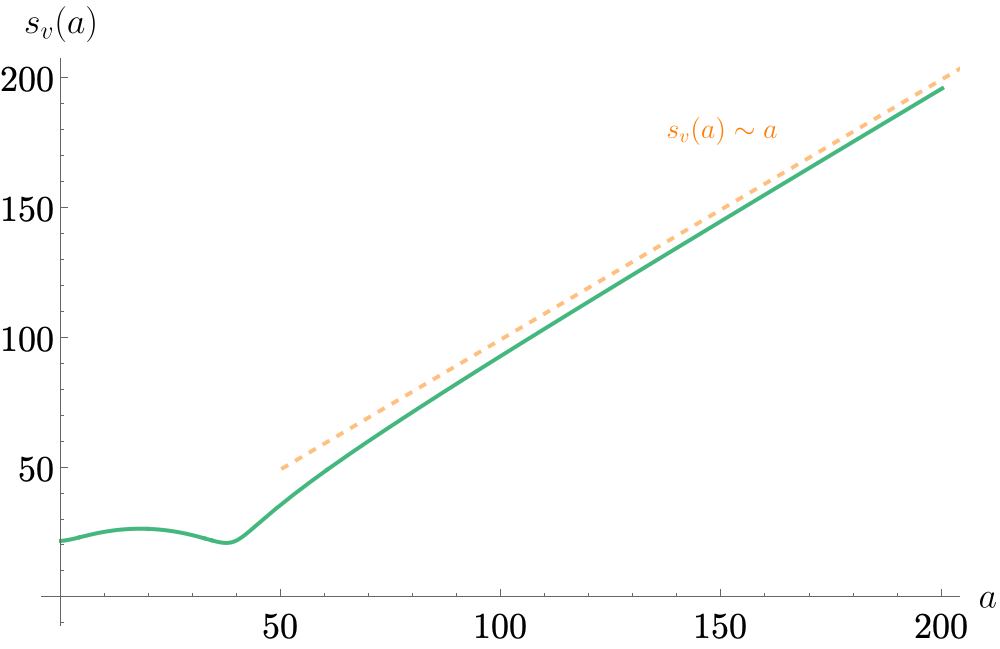} 
    \includegraphics[width=7cm]{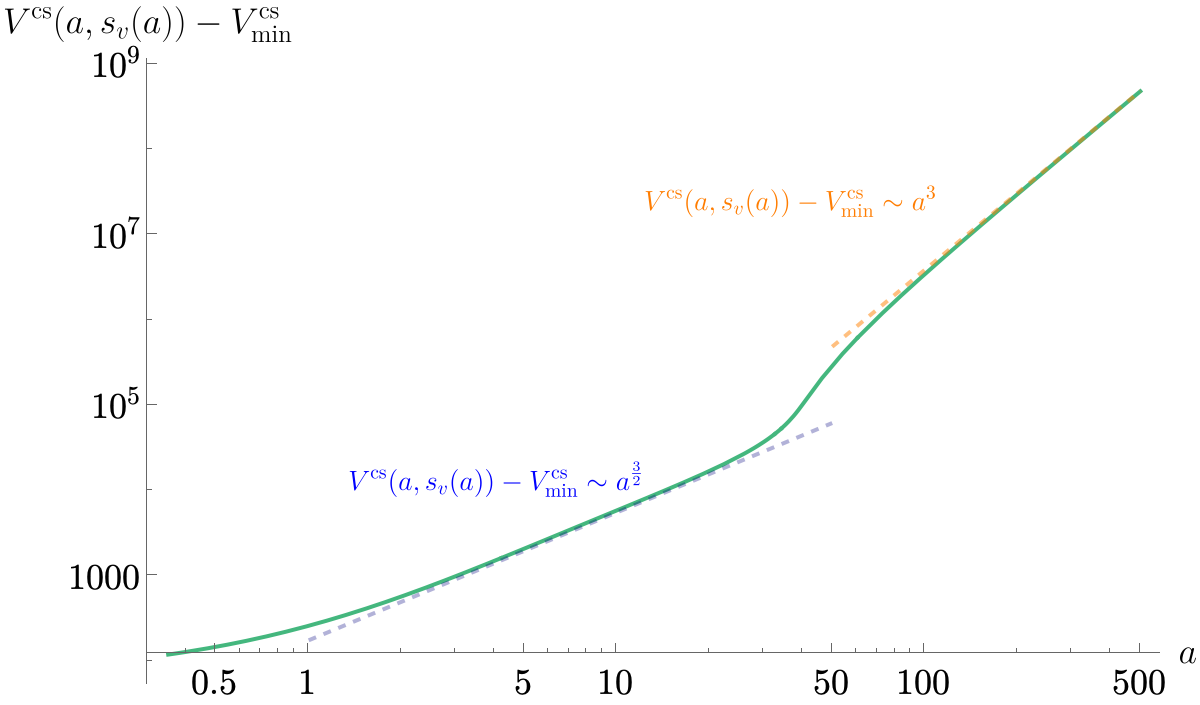} 
    \caption{\footnotesize On the left, the plot of the backreaction of the axion vev onto the saxion minimum dictated by \eqref{eq:Ex_sv_def}, for the effective theory defined towards a large complex structure boundary, specified by the parameters \eqref{eq:LCS_Ex_I_par}; on the right, the log-log plot of the associated complex structure scalar potential, to which we have subtracted its value at the minimum $V^{\rm cs}_{\text{min}}$.  
    For very large values of the axion $a$, $V^{\rm cs}(a,s_v(a)) \sim a^{3}$, while for smaller values of $a$ $V^{\rm cs}(a,s_v(a)) \sim a^{\frac32}$.
    \label{Fig:LCS_Ex_I_valley_V}}
\end{figure}

Figure~\ref{Fig:LCS_Ex_I_back} depicts for the axion valley, defined via \eqref{eq:Ex_sv_def}, on the left the scalar potential and on the right the uplift de Sitter coefficient.
To further exhibit the behavior of the axion valley, in Figure~\ref{Fig:LCS_Ex_I_valley_V} we have further plotted the backreacted saxion minimum along the valley as a function of the axion, as well as the scalar potential along the valley where we adjusted to zero the scalar potential at the minimum ending the valley.

As with the previous example, the backreaction effects become milder towards the penumbra region of the moduli space.
However, unlike the previous example, the saxion backreaction, depicted on the left of Figure~\ref{Fig:LCS_Ex_I_valley_V}, is not monotonic in the axion within the penumbra region.
Still, for values of the axion $a \gtrsim 100$, the backreaction is linear in the axion.
In turn, while the complex structure scalar potential grows as $V^{\rm cs}(a,s_v(a)) \sim a^{3}$ asymptotically, as depicted on the right of Figure~\ref{Fig:LCS_Ex_I_valley_V}, it behaves only as $V^{\rm cs}(a,s_v(a)) \sim a^{\frac32}$ towards the penumbra, with the axion vevs $a \lesssim 100$.

\begin{figure}[H]
    \centering
    \includegraphics[width=7cm]{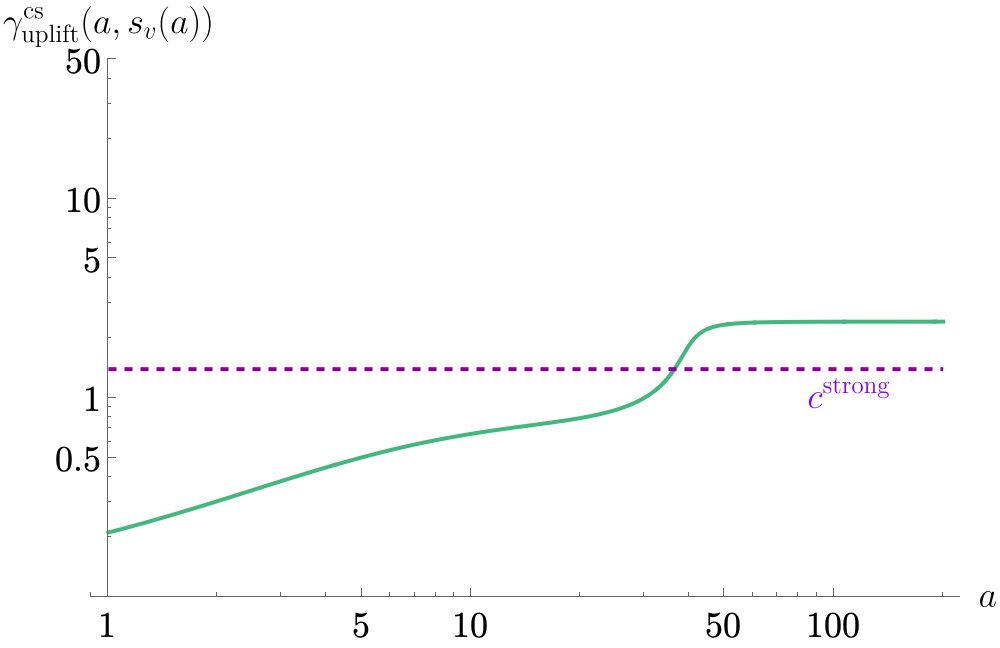} \qquad
    \includegraphics[width=7cm]{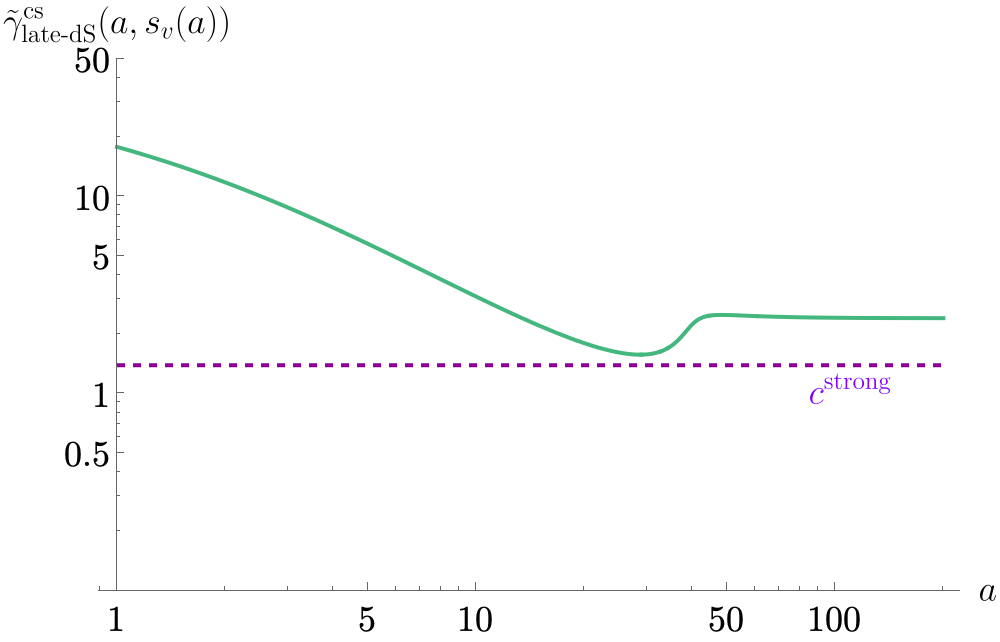} 
    \caption{\footnotesize On the left, the log-log plot of the uplift de Sitter coefficient \eqref{eq:HT_dScoeff_cs} and, on the right, the log-log plot of the late de Sitter coefficient \eqref{eq:HT_dScoeff_cs_corr} along the scalar potential valley, specified by the parameters \eqref{eq:LCS_Ex_I_par} and plotted in Figure~\ref{Fig:LCS_Ex_I_back}.
    \label{Fig:LCS_Ex_I_valley_gamma}}
\end{figure}

The uplift de Sitter coefficient and the late de Sitter coefficient along the axion valley are depicted in Figure~\ref{Fig:LCS_Ex_I_valley_gamma}.
Towards the penumbra, the uplift de Sitter coefficient \eqref{eq:HT_dScoeff_cs_corr} may be well below the bounds set by the strong de Sitter conjecture \eqref{eq:Sw_Infl_c_strong}, but the late de Sitter coefficient \eqref{eq:HT_dScoeff_cs_corr}, although getting closer to the strong de Sitter bound in the penumbra region, it still stays above these bounds.

However, different choices of parameters may lead to a late de Sitter coefficient that, along the scalar potential valley, may avoid the strong de Sitter conjecture \eqref{eq:Sw_Infl_c_strong}.

\begin{figure}[H]
    \centering
    \includegraphics[width=7cm]{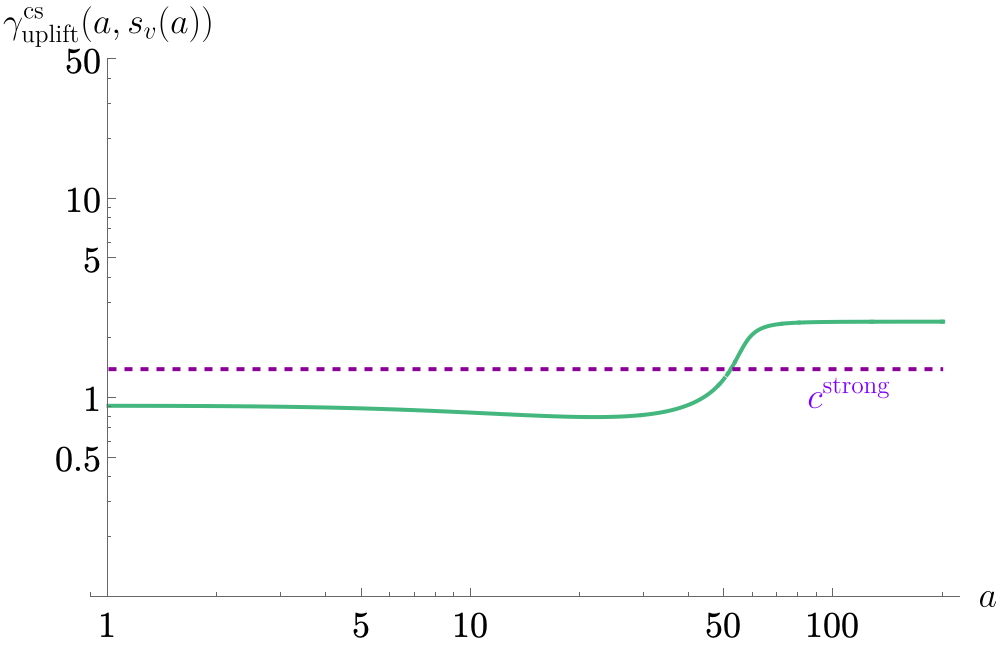} \qquad
    \includegraphics[width=7cm]{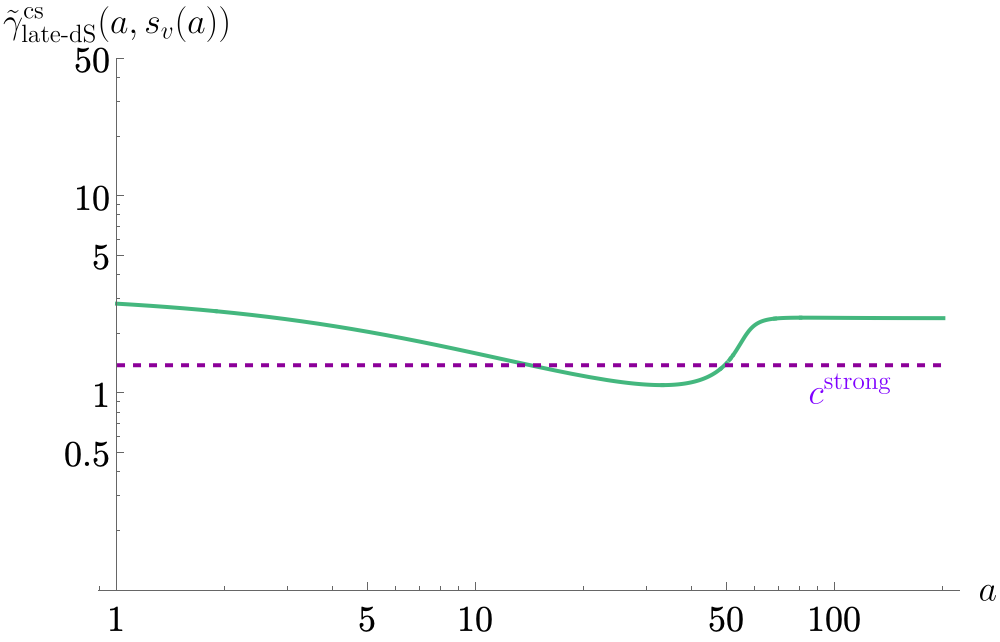} 
    \caption{\footnotesize On the left, the log-log plot of the uplift de Sitter coefficient \eqref{eq:HT_dScoeff_cs} and, on the right, the log-log plot of the late de Sitter coefficient \eqref{eq:HT_dScoeff_cs_corr} along the associated scalar potential valley, specified by the parameters \eqref{eq:LCS_Ex_Ic_par}.
    \label{Fig:LCS_Ex_Ic_valley_gamma}}
\end{figure}

For instance, the effective theory specified by the parameters
\begin{equation}
    \label{eq:LCS_Ex_Ic_par}
    e_0 = e_1 = m^0 = 1\,, \quad m^1 = 10\,, \quad m = 1\,,\quad n = 6\,, \quad \beta = 10^2 \,, \quad \chi = 10^3 \,, \quad \xi = 0\,,
\end{equation}
leads to a scalar potential valley along which the uplift de Sitter coefficient, and late de Sitter coefficient are as depicted in Figure~\ref{Fig:LCS_Ex_Ic_valley_gamma}.
Indeed, similar effective field theories are attractive from the inflationary perspective: the penumbral regions where the late de Sitter coefficient \eqref{eq:HT_dScoeff_cs_corr} is small enough may lead to viable axion monodromy slow-roll inflation.

\subsection{An example of an axion valley towards a Tyurin boundary}
\label{sec:Infl_Examples_Tyurin}

\begin{figure}[H]
    \centering
    \includegraphics[width=7cm]{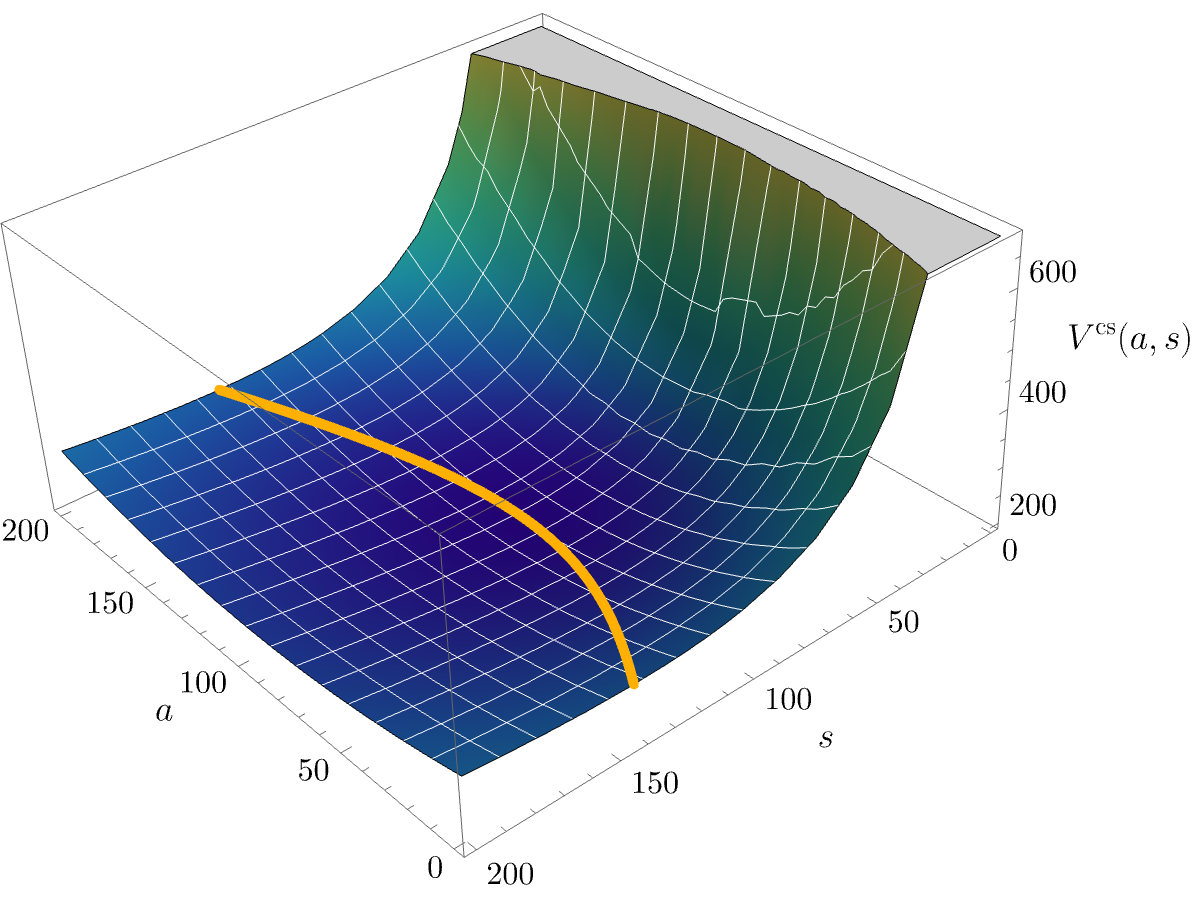} 
    \includegraphics[width=7cm]{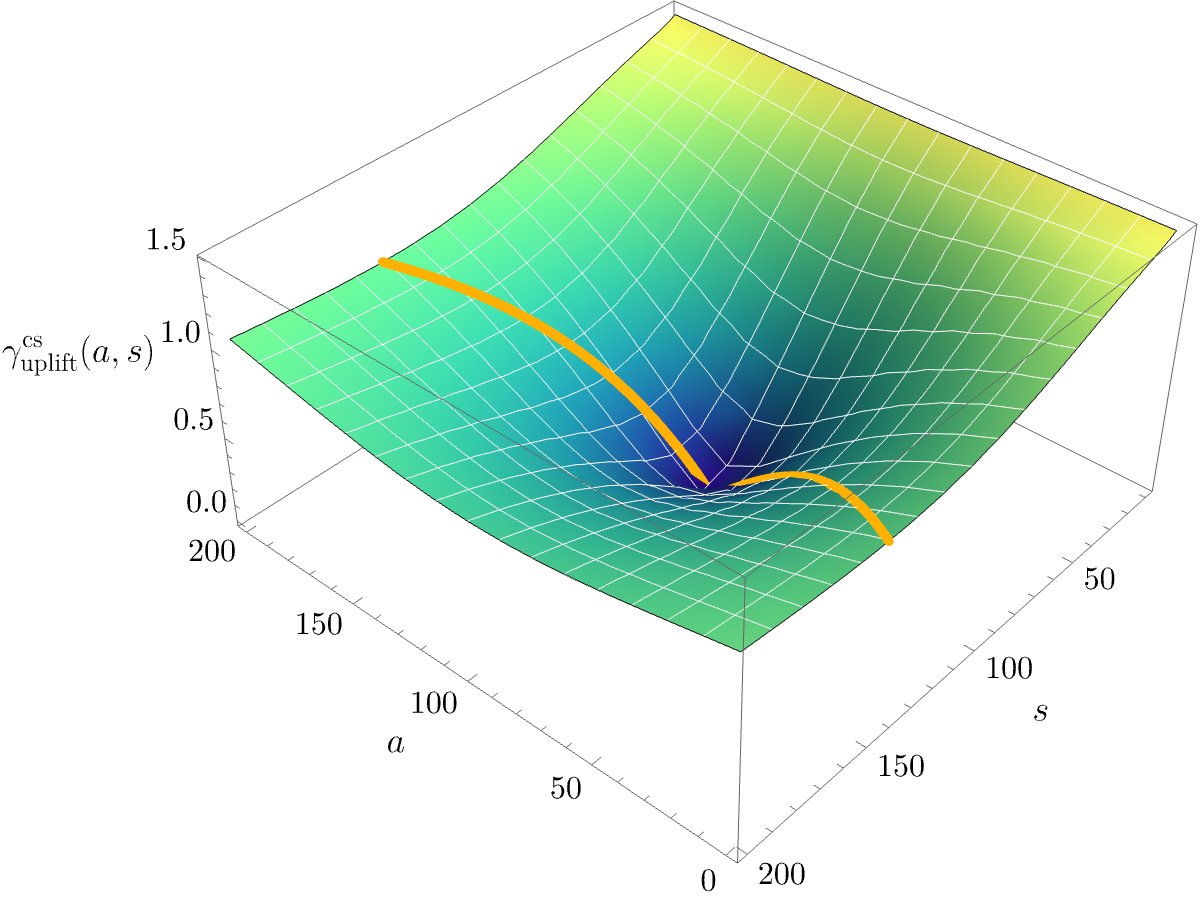} 
    \caption{\footnotesize The complex structure scalar potential (left) defined \eqref{eq:HT_Vfluxquad_nil} 
    and the associated uplift de Sitter coefficient (right) \eqref{eq:HT_dScoeff_cs}
    towards a large complex structure boundary, obtained employing the log-monodromy matrix \eqref{eq:IIB_1mod_II_N} and the ${\rm sl}(2)$-approximated matrix \eqref{eq:IIB_1mod_II_Z}, particularized to the choice of geometric parameters \eqref{eq:Tyurin_Ex_I_par}. 
    The orange, thick line denotes the backreaction of the axion vev onto the saxion minimum.
    \label{Fig:Tyurin_Ex_I_back}}
\end{figure}

The type IIB effective field theories defined towards a Tyurin boundary introduced in Section~\ref{sec:IIB_Vbound_Tyurin} may also host scalar potential valleys extending from the asymptotic regions of the moduli space towards the penumbra.
Indeed, let us consider the same model as in Section~\ref{sec:dS_uplift_Tyurin}, specified by the geometric data \eqref{eq:Tyurin_Ex_I_par}.
The scalar potential valley, as defined in \eqref{eq:Ex_sv_def}, is plotted in Figure~\ref{Fig:Tyurin_Ex_I_back}, stretching on the scalar potential and uplift de Sitter coefficient surfaces.

\begin{figure}[H]
    \centering
    \includegraphics[width=7cm]{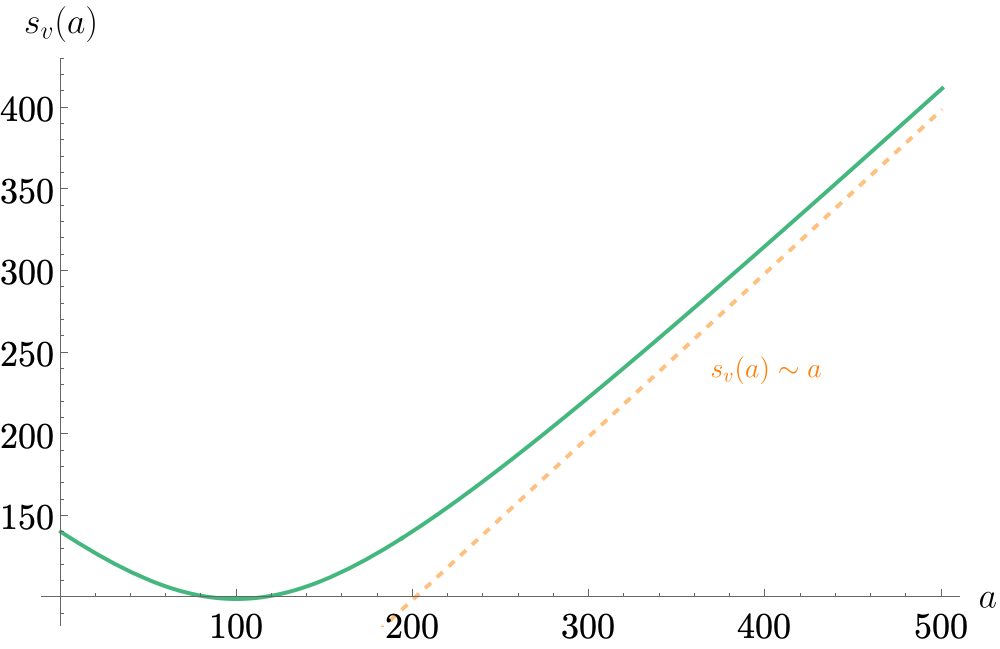} 
    \includegraphics[width=7cm]{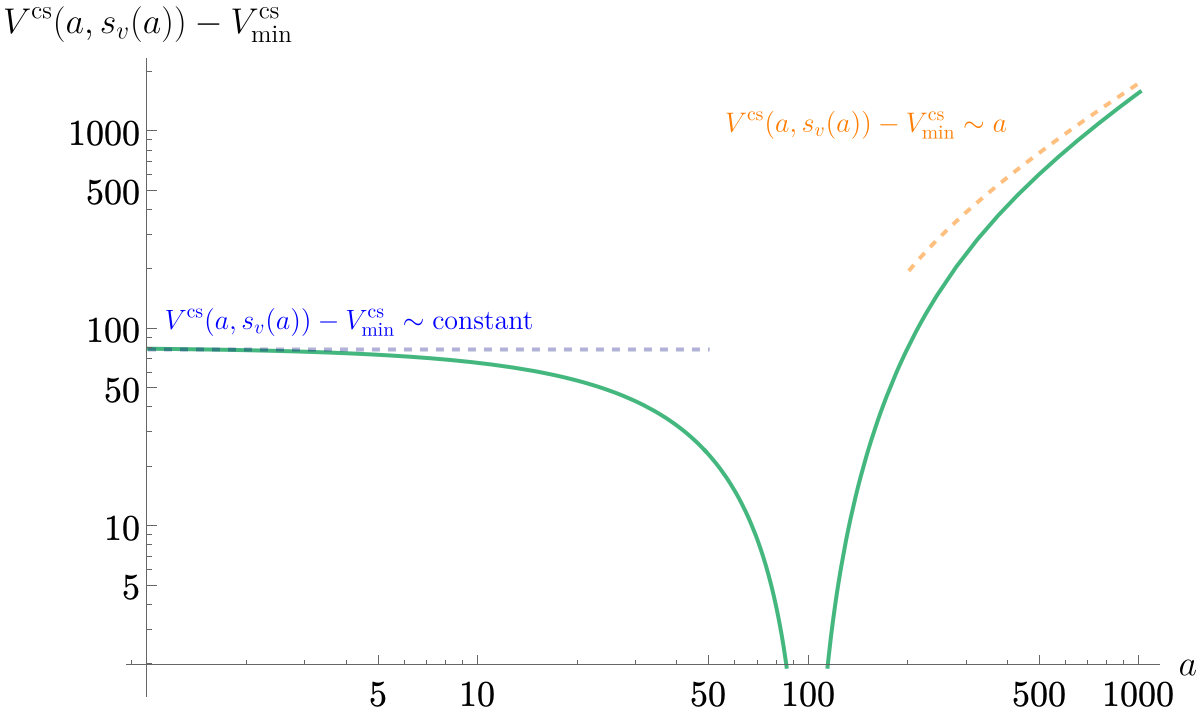} 
    \caption{\footnotesize On the left, the plot of the backreaction of the axion vev onto the saxion minimum dictated by \eqref{eq:Ex_sv_def}, for the effective theory defined towards a Tyurin boundary, specified by the parameters \eqref{eq:Tyurin_Ex_I_par}; on the right, the log-log plot of the associated complex structure scalar potential, to which we have subtracted its value at the minimum $V^{\rm cs}_{\text{min}}$.  
    For very large values of the axion $a$, $V^{\rm cs}(a,s_v(a)) \sim a$; for smaller values of the axion, below the minimum $a_{\rm min}$, $V^{\rm cs}(a,s_v(a))$ is approximately constant.
    \label{Fig:Tyurin_Ex_I_valley_V}}
\end{figure}

On the left of Figure~\ref{Fig:Tyurin_Ex_I_valley_V} is depicted the saxion backreaction as a function of the axion: while it exhibits the expected linear behavior asymptotically, it displays a milder behavior as the axion minimum, located at $a_{\rm min} \simeq 100$, is approached.
Indeed, the scalar potential grows linearly with the axion in the asymptotic region, while the penumbra hosts region where the scalar potential grows slower, being almost constant for small values of the axion, as depicted on the right of Figure~\ref{Fig:Tyurin_Ex_I_valley_V}.

\begin{figure}[H]
    \centering
    \includegraphics[width=7cm]{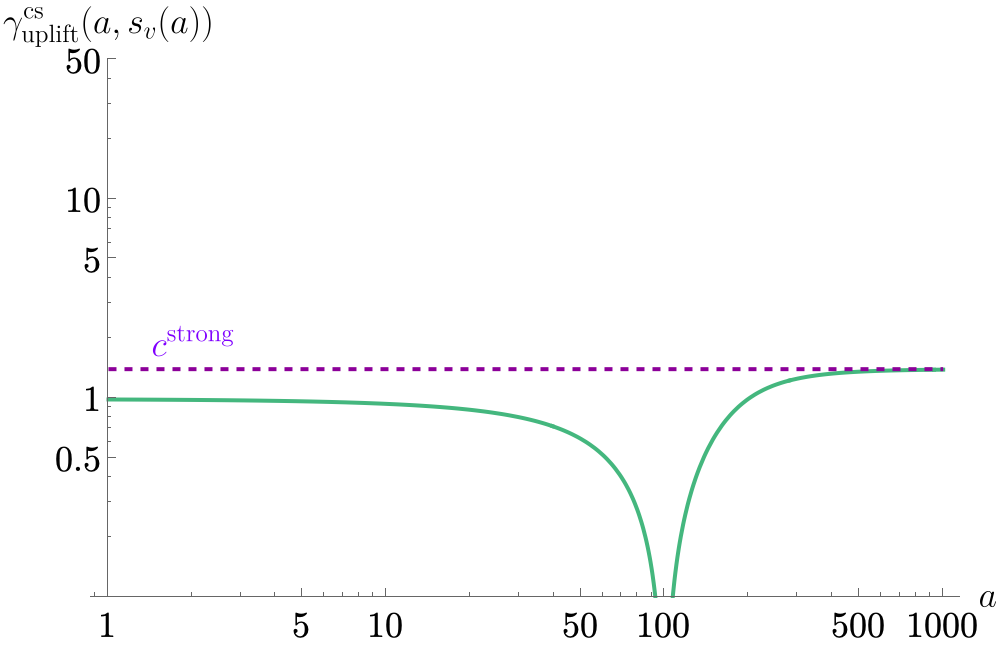} \qquad
    \includegraphics[width=7cm]{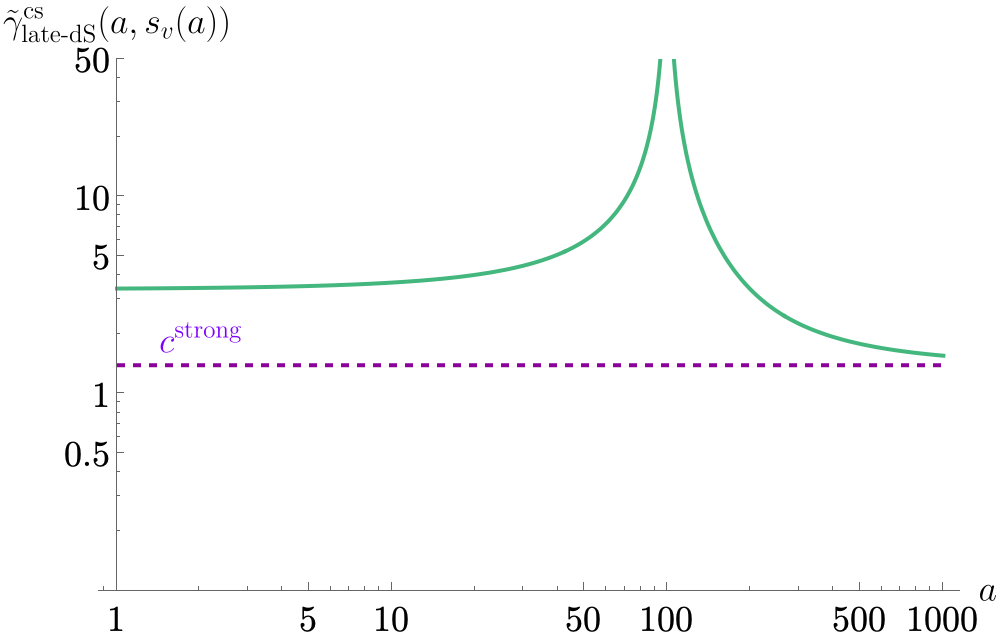} 
    \caption{\footnotesize On the left, the log-log plot of the uplift de Sitter coefficient \eqref{eq:HT_dScoeff_cs} and, on the right, the log-log plot of the late de Sitter coefficient \eqref{eq:HT_dScoeff_cs_corr} along the scalar potential valley plotted in Figure~\ref{Fig:Tyurin_Ex_I_back}.
    \label{Fig:Tyurin_Ex_I_valley_gamma}}
\end{figure}

In Figure~\ref{Fig:Tyurin_Ex_I_valley_gamma} we have plotted the uplift de Sitter coefficient and late de Sitter coefficient \eqref{eq:HT_dScoeff_cs}, \eqref{eq:HT_dScoeff_cs_corr} along the axion valley for the model.
While the uplift de Sitter coefficient, plotted on the left of Figure~\ref{Fig:Tyurin_Ex_I_valley_gamma}, can be very small in the penumbra region, in the vicinity of the axion minimum, the late de Sitter coefficient, depicted on the right, remains above the strong de Sitter bound \eqref{eq:Sw_Infl_c_strong}.

\subsection{Candidate valleys from a family of toy models}
\label{sec:Infl_Examples_Toy}

Although the penumbra region of the moduli space may host possible valleys exhibiting mild backreactions, these valleys may not correspond to viable paths realizing axion monodromy inflation, while consistently satisfying the slow-roll conditions.
In fact, as observed in the examples portrayed in Sections~\ref{sec:Infl_Examples_LCS} and~\ref{sec:Infl_Examples_Tyurin}, the late de Sitter coefficient \eqref{eq:HT_dScoeff_cs_corr} is not necessarily small.
In this section we will be taking a different approach: we inquire which families of effective theories may deliver valleys exhibiting small de Sitter coefficient, by considering simplified toy models.
In particular, we require that the late de Sitter coefficient \eqref{eq:HT_dScoeff_cs_corr} becomes smaller towards asymptotic regions of the moduli space, allowing inflationary paths to start well before the penumbra region is reached. 


As with the type IIB effective theories examined in the previous sections, the family of effective theories that we consider here hosts models that are characterized by a single dynamical complex field $z = a + \im s$.
Collecting the real fields as $\varphi^A = (a,s)$, we shall assume the field space metric to be $G_{AB} = \frac{1}{2s^2} \delta_{AB}$. 
Furthermore, we assume the axion $a$ and saxion $s$ to be subjected to the following scalar potential 
\begin{equation}
    \label{eq:Ex_V}
    V(a,s) = \frac{(e - a m)^2}{2s^p} + \frac12 m^2 s^q + V_0 \,,
\end{equation}
with the integral parameters $e, m \in \mathbb{Z}$, and $p, q \in \mathbb{Z}_{>0}$, and $V_0 \in \mathbb{R}$ being an arbitrary offset.

Indeed, the field space metric that we have just introduced mimics the one that emerges in string theory effective theories defined towards infinite-field distance boundaries, such as those of the models studied in Sections~\ref{sec:Infl_Examples_LCS} and~\ref{sec:Infl_Examples_Tyurin}.
Moreover, for some appropriate choices of powers $p$, $q$, the scalar potential \eqref{eq:Ex_V} is a subcase of the more general scalar potentials emerging towards such boundaries, where the parameters $e$ and $m$ are identified as background fluxes.

The scalar potentials defined in \eqref{eq:Ex_V} are endowed with a shift symmetry: in fact, the scalar potentials \eqref{eq:Ex_V} are invariant under axion shifts $a \to a + n$, for some $n \in \mathbb{Z}$, provided that the parameter $e$ is shifted as $e \to e + n m$.
Such a symmetry is spontaneously broken when the parameter $e$ is frozen; indeed, for fixed parameter $e$, the scalar potentials \eqref{eq:Ex_V} exhibit a single, axion minimum located at $a_{\rm min} = \frac{e}{m}$.

We would like to investigate whether any of the theories in the family so constructed may accommodate regions where slow-roll inflation is not strictly forbidden.
Hence, as a first, exploratory analysis, we may check whether there are some regions of the two-dimensional field space $(a,s)$ where the first slow-roll parameter \eqref{eq:Sw_Infl_eps} is not large.
For simplicity, we may assume that any given point in the field space is threaded by a geodesic inflationary path, so that \eqref{eq:Sw_Infl_gamma_eps} leads to the simpler identification $\varepsilon = \frac{\gamma^2}{2}$.

\begin{figure}[H]
    \centering
    \includegraphics[width=14cm]{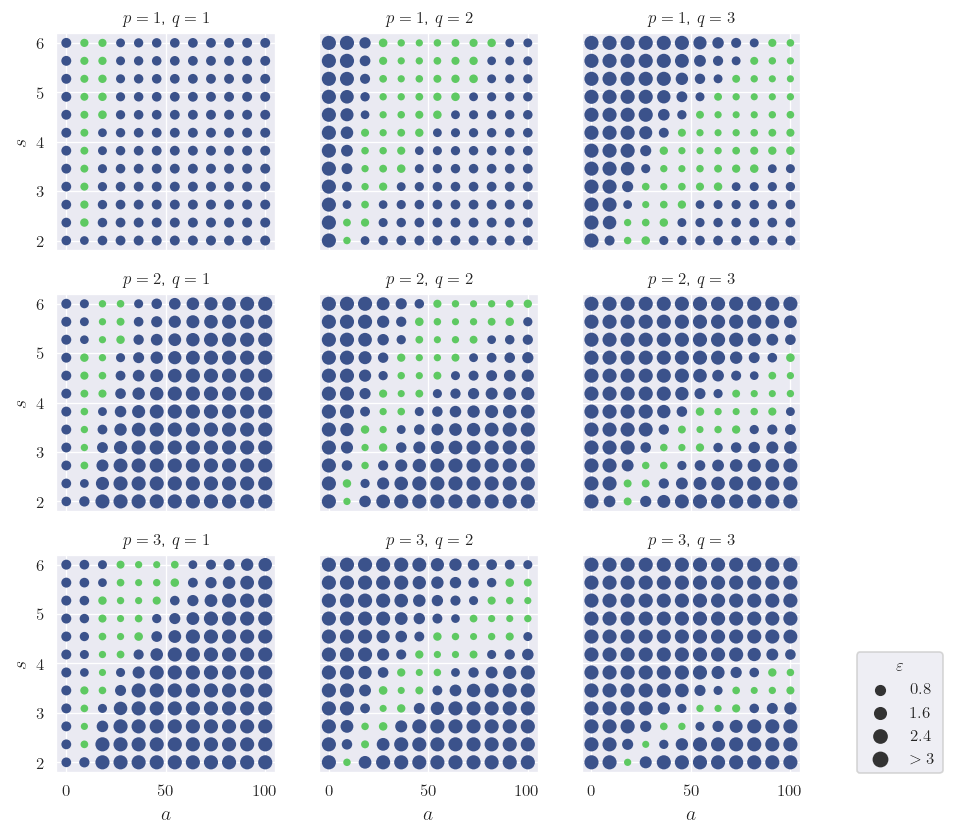}
    \caption{\footnotesize The first slow-roll parameter $\varepsilon = \frac{\gamma^2}{2}$ across the field space for some values of $p$, $q$.
    The size of the dots denotes the magnitude of $\varepsilon$ at the given field space point; in particular, the green dots are those for which $\varepsilon < \frac12$.
    \label{Fig:Toy_gamma_scan}}
\end{figure}

Figure~\ref{Fig:Toy_gamma_scan} shows a scan of the values of the first slow-roll parameter $\varepsilon = \frac{\gamma^2}{2}$ across the field space, choosing $e = m = 1$ and $V_0 = 0$, for some values of $p$, $q$.
Remarkably, for some appropriate vevs of the axion and the saxion, the first slow-roll parameter can be quite small; indeed, in Figure~\ref{Fig:Toy_gamma_scan}, the green dots correspond to field space points at which $\varepsilon < \frac12$.


But how can eventual inflationary paths look like?
Ideally, we would like to treat one of the two fields as the inflaton.
Since all the scalar potentials in the family \eqref{eq:Ex_V} are endowed with a quadratic axion coupling, the axion $a$ may play the role of the inflaton.
To this end, it is crucial that the partner field, the saxion $s$, can be consistently fixed at its minimum.
However, backreaction effects may prevent this from happening.
In fact, solving \eqref{eq:Ex_sv_def} leads to the valley
\begin{equation}
    \label{eq:Ex_sv}
    s_v(a) = \left[ \frac{p}{q m^2} (e - a m)^2 \right]^{\frac{1}{p+q}}\,,
\end{equation}
which exhibits how the saxion minimum is displaced due to a change of the axion vev.
In particular, for large values of the axion, the backreaction onto the saxion vev \eqref{eq:Ex_sv} behaves as $s_v(a) \sim a^{\frac{2}{p+q}}$.
However, it is worth remarking that, although \eqref{eq:Ex_sv} indicates that one cannot vary the axion while keeping the saxion fixed, the backreaction \eqref{eq:Ex_sv} \emph{flattens} for large values of $p$, $q$, thus becoming less, and less relevant.

In order for the valleys \eqref{eq:Ex_sv} to possibly host inflationary paths, it is crucial that the first slow-roll parameter \eqref{eq:Sw_Infl_eps} is small, at least on a portion of the valley.
In order to examine how the first slow-roll parameter varies along the valley, it is convenient to introduce the saxion-valley de Sitter coefficient 
\begin{equation}
    \label{eq:Ex_gammav}
    \gamma_v(a) := \gamma |_{s = s_v(a)} = \frac{\sqrt{G^{aa}} |\partial_a V| }{V} \,.
\end{equation}
Clearly, in general, it varies along the axion valley \eqref{eq:Ex_sv}; indeed, is immediate to find, employing \eqref{eq:Ex_sv}, that $\gamma_v(a) \sim a^{\frac4{p+q}-2}$ for large axion vevs; in particular, for $p + q > 2$, $\gamma_v(a) \to 0$ for large axion $a$.

\begin{figure}[H]
    \centering
    \includegraphics[width=12cm]{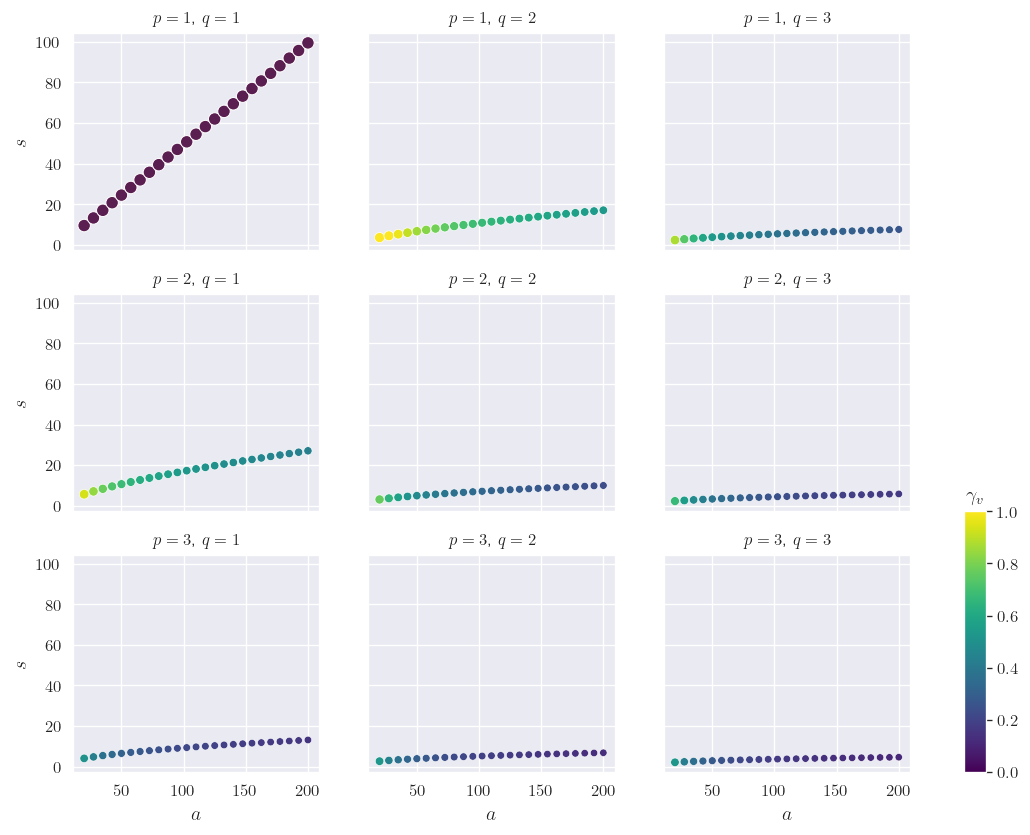}
    \caption{\footnotesize The de Sitter coefficient \eqref{eq:Sw_Infl_gamma} along the saxion backreaction of the scalar potential \eqref{eq:Ex_V}, with $e = m = 1$, for some values of $p$, $q$. 
    Notice that in the case for which $p=q=1$, $\gamma_v \sim \sqrt{2}$ along all the backreaction.
    \label{Fig:Toy_gamma_back}}
\end{figure}

For concreteness, in Figure~\ref{Fig:Toy_gamma_back} we have plotted the backreaction \eqref{eq:Ex_sv} of the axion vev onto the saxion minimum, choosing $e = m = 1$, and setting $V_0 =0$.
In particular, as expected, for $p = q = 1$, the de Sitter coefficient \eqref{eq:Sw_Infl_gamma} remains constant, as $\gamma_v \sim \sqrt{2}$.
Conversely, in all the other remaining cases, the de Sitter coefficient decreases along the valley towards larger axion vevs.

\section{Machine learning-driven analysis of penumbra inflationary regions}
\label{sec:ML}

In the previous sections, we illustrated some selected examples, either towards large complex structure or Tyurin boundaries, that support de Sitter uplifts, and the associated valleys that they are endowed with.
However, the examples shown in Section~\ref{sec:Infl_Examples} are not the sole ones having these features: indeed, scalar potential valleys, stretching along de Sitter uplift directions of the scalar potentials, occur for \emph{large} sets of effective field theories.
In this section we will show how effective field theories hosting such axion valleys can be detected by employing simple machine learning techniques.

The main steps of the analysis, common to the large complex structure, and Tyurin boundaries are the following:

\noindent\textbf{Creation of a database.} Firstly, we choose some pairs of values of the axion and saxion $(a_{(\textsf{I})}, s_{(\textsf{I})})$, with $\textsf{I} = 1, \ldots, \textsf{N}$ -- in particular, below, we shall set $\textsf{N} = 9$.
For each element of the pair $(a_{(\textsf{I})}, s_{(\textsf{I})})$, we scan large subsets of the parameter space: given a point in such a parameter set, we compute the scalar potential specified by those parameters, as well as the uplift de Sitter coefficient \eqref{eq:HT_dScoeff_cs} at the field space point $(a_{(\textsf{I})}, s_{(\textsf{I})})$. 
Then, if at that given point in the parameter space $\gamma^{\text{cs}}_{\text{uplift}} < c$, for some small $c > 0$, we assign to that point the label `$\textsf{1}$', or `$\textsf{True}$', meaning that the given parameter space point might deliver a valley, passing through the space point $(a_{(\textsf{I})}, s_{(\textsf{I})})$, with small enough uplift de Sitter coefficient; otherwise, that parameter space point is associated with the label `$\textsf{0}$', or `$\textsf{False}$'.
Conventionally, we shall set the constant $c$ equal to $1$.

\noindent\textbf{Machine learning the effective theories hosting valleys.} The set of databases created as described above can then be fed to a machine learning algorithm to learn which parameters can lead to valleys distinguished by an uplift de Sitter coefficient that is small enough.
Or, more appropriately, such a machine learning algorithm would allow us to \emph{exclude} the parameter regions where an eventual valley passing through the given field space point $(a_{(\textsf{I})}, s_{(\textsf{I})})$ cannot be small enough.
Concretely, the regions can be learned by feeding the databases constructed as above to a simple $k$-nearest neighbors algorithm
-- see \cite{Ruehle:2020jrk,Lanza:2022zyg} for a physics-oriented explanation of this algorithm.
Such an algorithm allows one to \emph{learn} what are the parameter regions whose points are more likely to be associated with the label `$\textsf{1}$', or `$\textsf{0}$', namely which parameter regions are characterized, or not by small enough uplift de Sitter coefficient for at the field space point $(a_{(\textsf{I})}, s_{(\textsf{I})})$.
Clearly, such a procedure can be repeated for any of the field space points $(a_{(\textsf{I})}, s_{(\textsf{I})})$ originally chosen.

Below we shall apply this analysis in order to learn which effective theories defined either towards an LCS, or a Tyurin boundary may host candidate axion valleys with small enough uplift de Sitter coefficient.

\subsection{Axion valley regions towards LCS boundaries}
\label{sec:ML_LCS}

The type IIB effective field theories defined towards an LCS boundary introduced in Section~\ref{sec:IIB_Vbound_LCS} are characterized by a ten-dimensional parameter space, given by the four fluxes $e_I, m^I$, the real geometric parameters $m,n,\beta,\chi$ and the complex parameter $\xi$.
For simplicity, we shall focus only on the two-dimensional subspace spanned by the geometric parameters $\beta$ and $\chi$.
We will fix the fluxes as $e_0 = e_1 = m^0 =m^1 = 1$, and the remaining geometric parameters as $m = 1$, $n = 6$, $\xi = 0$.

\begin{figure}[H]
    \centering
    \includegraphics[width=13cm]{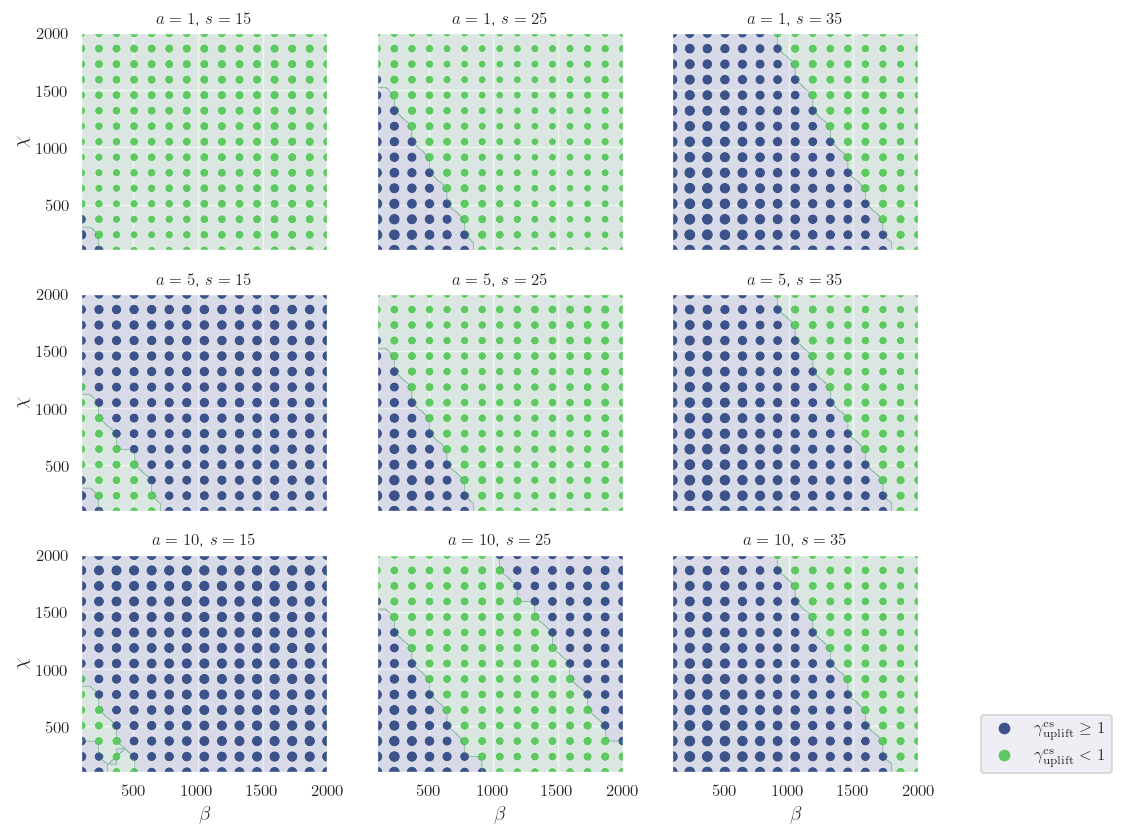}
    \caption{\footnotesize The realization of the first slow-roll condition in the parameter space spanned by $\beta$ and $\chi$ for some values of the axion $a$ and saxion $s$.
    We chose the fluxes  $e_0 = e_1 = m^0 =m^1 = 1$, and the remaining parameters are set to $m = 1$, $n = 6$, $\xi = 0$.
    The green dots denote points for which $\gamma^{\text{cs}}_{\text{uplift}} (a_{(\textsf{I})}, s_{(\textsf{I})}) < 1$, while the blue dots those for which $\gamma^{\text{cs}}_{\text{uplift}} (a_{(\textsf{I})}, s_{(\textsf{I})}) \geq 1$, with bigger sizes associated to bigger values of $\gamma^{\text{cs}}_{\text{uplift}} (a_{(\textsf{I})}, s_{(\textsf{I})})$. The associated regions are obtained via a standard $k$-nearest neighbors machine learning algorithm, where we chose $k = 5$.
    \label{Fig:ML_LCS_kNN}}
\end{figure}

We can construct a database of effective field theories by following the procedure above.
Firstly, we pick nine pairs of axion and saxion vevs $(a_{(\textsf{I})}, s_{(\textsf{I})})$, where $a_{(\textsf{I})} = \{ 1, 5, 10\}$, $s_{(\textsf{I})} = \{ 35, 50, 65\}$.
For each of these nine pairs, we create a grid composed by $15$ possible values of the parameter $\beta$, and an equal number of possible values of the paratemer $\chi$, both chosen in the interval $[1,2000]$.
For each point in the grid, we compute the scalar potential, and then the uplift de Sitter coefficient, evaluated at the space point $(a_{(\textsf{I})}, s_{(\textsf{I})})$.
The result is plotted in Figure~\ref{Fig:ML_LCS_kNN}: each of the plots is associated to a specific pair of field space points $(a_{(\textsf{I})}, s_{(\textsf{I})})$. 
The green dots are associated to parameter space points where $\gamma^{\text{cs}}_{\text{uplift}} (a_{(\textsf{I})}, s_{(\textsf{I})}) < 1$, while the blue dots are associated to parameter space points for which $\gamma^{\text{cs}}_{\text{uplift}} (a_{(\textsf{I})}, s_{(\textsf{I})}) \geq 1$, with the size being an indication of the specific value.

Then, feeding the databases so created to a standard $k$-nearest neighbors machine learning algorithm the parameter space regions where either $\gamma^{\text{cs}}_{\text{uplift}} (a_{(\textsf{I})}, s_{(\textsf{I})}) < 1$, or $\gamma^{\text{cs}}_{\text{uplift}} (a_{(\textsf{I})}, s_{(\textsf{I})}) \geq 1$ can be learned.
It is worth mentioning that we have constructed a database with a grid since the algorithm leading to its creation is quite efficient -- the creation of the aforementioned database requires roughly $0.1 s$; alternatively, we could have also chosen random parameters space points.
Moreover, the regions learned, and plotted in Figure~\ref{Fig:ML_LCS_kNN} indicate that larger axion and saxion vevs require \emph{larger}, more extreme values of the geometric parameter to deliver axion valley, indeed suggesting that the penumbra region is more likely to host axion valleys fit for the realization of slow-roll inflation.

\subsection{Axion valley regions towards Tyurin boundaries}
\label{sec:ML_Tyurin}

The type IIB effective field theories close to a Tyurin boundary introduced in Section~\ref{sec:IIB_Vbound_Tyurin} are endowed with an eight-dimensional parameter space, constituted by the four fluxes $e_I, m^I$, and the four real geometric parameters $m,n,c,d$.
As for the analysis carried in the previous section for LCS boundaries, we focus solely on the two-dimensional subspace spanned by the geometric parameters $c$ and $d$, while fixing the fluxes as $e_0 = e_1 = m^0 =m^1 = 1$, and the remaining geometric parameters as $m = n = 1$.

\begin{figure}[H]
    \centering
    \includegraphics[width=13cm]{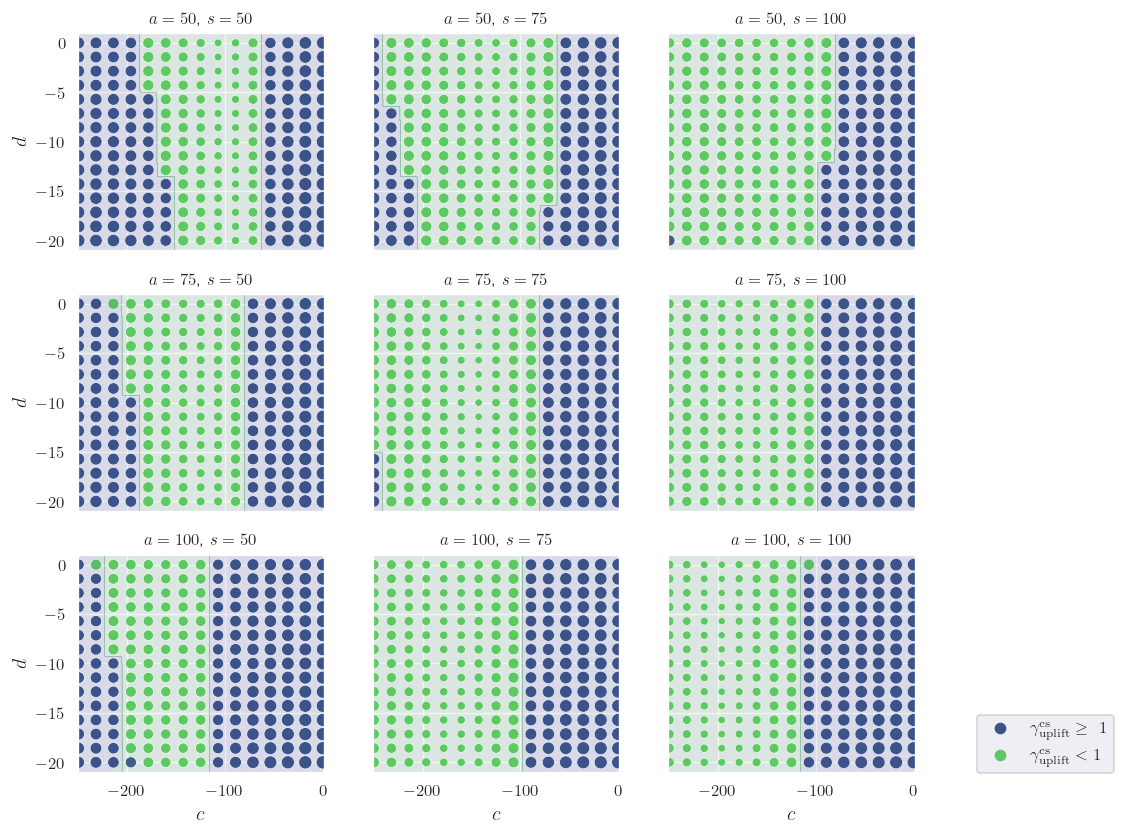}
    \caption{\footnotesize The realization of the first slow-roll condition in the parameter space spanned by $c$ and $d$ for some values of the axion $a$ and saxion $s$.
    We chose the fluxes  $e_0 = e_1 = m^0 =m^1 = 1$, and the remaining parameters are set to $m = n = 1$.
    The green dots denote points for which $\gamma^{\text{cs}}_{\text{uplift}} (a_{(\textsf{I})}, s_{(\textsf{I})}) < 1$, while the blue dots those for which $\gamma^{\text{cs}}_{\text{uplift}} (a_{(\textsf{I})}, s_{(\textsf{I})}) \geq 1$, with bigger sizes associated to bigger values of $\gamma^{\text{cs}}_{\text{uplift}} (a_{(\textsf{I})}, s_{(\textsf{I})})$. The associated regions are obtained via a standard $k$-nearest neighbors machine learning algorithm, where we chose $k = 5$.
    \label{Fig:ML_Tyurin_kNN}}
\end{figure}

The analysis proceeds as with the LCS effective theories examined in Section~\ref{sec:ML_LCS}: we choose the axion and saxion vevs pairs $(a_{(\textsf{I})}, s_{(\textsf{I})})$, with $a_{(\textsf{I})}, s_{(\textsf{I})} = \{ 50, 75, 100\}$.
Then, for each of the space point pairs, we introduce a grid scanning the parameter subspace $(c,d)$, with $c \in [-250,-1]$, $d \in [-20,-1]$, and for each of the points on the grid we compute the uplift de Sitter coefficient, evaluated at the given field space point $(a_{(\textsf{I})}, s_{(\textsf{I})})$.
In Figure~\ref{Fig:ML_Tyurin_kNN} such databases are plotted, with the green dots denoting the parameters for which $\gamma^{\text{cs}}_{\text{uplift}} (a_{(\textsf{I})}, s_{(\textsf{I})}) < 1$, while the blue dots those for which $\gamma^{\text{cs}}_{\text{uplift}} (a_{(\textsf{I})}, s_{(\textsf{I})}) \geq 1$.
The associated regions are learned via a standard $k$-nearest neighbors algorithm.
Indeed, as noticed for the effective theories defined towards an LCS boundary, larger axion and saxion vevs require more extreme values of the parameters in order to realize viable axion valleys.

\section{Conclusions and outlook}
\label{sec:Conclusions}

In this work we have presented some rather interesting properties of four-dimensional effective field theories from type IIB Calabi-Yau string compactifications defined in `\emph{penumbral}' regions crossing over from the interior of moduli space (the umbra) towards boundaries at infinite field distance.
Within four-dimensional EFTs with one dynamical complex structure modulus, we find that these regions  host uplifting vacua in the complex structure sector which may be suitable as uplifts for the construction of four-dimensional de Sitter vacua, and which we hence called `de Sitter uplifts'.
Such de Sitter uplifts are obtained by employing three key assumptions: the vevs of the axion and saxion fields are only moderately large, the geometric parameters specifying the effective theory may be appropriately tuned, and the remaining moduli fields are disregarded presuming stabilization by other methods (extra three-form fluxes for the remaining complex structure moduli and one of the extant methods of K\"ahler moduli stabilization \cite{Kachru:2003aw,Saltman:2004jh,Balasubramanian:2005zx,Cicoli:2023opf}).
We note that the de Sitter uplifts that we find are expected to survive under the inclusion of further subleading contributions to the scalar potential (barring unexpectedly huge coefficients), as these will be suppressed by additional inverse powers of the moderately large saxion vev. By the same logic, instanton corrections scaling as $\exp(-2\pi s)$ are completely negligible in this regime.

The de Sitter uplifts found this way are also associated to axion valleys, along which the saxion field attains its local minimum, and characterized by small a uplift de Sitter coefficient and modest backreaction. These axion valleys are long-range in that they extend over multiple axion periods. In this sense, they show axion monodromy.
Whether these valleys may be viable paths for realizing axion monodromy inflation is  still an open question: a preliminary feature that these valley ought to exhibit is a small enough late de Sitter coefficient.
While we do find examples sharing such a feature, a more detailed analysis has to be performed in order to check how small the late de Sitter coefficient can become along these candidate valleys.
We leave this including the necessary accompanying analysis of other slow-roll inflationary constraints along the valley for future work.

While we have illustrated how machine learning can help make the search for effective field theories hosting candidate axion valleys in the penumbral region systematic, a thorough machine learning analysis, encompassing all the necessary constraints for these valley candidates to be consistent axion valleys, is postponed to future work.
Indeed, a machine learning driven analysis may be necessary in order to further extend the reach of our investigations.
Analogous models with multiple dynamical complex structure moduli in the penumbral region require additional geometrical parameters that make a systematic scan too involved for the limited numerical methods employed in this work.

It would be also intriguing to check for different sectors of the moduli space as a well as for different families of effective field theories whether they could exhibit analogous uplifts, and associated candidate axion valleys.
For instance, it is tantalizing to examine whether the de Sitter uplifts studied in this work survive after assuming that the K\"ahler moduli sector is dynamical; or whether more general F-theory four-dimensional effective field theories, equipped with a dynamical axio-dilaton, and with fluxes constrained by the tadpole cancellation condition, can also host these uplifts.

\noindent\textcolor{colorloc4}{\textbf{Acknowledgments.}} We are grateful to helpful comments from Thomas Grimm and Timo Weigand. The work of SL and AW is supported in part by Deutsche Forschungsgemeinschaft under Germany’s Excellence Strategy EXC 2121 Quantum Universe 390833306, by Deutsche Forschungsgemeinschaft through the Collaborative Research Center 1624 “Higher Structures, Moduli Spaces and Integrability”, and by Deutsche Forschungsgemeinschaft through a German-Israeli Project Cooperation (DIP) grant “Holography and the Swampland”.

\appendix

\section{An overview of the type IIB \texorpdfstring{$\mathcal{N}=1$}{N=1} four-dimensional effective theories}
\label{sec:IIB_summ}

In order to keep this work self-contained, and to clarify the notation employed in Section~\ref{sec:IIB} of the main text, in this appendix we briefly review some of the most important features of the four-dimensional theories that are obtained after the compactification of type IIB string theory over an orientifolded Calabi-Yau three-fold.

The compactification of type IIB string theory over an Calabi-Yau three-fold, equipped with $\text{O}3$, $\text{O}7$-orientifolds, leads to four-dimensional theories equipped with $\mathcal{N}=1$ local supersymmetry \cite{Grimm:2004uq}.
In this work, we consider only the closed string sector, further restricting our focus to the following, bosonic fields: the graviton $g_{\mu\nu}$, being part of the gravitational tensor multiplet, and a set of complex scalar fields $\varphi^\alpha$, with $\alpha = 1, \ldots, N$, supersymmetrically embedded within chiral multiplets.
The scalar fields $\varphi^\alpha$ are identified with the moduli of the Calabi-Yau manifold upon which type IIB string theory is compactified, and may be split into three-families, according to their geometric origin.

Firstly, the effective theory is endowed with the axio-dilaton $\tau = C_0 + \im e^{-\phi} $,
where $\phi$ is the ten-dimensional dilaton, related to the string coupling as $g_{\rm s} = e^{\phi}$, and $C_0$ is the RR zero-form.
Secondly, the effective theory is populated by the real scalar fields $v_\lambda$, with $\lambda = 1, \ldots, h^{1,1}_+$, which are identified as the K\"ahler moduli of the Calabi-Yau manifold; indeed, these can be most readily obtained by expanding the K\"ahler two-form in a basis of $h^{1,1}_+$ divisors $D^\lambda$ as $J = v_\lambda [D^\lambda]$, with $[D^\lambda]$ being the Poincar\'e dual of the divisor $D^\lambda$.
The real scalar fields $v_\lambda$ pair with those descending from the dimensional reduction of the RR four-form over the basis of divisors $D^\lambda$ as $a^\lambda = \int_{D^\lambda} C_4$.
Together, the real scalar fields $v_\lambda$ and $a^\lambda$ form the complex scalar fields $u^\lambda = a^\lambda + \im s^\lambda$ where we have introduced $s^\lambda = \frac12 \int_{D^\lambda} J\wedge J  = \frac12 \kappa^{\lambda\rho\sigma} v_\lambda v_\sigma$, with $\kappa^{\lambda\rho\sigma}$ the Calabi-Yau intersection numbers.

Finally, the four-dimensional effective theory contains the complex scalar fields $z^i$, with $i = 1, \ldots, h^{2,1}_-$, parametrizing the variations of the Calabi-Yau complex structure.
These can be obtained by expanding the Calabi-Yau holomorphic three-form $\Omega$ in a basis of $2 (h^{2,1}_- + 1)$ three-forms $\gamma_{\mathcal{I}}$, with $\mathcal{I} = 1, \ldots, 2 (h^{2,1}_- + 1)$ as $\Omega = \Pi^{\mathcal{I}}(z) \gamma_{\mathcal{I}}$, with $\Pi^{\mathcal{I}}(z)$ being the holomorphic \emph{periods} entailing the complex structure deformations.

It is convenient to consider a \emph{symplectic} basis of three-forms $\gamma_{\mathcal{I}} = (\alpha_I, \beta^J)$, with $I, J = 0, 1, \ldots, h^{2,1}_-$.
The associated symplectic pairings are defined by the following matrix
\begin{equation}
	\label{eq:IIB_eta}
	\eta = \begin{pmatrix}
		\int \alpha_I \wedge \alpha_J & \int \alpha_I \wedge \beta^J\\  \int \beta^I \wedge \alpha_J  & \int \beta^I \wedge \beta^J
	\end{pmatrix} = \begin{pmatrix}
		0 & \mathds{1}\\ -\mathds{1} & 0
	\end{pmatrix}
\end{equation}
expressed in $(h^{2,1}+1) \times (h^{2,1}+1)$ blocks. 
Accordingly, over such a symplectic basis, the periods $\Pi^{\mathcal{I}}(z)$ split as
\begin{equation}
	\label{eq:IIB_Pi_sym}
	{\bm \Pi}(z) = \begin{pmatrix}
		X^I(z) \\ -\mathcal{F}_I(z)
	\end{pmatrix}\,,
\end{equation}
and the three-form $\Omega$ can be expanded as $\Omega = X^I \alpha_I - \mathcal{F}_I \beta^I$.

We further recall that the periods \eqref{eq:IIB_Pi_sym} ought to holomorphically depend only on the complex structure scalar fields $z^i$.
Hence, to remove the redundancies, firstly, we will regard the quantities $\mathcal{F}_I$ as  holomorphic functions of the coordinates $X^I$; in some cases, the functions $\mathcal{F}_I$ can be additionally thought of as being derivatives of a single quantity, the \emph{prepotential} $\mathcal{F}$.
In such cases, the prepotential is a homogenous function of degree two in the $X^I$, and the quantities $\mathcal{F}_I$ are obtained via $\mathcal{F}_I = \frac{\partial \mathcal{F}}{\partial X^I}$. 
Secondly, we shall perform a gauge-fixing of the coordinates $X^I$ by setting one coordinate -- $X^0$, for instance -- to a constant.

The effective field theory capturing the interactions among the moduli sectors introduced above is
\begin{equation}
	\label{eq:IIB_action}
	S=\int \left(\frac{1}{2} M^2_{\rm P} R * 1-M^2_{\rm P} K_{\alpha \bar \beta}\,{\rm d} \varphi^\alpha\wedge * {\rm d}\bar\varphi^{\bar \beta} -  V * 1 \right)\, .
\end{equation}
Here, $M_{\rm P} $ is the four-dimensional Planck mass, and $R$ is the Ricci scalar. 
Furthermore, we have collected the moduli as $\varphi^\alpha = (z^i, u^\lambda, \tau)$, and introduced $K_{\alpha\bar{\beta}} = \frac{\partial^2 K}{\partial \varphi^\alpha \partial \bar{\varphi}^{\bar\beta}} $, with $K$ the K\"ahler potential.
Additionally, the scalar potential $V$ generically depends on all the moduli populating the effective theory.

In this work, we assume that the K\"ahler potential is given by distinct contributions from the three moduli sectors, as
\begin{equation}
	\label{eq:IIB_K}
	K = K^{\rm cs} + K^{\rm ks} - \log \left[- \im (\tau -\bar \tau) \right]\, .
\end{equation}
Here, $K^{\rm ks}$ solely depends on the complex structure moduli, and can be fully expressed in terms of period components as follows 
\begin{equation}
	\label{eq:IIB_Kcsb}
	K^{\rm cs} = -\log \left(  \im \int_{\hat Y} \Omega \wedge \bar\Omega \right)=- \log \im\, {\bm \Pi}^T \eta \bar{\bm \Pi} = - \log \im(\bar{X}^I \mathcal{F}_I -  X^I \bar{\mathcal{F}}_I)\, ,
\end{equation}
whereas the K\"ahler moduli are separately encoded within the K\"ahler potential contribution
\begin{equation}
	\label{eq:IIB_Kk}
	K^{\rm ks} = -2 \log \int_{\hat Y} J \wedge J \wedge J = - 2 \log \kappa^{\lambda\rho\sigma} v_\lambda v_\rho v_\sigma\,.
\end{equation}
and it satisfies the no-scale condition $K^{\lambda\bar\rho}_{\rm ks} K^{\rm ks}_{\lambda} K^{\rm ks}_{\bar\rho} = 3$ \cite{Grimm:2004uq}.

Moreover, the scalar potential involving the scalar fields $\varphi^\alpha$ stems from a holomorphic superpotential $W(\varphi)$ via the formula \cite{Cremmer:1982en}
\begin{equation}
	\label{eq:IIB_Vgen}
	V = \frac{1}{M_{\rm P}^2} e^K (K^{\alpha \bar\beta} D_\alpha W \bar{D}_{\bar\beta} \bar W - 3 W \bar{W})\,,
\end{equation}
where we have introduced the K\"ahler covariant derivative $D_\alpha = \frac{\partial}{\partial\varphi^\alpha} + \frac{\partial K}{\partial \varphi^\alpha}$.
In this work, we further consider effective theories for which the superpotential only depends on the complex structure moduli $z^i$ and the dilaton $\tau$, and can be obtained from a background RR three-form flux $F_3$ as follows \cite{Gukov:1999ya}
\begin{equation}
	\label{eq:IIB_WGVW}
	W (t,\tau) = M_{\rm P}^3 \int_{\hat{Y}} \Omega \wedge F_3 = M_{\rm P}^3\, {\bf f}^T\, \eta\, {\bm \Pi} (t)\,, \qquad F_3 =  {\bf f} \, {\bm \gamma} \, .
\end{equation}
Employing the superpotential \eqref{eq:IIB_WGVW}, the scalar potential \eqref{eq:IIB_Vgen} can be also written as 
\begin{equation}
	\label{eq:IIB_Vfluxquad}
	V_{\bf f} =: M^4_{\rm P} e^{\hat{K}}\, V_{\bf f}^{\rm cs} = \frac12 M^4_{\rm P} e^{\hat{K}}\, {\bf f}^T\, \mathcal{T}(a,s) \, {\bf f}\,, \qquad \text{with}\quad {\bf f} = \begin{pmatrix}
		m^I \\ - e_I
	\end{pmatrix}\,.
\end{equation}
Here, we have singled out the contribution $V_{\bf f}^{\rm cs}$ that solely depends on the complex structure moduli; indeed, the remaining moduli, the axio-dilaton and the K\"ahler moduli, only enter the scalar potential \eqref{eq:IIB_Vfluxquad} via the prefactor $e^{\hat{K}}$ .  
Remarkably, \eqref{eq:IIB_Vfluxquad} displays the quadratic dependence on the background fluxes ${\bf f}$, with the positive semi-definite matrix\footnote{Notice that the matrix $\mathcal{T}$ defined in \eqref{eq:IIB_Tmat} differs by a sign in comparison with the matrix defined in (3.17) of \cite{Grimm:2022xmj}, namely $\mathcal{T}^{\text{(here)}} = -\mathcal{T}^{\text{(there)}}$.}
\begin{equation}
	\label{eq:IIB_Tmat}
	\mathcal{T}(a,s) = -\eta\, \mathbb{T}\, \eta \,, \qquad \text{where}\,\quad \mathbb{T}^{\mathcal{I}\mathcal{J}} := 2\, e^{K} {\rm Re}\left(K^{i \bar\jmath}_{\rm cs} D_i \Pi^{\mathcal{I}} \bar D_{\bar\jmath} \bar\Pi^{\mathcal{J}} + \Pi^{\mathcal{I}} \bar\Pi^{\mathcal{J}}\right) \,.
\end{equation}

\section{The Hodge-theoretical origin of the type IIB scalar potential}
\label{sec:Hodge}

In this appendix we overview the main aspects of Hodge Theory that have been employed in the main text. In particular, we highlight how the asymptotic Hodge Theory can be exploited to compute the scalar potential towards specific field space boundaries, such as those listed in Section~\ref{sec:IIB_Vbound}.
For more detailed reviews on the subject, we refer to the works \cite{Grimm:2018ohb,Grimm:2018cpv,Grimm:2019ixq,Grimm:2020ouv,Grimm:2021ckh,Grimm:2022sbl};
here, we will be mostly following \cite{Green:2008}, and the notations and conventions of \cite{Grimm:2022xmj}.

\subsection{An overview of Hodge Theory}

Within the four-dimensional type IIB effective field theories under investigation in this work, the asymptotic Hodge theory offers powerful tools to estimate the behavior of physical quantities towards \emph{singularities}, or \emph{boundaries} of the complex structure moduli space $\mathcal{M}_{\rm cs}$.
In analogy with the analysis of Section~\ref{sec:IIB_Vgen}, we will focus on the case where ${\rm dim}_{\mathbb{C}} \mathcal{M}_{\rm cs} = 1$, and we denote with $z$ the complex coordinate spanning the complex structure moduli space. 
We shall assume that the boundary of interest is the locus $\zeta = 0$ and, as in the main text, it is convenient to further introduce the coordinate
\begin{equation}
    \label{eq:AppHT_t}
    z := a + \im s = \frac{1}{2 \pi \im} \log \zeta\,,
\end{equation}
where the real coordinates $a$ and $s$ denote, respectively, the axion and the saxion field.
Additionally, we assume that the axion spans the domain $0 \leq a < 1$, with the identification $a \sim a + 1$, with the axion acquiring values $s > 0$. 
Then, in these new coordinates, the aformentioned boundary is reached as $z \to \im \infty$, namely as $s \to \infty$ irrespective of the axion value.

Each boundary is distinguished by a characteristic \emph{monodromy matrix} $T$ which, in the one-dimensional case under examination, is a $4 \times 4$ matrix. 
The monodromy matrix $T$ can be decomposed into a semi-simple part $T^{(s)}$ of finite order and a unipotent part of infinite order $T^{(u)}$ as $T = T^{(s)} T^{(u)}$, with $T^{(s)}$ and $T^{(u)}$ such that
\begin{equation}
	\label{eq:AppHT_Tus}
	\begin{aligned}
		&(T^{(s)})^{m-1} \neq \mathds{1} \,, &\qquad&  (T^{(s)})^{m} = \mathds{1}\,,
		\\
		&(T^{(u)} - \mathds{1})^{n} \neq 0\,, &\qquad&  (T^{(u)} - \mathds{1})^{n} = 0\,,
	\end{aligned}
\end{equation}
for some $m, n \in \mathbb{N}$. 
In turn, the semi-simple part defines the \emph{log-monodromy matrix} ias follows
\begin{equation}
	\label{eq:AppHT_N}
	N := \frac{1}{m} \log T^{m} = \log T^{(u)} \,.
\end{equation}
As an immediate consequence of its definition \eqref{eq:AppHT_N}, the log-monodromy matrix $N$ is nilpotent, obeying $N^n \neq 0$, $N^{n+1} = 0$ for some $n \in \mathbb{N}$; for the boundaries examined in this work, $0 \leq n \leq 3$.

In addition to the ingredients just introduced, in \cite{MR0382272,MR840721} finer Hodge-theoretical structures were introduced to better capture the behavior of key quantities towards the moduli space boundaries.
Consider the middle cohomology $H^3(Y, \mathbb{C})$ of the Calabi-Yau three-fold $Y$, within which the periods \eqref{eq:IIB_Pi_sym} reside. This can be decomposed into Dolbeault cohomology groups as 
\begin{equation}
	\label{eq:AppHT_Hdec}
	H^3(Y, \mathbb{C}) = \bigoplus\limits_{p=0}^3 H^{3-p,p} = H^{3,0} \oplus H^{2,1} \oplus H^{1,2} \oplus H^{0,3}\,,
\end{equation}
with $\overline{H^{p,q}} = H^{q,p}$.
The decomposition \eqref{eq:AppHT_Hdec} constitutes a \emph{pure} Hodge structure, that is however not enough to describe how the periods change across the moduli space.
Indeed, in \cite{MR0382272,MR840721} it has been proposed to rather consider \emph{mixed} Hodge structures.
These rely on the following filtration, defined out of the Dolbeault cohomology groups \eqref{eq:AppHT_Hdec}:
\begin{equation}
	\label{eq:AppHT_Fdec}
	\begin{aligned}
		&F^3 = H^{3,0}\,,  &\qquad&  F^2 = H^{3,0} \oplus H^{2,1}\,,
		\\
		&F^1 = H^{3,0} \oplus H^{2,1} \oplus H^{1,2}\,,  &\qquad&  F^0 = H^{3,0} \oplus H^{2,1} \oplus H^{1,2} \oplus H^{0,3}\,,
	\end{aligned}	
\end{equation}
Such a filtration can then be decomposed according to the \emph{Deligne splitting} as
\begin{equation}
    \label{eq:AppHT_Fdec_I}
    F^p = \bigoplus_{r\geq p} \bigoplus_s I^{r,s}\,,
\end{equation}
with $I^{r,s}$ introducing a finer splitting with respect to the decomposition \eqref{eq:AppHT_Hdec}.
In this work we will not enter the details about how to explicitly compute the splitting \eqref{eq:AppHT_Fdec_I}.

In general, the periods \eqref{eq:IIB_Pi_sym} can acquire complicated expressions in terms of the moduli across $\mathcal{M}_{\rm cs}$.
However, close to a moduli space boundary, \emph{Schimd’s nilpotent orbit theorem} holds \cite{MR0382272}.
This pivotal theoreom states that, close enough to a moduli space boundary, reached as $z \to \im \infty$, the filtration \eqref{eq:AppHT_Fdec} can be decomposed as
\begin{equation}
	\label{eq:AppHT_Fnil}
	F^p  = e^{z N} e^{\Gamma(\zeta)} F^p_0\,.
\end{equation}
Here, $\Gamma(\zeta)$ is the \emph{instanton map}, such that $\Gamma(0)=0$, and $F^p_0$ is the \emph{limiting Hodge filtration}.

The relation \eqref{eq:AppHT_Fnil} involves vector spaces, and focusing on the case $p = 3$ tells how the periods \eqref{eq:IIB_Pi_sym} rearrange towards a moduli space boundary.
Indeed, defining ${\bf a}_0 \in F^p_0$, the periods \eqref{eq:IIB_Pi_sym} can be expanded
\begin{equation}
	\label{eq:AppHT_Per_nil}
	{\bm \Pi}  = e^{z N} e^{\Gamma(\zeta)} {\bf a}_0 = e^{z N} \left({\bf a}_0 + \zeta {\bf a}_{1}  + \zeta^2 {\bf a}_{2} + \ldots \right) \simeq e^{z N} {\bf a}_0\equiv {\bm \Pi}_{\rm nil}\,,
\end{equation}
where the leading part constitutes the nilpotent orbit-approximated periods ${\bm \Pi}_{\rm nil}$; with respect to ${\bm \Pi}_{\rm nil}$, other terms are exponentially suppressed in the saxion.
Moreover, we notice that the computation of ${\bm \Pi}_{\rm nil}$ only requires the log-monodromy matrix $N$ and the vector ${\bf a}_0$ enter, being the only boundary data needed.

\subsection{The near-boundary form of the \texorpdfstring{$\mathcal{N}=1$}{N=1} type IIB scalar potential}

Schimd’s nilpotent orbit theorem \eqref{eq:AppHT_Fnil} does not solely help in the estimation of periods \eqref{eq:IIB_Pi_sym} in the near-boundary regime. 
Indeed, the theorem \eqref{eq:AppHT_Fnil} is useful for the estimation of the asymptotic behavior of \emph{Hodge norms} as well.
Given $\omega \in H^3(Y,\mathbb{C})$, its Hodge norm is defined as 
\begin{equation}
\label{eq:AppHT_norm}
    \| \omega \|^2 := \int_Y \omega \wedge \star \,\bar{\omega}\,.
\end{equation}

In order to arrive at estimations for the norms \eqref{eq:AppHT_norm}, we need to introduce additional structures.
Firstly, it is crucial to notice that the nilpotent matrix \eqref{eq:AppHT_N} may be regarded as a part of an ${\rm sl}(2)$-triple
\begin{equation}
	\label{eq:AppHHT_sl2_triples}
	\{N := N^-, N^+, N^0\}\in\mathfrak{sp}(2h^{2,1}+2,\mathbb{R})\,,
\end{equation}
for which the log-monodromy matrix \eqref{eq:AppHT_N} plays the role of lowering operator.
Accordingly, one can `rotate' the Deligne splitting \eqref{eq:AppHT_Fdec_I} in such a way that the rotated
groups $\tilde{I}^{p,q}$ are eigenspaces for the $N^0$-operator:\footnote{This procedure is shown in \cite{Green:2008,Grimm:2022xmj} for the boundaries examined in Section~\ref{sec:IIB_Vbound}. In particular, there it has been illustrated how to perform such a rotation while maintaining the integrality of the elements of the log-monodromy matrix.}
\begin{equation}
    N^0\tilde{I}^{p,q}=(p+q-3)\tilde{I}^{p,q}.
\end{equation}

Now, consider the region $\Sigma = \{ |a| < \delta,\ \  s > 1 \}$, encompassing the moduli space boundary. 
It can be shown that, in this region, the Hodge norm \eqref{eq:AppHT_norm} behaves as \cite{MR840721}
\begin{equation}
\label{eq:AppHT_norm_growth}
    \| \omega \|^2 \sim s^{\ell} \,,
\end{equation}
for large saxion $s$, and where $\ell = p+q-3$ is the ${\rm sl}(2)$-weight of the form $\omega$.

The bilinear leading to the norm \eqref{eq:AppHT_norm} can be concretely constructed as follows.
Firstly, let us rewrite the norm \eqref{eq:AppHT_norm} as follows
\begin{equation}
\label{eq:AppHT_norm_b}
    \| \omega \|^2 =: {\bm \omega}^T \mathcal{T} {\bm \omega}\,.
\end{equation}
for some operator $\mathcal{T}$, and ${\bm \omega}$ denoting the vector components of the form $\omega$. 
Furthermore, it can be shown that the operator $\mathcal{T}$ can be written as $\mathcal{T}  = \eta C$,
where $C$ is the \emph{Weil operator} $C$.
The latter acts on the symplectic basis $\{\alpha_I, \beta^I\}$ introduced in Section~\ref{sec:IIB_summ} as
\begin{equation}
	\label{eq:AppHT_Csymb}
	\star \begin{pmatrix}
		\alpha_I \\ \beta^I 
	\end{pmatrix} = C \begin{pmatrix}
	\alpha_I \\ \beta^I
\end{pmatrix}\,.
\end{equation}
Importantly, for large enough saxion $s$, the Weil operator can be efficiently approximated by employing the ${\rm sl}(2)$-orbit approximation.
This approximation scheme generically holds for larger values of the saxion with respect to the nilpotent orbit approximation, and it differs from the latter by subleading, polynomial terms \cite{Grimm:2021ckh}.
Within such an approximation scheme, the ${\rm sl}(2)$-orbit approximated Weil operator is given by
\begin{equation}
	\label{eq:AppHT_Csl2}
	C_{\mathfrak{sl}(2)} = e^{a N^-} e(s)^{-1} C_\infty e(s) e^{-a N^-}  \,, \qquad {\rm with}\quad e(s) := \exp \left(\frac12 \log s\, N^0 \right)\,,
\end{equation}
with the operator $C_{\infty}$ acting on the Dolbeaut cohomology as $C_{\infty}\omega^{p,q}=\im^{p-q}\omega^{p,q}$.
Hence, the computation of the approximated operator \eqref{eq:AppHT_Csl2} solely relies on the boundary data.
Then, consequently, the matrix $\mathcal{T}$ introduced in \eqref{eq:IIB_Tmat} and defining the Hodge norm \eqref{eq:AppHT_norm} gets approximated as $\mathcal{T} _{\mathfrak{sl}(2)} = \eta C_{\mathfrak{sl}(2)}$.

As has been observed in several works, such as \cite{Grimm:2018ohb,Grimm:2018cpv,Grimm:2019ixq,Grimm:2020ouv,Grimm:2021ckh,Grimm:2022sbl,Grimm:2022xmj}, many key physical quantities appearing in string theory-originated effective theories can be understood as Hodge norms.
In particular, within the four-dimensional type IIB effective theories described in Section~\ref{sec:IIB_summ}, the scalar potential \eqref{eq:IIB_Vgen} offers a natural interpretation as a Hodge norm.
In fact, recalling its microscopic origin in terms of a RR three-form $F_3$ defined in \eqref{eq:IIB_WGVW}, the scalar potential can be written as a Hodge norm for the generating flux vector ${\bf f}$:
\begin{equation}
	\label{eq:AppHT_Vflux}
	V_{\bf f} = \frac12 M^4_{\rm P} e^{\hat{K}}\int_Y F_3 \wedge \star F_3 = \frac12 M^4_{\rm P} e^{\hat{K}} \| {\bf f} \|^2\,.
\end{equation}
As such, for large saxion $s$ and finite value of the axion $a$, its behavior can be estimated via \eqref{eq:AppHT_norm_growth}, with $\ell$ being the ${\rm sl}(2)$-weight of the flux vector ${\bf f}$.

\bibliographystyle{jhep}
\bibliography{references.bib}

\end{document}